\documentclass[12pt]{article}

\usepackage[margin=1in]{geometry}

\usepackage{setspace}

\usepackage{graphicx}

\usepackage{enumitem}

\usepackage{microtype}
\usepackage{subfigure}
\usepackage{booktabs}
\usepackage{comment}
\usepackage{wrapfig}
\usepackage{arydshln}
\usepackage{enumitem}
\usepackage{multirow}
\usepackage{url}
\usepackage{natbib}
\usepackage{authblk}

\RequirePackage[colorlinks,citecolor=blue,linkcolor=blue,urlcolor=blue,pagebackref]{hyperref}

\usepackage{amsmath}
\usepackage{amssymb}
\usepackage{mathtools}
\usepackage{amsfonts}
\usepackage{bm}
\usepackage{bbm}

\usepackage{algorithm}
\usepackage{algorithmic}

\def\eqref#1{equation~\ref{#1}}

\def\1{\bm{1}}

\def\rmd{{\mathrm{d}}}

\DeclareMathAlphabet{\mathsfit}{\encodingdefault}{\sfdefault}{m}{sl}
\SetMathAlphabet{\mathsfit}{bold}{\encodingdefault}{\sfdefault}{bx}{n}

\def\calF{{\mathcal{F}}}

\def\calN{{\mathcal{N}}}

\def\calR{{\mathcal{R}}}

\def\calT{{\mathcal{T}}}

\def\calX{{\mathcal{X}}}
\def\calY{{\mathcal{Y}}}
\def\calZ{{\mathcal{Z}}}

\def\bbE{{\mathbb{E}}}

\def\bbP{{\mathbb{P}}}

\newcommand{\R}{\mathbb{R}}

\newcommand{\Var}{\mathrm{Var}}

\newcommand{\p}[1]{\left(#1\right)}
\newcommand{\sqb}[1]{\left[#1\right]}

\newcommand{\bigsqb}[1]{\big[#1\big]}

\newcommand{\Bigp}[1]{\Big(#1\Big)}
\newcommand{\Bigsqb}[1]{\Big[#1\Big]}

\newcommand{\Biggsqb}[1]{\Bigg[#1\Bigg]}

\usepackage{amsthm}

\theoremstyle{plain}

\newtheorem{theorem}{Theorem}[section]
\newtheorem{lemma}[theorem]{Lemma}

\newtheorem{corollary}[theorem]{Corollary}
\newtheorem{proposition}[theorem]{Proposition}

\newtheorem{assumption}{Assumption}[section]

\newtheorem*{remark}{Remark}

\usepackage[textsize=tiny]{todonotes}
\usepackage{multirow}
\usepackage{wrapfig}
\usepackage{subfigure}

\renewcommand{\eqref}[1]{(\ref{#1})}

\usepackage{xcolor}
\newcount\Comments  
\newcommand{\kibitz}[2]{\ifnum\Comments=1\textcolor{#1}{#2}\fi}

\usepackage{enumitem}
\usepackage{subfigure}

\usepackage[capitalize,noabbrev]{cleveref}

\allowdisplaybreaks

\title{ScoreMatchingRiesz: Score Matching for Debiased Machine Learning and Policy Path Estimation}

\author{Masahiro Kato\thanks{Email: \texttt{mkato-csecon@g.ecc.u-tokyo.ac.jp}}$\,$}

\affil{The University of Tokyo}

\date{\today}

\begin{document}

\maketitle 

\begin{abstract}
We propose \emph{ScoreMatchingRiesz}, a family of Riesz representer estimators based on score matching. The Riesz representer is a key nuisance component in debiased machine learning, enabling $\sqrt{n}$-consistent and asymptotically efficient estimation of causal and structural targets via Neyman-orthogonal scores. We formulate Riesz representer estimation as a score estimation problem. This perspective stabilizes representer estimation by allowing us to leverage denoising score matching and telescoping density ratio estimation. We also introduce the policy path, a parameter that captures how policy effects evolve under continuous treatments. We show that the policy path can be estimated via score matching by smoothly connecting average marginal effect (AME) and average policy effect (APE) estimation, which improves the interpretability of policy effects.
\end{abstract}

\section{Introduction}
Estimating the Riesz representer is a central problem in debiased machine learning for causal inference and structural parameter estimation \citep{Chernozhukov2022automaticdebiased}. For example, in average treatment effect (ATE) estimation, the Riesz representer corresponds to inverse propensity score weights, and accurate estimation of this representer is essential for accurate ATE estimation. \emph{Score matching} refers to methods for estimating the score, the derivative of a log density with respect to a variable. Score matching has recently attracted renewed attention through the diffusion model literature \citep{Song2020generativemodeling}.

Riesz representer estimation and score matching are closely related, but leveraging this link for debiased machine learning requires more than simply importing existing score estimators. For example, in average marginal effect (AME) estimation, the Riesz representer coincides with the (negative) score of the joint density of $(D,Z)$ with respect to the treatment coordinate $D$ given covariates $Z$. Moreover, Hyv\"arinen score matching \citep{Hyvarinen2005estimationof} yields an objective that is closely related to Riesz regression \citep{Chernozhukov2021automaticdebiased}. However, Hyv\"arinen-type objectives can be numerically challenging in high dimensions because they involve divergence computations. In applications where the representer is written in terms of density ratios \citep{Kato2025directbias}, one may instead turn to density ratio estimation (DRE), but flexible DRE is known to suffer from severe overfitting \citep{Kato2021nonnegativebregman,Rhodes2020telescopingdensityratio}. While denoising score matching (DSM) \citep{Vincent2011aconnection} and DRE-$\infty$ \citep{Choi2022densityratio} have been developed to stabilize score and ratio estimation, their use as \emph{Riesz representer estimators} within Neyman orthogonal scores calls for an explicit construction and an analysis.

Motivated by these relationships, this paper develops score matching-based methods for Riesz representer estimation, which we call \emph{ScoreMatchingRiesz}. The key idea is to use score models as a Riesz representer estimation module: we construct the Riesz representer either directly from a score field or via density ratios recovered by integrating a score along a bridge of intermediate distributions. Our framework involves two types of scores. The \emph{data score} is the derivative of the log density with respect to the data variable, and it is estimated by diffusion-style score matching. The \emph{time score} is the derivative of the log density along a continuum of intermediate distributions, and it is estimated via infinitesimal classification-based score matching (the continuum limit underlying DRE-$\infty$). We refer to ScoreMatchingRiesz designed for data-score estimation as Data-ScoreMatchingRiesz and that designed for time-score estimation as Time-ScoreMatchingRiesz.

In AME estimation, the Riesz representer can be estimated directly from the data score. In other applications such as APE and ATE estimation, however, estimating the Riesz representer requires estimating density ratios, and time-score estimation provides a general way to recover these ratios through a bridge between endpoint distributions. In some restricted settings, the data-score construction is sufficient on its own. For example, in APE estimation under a known pushforward policy, the relevant density ratio can be expressed as an integral of the data score along a known path, which yields the representer without explicitly learning an endpoint-to-endpoint ratio. Joint training of data and time scores can further stabilize the estimation process across these regimes. See Table~\ref{tab:comparison}.

This score-based representer construction also provides a concrete link between AME and APE through a continuum of policy effects. In this study, we define the \emph{policy path} as a parameter of interest and propose an estimator. In shift-policy settings, the APE representer can be constructed from the AME representer by integrating the score along the policy path. Using this relationship, we estimate intermediate policy effects as the policy shifts continuously.

\begin{table}[t]
    \caption{Specific forms of ScoreMatchingRiesz (SMR).}
    \label{tab:comparison}
    \centering
    \scalebox{1.0}{
    \begin{tabular}{c|c|c|c|c}
        & \multirow[c]{2}{*}{ATE} & \multirow[c]{2}{*}{AME} & APE with a known  & APE with a non-\\
        & &  & pushforward policy & pushforward policy \\
        \hline
        Data-SMR &  & \checkmark & \checkmark & \\
        Time-SMR & \checkmark & \checkmark & \checkmark & \checkmark \\
    \end{tabular}
    }
\end{table}

In summary, our contributions are as follows:
\begin{itemize}[topsep=0pt, itemsep=0pt, partopsep=0pt, leftmargin=*]
    \item We develop \emph{ScoreMatchingRiesz}, a unified score matching-based framework that estimates Riesz representers from data and time scores, with three implementations: Data- and Time-ScoreMatchingRiesz.
    \item We generalize score matching as a variant of generalized Riesz regression based on Bregman divergence minimization and provide an interpretation of the loss and model selection from a covariate-balancing viewpoint.
    \item We define the policy path as a parameter of interest that bridges AME and APE, and we propose an estimation method based on score integration along the path.
    \item We establish convergence rate results for the proposed Riesz representer estimators, which are sufficient for debiased machine learning via Neyman orthogonal scores.
\end{itemize}
There are additional contributions, mainly in the appendices, due to space constraints. For example, we discuss when AME and APE are seamlessly connected via scores. In Appendix~\ref{app:empirical_finance}, we provide an application of our method to the local projection in economic analysis \citep{Jorda2005estimationand}.

\paragraph{Related work.}
Constructing (asymptotically) efficient estimators is a central task in causal inference \citep{VanderVaart1998asymptoticstatistics}, and various unified perspectives on this task have been established. Among them, the debiased machine learning framework formulates the problem through Neyman orthogonal score equations together with cross-fitting \citep{Chernozhukov2018doubledebiased}. Various methods for Riesz representer estimation have been proposed, including Riesz regression \citep{Chernozhukov2021automaticdebiased}, nearest neighbor matching \citep{Kato2025nearestneighbor,Lin2023estimationbased}, and covariate balancing \citep{Zhao2019covariatebalancing,BrunsSmith2025augmentedbalancing}.

The goal of DRE is to estimate the ratio between two probability density functions. Many methods have been proposed for DRE \citep{Sugiyama2012densityratio}, but flexible DRE is known to suffer from severe overfitting, called train loss hacking \citep{Kato2021nonnegativebregman} or the density-chasm \citep{Rhodes2020telescopingdensityratio}. One approach to mitigating this overfitting is to avoid estimating a single log density ratio in one step and instead introduce a sequence of intermediate density ratios \citep{Rhodes2020telescopingdensityratio}. Building on this line of work, \citet{Choi2022densityratio} proposes an infinitesimal classification approach, which takes a continuum limit of telescoping ratios and learns the resulting score representation via score matching. We use this idea through Time-ScoreMatchingRiesz as a representer-estimation primitive when the Riesz representer is ratio-based.

As we and prior work note \citep{Kato2025directbias,Kato2026rieszrepresenter,Benmichael2021balancingact}, Riesz representer estimation admits complementary views via score matching and DRE. As a consequence, the instability issues known for Hyv\"arinen score matching and the overfitting issues documented for flexible DRE can also arise in Riesz representer estimation. We make these connections operational by constructing Riesz representers from data scores and time scores, and by providing convergence guarantees in the form needed for debiased machine learning.

For detailed related work, see Appendix~\ref{appdx:related}.

\section{Setup}
Let $W=(X,Y)$ consist of a regressor $X\in\calX\subseteq \R^d$ and a scalar outcome $Y\in\calY\subseteq \R$. Assume $(X,Y)\sim P_0$, and define the regression function $\gamma_0(x)\coloneqq \bbE[Y\mid X=x]$. We observe $\{(X_i,Y_i)\}_{i=1}^n$, where $(X_i,Y_i)$ are i.i.d.  copies of $(X,Y)$. Our target parameter is
\[
\theta_0 \coloneqq \bbE\bigsqb{m(W,\gamma_0)},
\]
where $m(W,\gamma)$ is a functional of the data and a regression function $\gamma\colon\calX\to\calY$. We assume that the map $\gamma \mapsto \bbE\bigsqb{m(W,\gamma)}$ 
is \emph{linear and continuous} on $L_2(P_X)$ (the $L_2$ space under the marginal law of $X$).

\subsection{Riesz representer and Neyman orthogonal score}
By the Riesz representation theorem, there exists a function $\alpha_0\in L_2(P_X)$ such that, for all $\gamma\in L_2(P_X)$, $\bbE[m(W,\gamma)] = \bbE[\alpha_0(X)\gamma(X)]$ holds. 
We call $\alpha_0$ the \emph{Riesz representer} associated with $m$. Let $\eta_0\coloneqq(\gamma_0,\alpha_0)$ denote the nuisance parameters. For a candidate pair $\eta=(\gamma,\alpha)$ and scalar $\theta$, define the Neyman orthogonal score $\psi(W;\eta,\theta) \coloneqq m(W,\gamma) + \alpha(X)\bigl(Y-\gamma(X)\bigr) - \theta$. 
Then $\bbE[\psi(W;\eta_0,\theta_0)]=0$ and the moment condition is orthogonal to first-order perturbations of $\eta$ at $\eta_0$ under regularity conditions \citep{Chernozhukov2018doubledebiased}.

\subsection{Estimation of the parameter of interest}
An estimator $\widehat{\theta}$ is obtained by plugging in nuisance estimates $\widehat{\eta}=(\widehat{\gamma},\widehat{\alpha})$ and solving 
\[\frac{1}{n}\sum_{i=1}^n \psi\p{W_i;\widehat{\eta},\widehat{\theta}} = 0;\] 
equivalently, the estimator is given by
\[
\widehat{\theta}
=
\frac{1}{n}\sum_{i=1}^n \Bigp{m(W_i,\widehat{\gamma})+\widehat{\alpha}(X_i)(Y_i-\widehat{\gamma}(X_i))}.
\]
Assume that $\widehat{\eta}$ satisfies the Donsker condition or is constructed via cross-fitting. Then, under suitable convergence rate conditions on $\widehat{\eta}$, we have $\sqrt{n}\p{\widehat{\theta}-\theta_0} \xrightarrow{d} \calN(0,V)$, where $V > 0$ is the efficient asymptotic variance. 

\paragraph{Notation and assumptions.}
For simplicity, we assume that the components of $X$ are continuous, except in specific cases such as the treatment $D$ in ATE estimation. We also assume that each continuous random variable $R$ admits a probability density function $p_0(r)$ under $P_0$. We write $R\sim p_0$. Expectations are taken under $P_0$ unless stated otherwise; when needed, we write $\bbE_{R\sim p}[\cdot]$ for $R$ with density $p$. For a pdf $p_0(x)$, we define the data score as $s^{\text{data}}_0(x) \coloneqq \nabla_x\log p_0(x)$. For a one-parameter family of pdfs $\{p_t(x)\}_{t\in\calT}$, we define the time score as $s^{\text{time}}_t(x) \coloneqq \partial_t\log p_t(x)$. We use $t$, $\delta$, and $\sigma$ for indexing the bridge, policy-shift, and the Gaussian noise level, but they become mathematically equivalent under certain assumptions.

\subsection{Applications}
We illustrate how the Riesz representer appears in applications by presenting examples of AME and APE estimation. Note that our method can also be applied to more general cases such as ATE estimation (see Appendix~\ref{appdx:ate}).

\paragraph{AME estimation.}
Let $X=(D,Z)$ with continuous $D\in[-1,1]$. The AME is
\[
\theta^{\mathrm{AME}}_0 \coloneqq \bbE\bigl[\partial_d \gamma_0(D,Z)\bigr],
\quad
m^{\mathrm{AME}}(W,\gamma)\coloneqq \partial_d \gamma(D,Z).
\]
Under standard integration-by-parts conditions (including suitable boundary behavior), the Riesz representer is the negative score of the joint density of $(D,Z)$ with respect to $d$ and given as
\[
\alpha^{\mathrm{AME}}_0(D,Z)\coloneqq -\partial_d \log p_0(d,Z)\big|_{d=D}.
\]

\paragraph{APE estimation.}
Let $P_1$ and $P_{-1}$ be two counterfactual distributions of $X$ with densities $p_1(x)$ and $p_{-1}(x)$, both absolutely continuous with respect to $P_0$ (density $p_0$). The APE is
\[
\theta^{\mathrm{APE}}_0 \coloneqq \bbE_{X\sim p_1}[\gamma_0(X)]-\bbE_{X\sim p_{-1}}[\gamma_0(X)].
\]
This can be written as $\theta^{\mathrm{APE}}_0=\bbE[m^{\mathrm{APE}}(W,\gamma_0)]$ with $m^{\mathrm{APE}}(W,\gamma)\coloneqq \gamma(X) \frac{p_1(X)-p_{-1}(X)}{p_0(X)}$. 
Hence, the Riesz representer is given as
\[
\alpha^{\mathrm{APE}}_0(X)\coloneqq \frac{p_1(X)-p_{-1}(X)}{p_0(X)}.
\]

\section{Recaps of Riesz Regression, Score Matching, and DRE-\texorpdfstring{$\infty$}{infinity}}
\label{sec:dreic}
This section reviews Riesz regression \citep{Chernozhukov2021automaticdebiased}, score matching \citep{Song2021denoisingdiffusion}, and DRE-$\infty$ \citep{Choi2022densityratio}. Score matching can be carried out via Hyv\"arinen score matching or denoising score matching (DSM). In AME estimation, Riesz regression coincides with Hyv\"arinen score matching and can also be interpreted as a special case of LSIF. Accordingly, we first introduce Riesz regression (Hyv\"arinen score matching) and then DSM and DRE-$\infty$.

\subsection{Definitions of Scores}
\label{subsec:score_matching}
Throughout this subsection, let $X\in\R^d$ be a continuously distributed random vector with density $p_0(x)$ under $P_0$.
Recall that we defined the data score of $p_0$ as
\[
\text{(Data score)}\ \ s^{\text{data}}_0(x)\coloneqq \nabla_x\log p_0(x)\in\R^d.
\]
The Riesz representer in AME estimation directly corresponds to the data score as
$\alpha^{\mathrm{AME}}_0(D,Z)=-\partial_d\log p_0(D,Z)$. 

We introduce a parametric family $\{p_t(x)\}_{t\in\calT}$ such that $p_t(x)=p_0(x)$ at $t=0$. This family is specified by the researcher based on applications. We refer to the parameter $t$ as ``time.'' Using this family, we define the time score as
\[
\text{(Time score)}\ \ s^{\text{time}}_t(x)\coloneqq \partial_t \log p_t(x)\in\R.
\]

\subsection{Hyv\"arinen Score Matching Equals Riesz Regression and LSIF in AME Estimation}
We first consider the estimation of the data score $s^{\text{data}}_0$. 
Let $s^{\text{data}}\colon\calX\to\R^d$ be a candidate score function (e.g., a neural network $s^{\text{data}}_{\theta}$).
A natural population criterion is the mean squared error (MSE) between $s$ and the true score $s^{\text{data}}_0$: $\calR^\dagger(s^{\text{data}}) \coloneqq \frac12\bbE_{X\sim p_0}\bigsqb{\|s^{\text{data}}(X)-s^{\text{data}}_0(X)\|_2^2}$. 
This objective is not directly computable because $s^{\text{data}}_0$ is unknown. However, under standard
integration-by-parts conditions, minimizing $\calR^\dagger$ is equivalent to minimizing
\begin{align}
\label{eq:hyvarinen_sm}
&\calR^{\text{HSM}}(s^{\text{data}})
 \coloneqq 
\bbE_{X\sim p_0}\sqb{\frac12\| s^{\text{data}}(X)\|_2^2+\nabla_x\cdot s^{\text{data}}(X)},
\end{align}
where $\nabla_x\cdot s^{\text{data}}(x)\coloneqq \sum_{j=1}^d \partial_{x_j}s^{\text{data}}_j(x)$.
This estimation approach is called Hyv\"arinen score matching, a special case of LSIF \citep{Kanamori2009aleastsquares} and Riesz regression \citep{Chernozhukov2021automaticdebiased} (See Appendix~\ref{appdx:hsm}).  

\subsection{DSM}
In high dimensions, directly optimizing \eqref{eq:hyvarinen_sm} can be numerically challenging
because it involves the divergence $\nabla_x\cdot s_{\theta}(X)$, which requires Jacobian-trace computations.
A widely used alternative is DSM \citep{Vincent2011aconnection},
which replaces the (possibly rough) target score $s^{\text{data}}_0$ by the score of a \emph{Gaussian-smoothed} density. We explain the details in Section~\ref{sec:datasmr}. In this study, we refer to DSM-based Riesz representer estimation as Data-ScoreMatchingRiesz.

\subsection{DRE-\texorpdfstring{$\infty$}{infty}}
Let $q(x)$ and $p(x)$ be two densities on $\calX$, and assume $q$ is absolutely continuous with respect to $p$ on $\calX$. Define the density ratio as
$r(x)\coloneqq q(x) / p(x)$. 
Assume we can draw i.i.d.  samples from $q$ and $p$.

\paragraph{Bridge distributions and telescoping ratios.}
Following \citet{Choi2022densityratio}, construct a continuum of \emph{bridge} distributions $\{p_t\}_{t\in[0,1]}$ by sampling
\[
X_t \coloneqq \beta^{(1)}(t)X_0+\beta^{(2)}(t)X_1,
\]
where $X_0\sim q$, $X_1\sim p$ are independent, and $\beta^{(1)},\beta^{(2)}:[0,1]\to[0,1]$ are $C^2$ functions satisfying $\beta^{(1)}(0)=1$, $\beta^{(2)}(0)=0$, $\beta^{(1)}(1)=0$, $\beta^{(2)}(1)=1$. Let $p_t$ denote the law of $X_t$ (and write $p_t(x)$ for its density when it exists).

Using intermediate ratios, we have for any integer $T\ge1$, $r(x)=\frac{q(x)}{p(x)}=\prod_{k=1}^T \frac{p_{(k-1)/T}(x)}{p_{k/T}(x)}$. 
Taking logs gives $\log r(x)=\sum_{k=1}^T \log \frac{p_{(k-1)/T}(x)}{p_{k/T}(x)}$. 
In the continuum limit, the following result holds:

\begin{proposition}[Continuum limit of telescoping ratios; c.f. Proposition~1 in \citet{Choi2022densityratio}]
As $T\to\infty$, we have
\[
\log r(x)
=
\int_{1}^{0} \partial_t \log p_t(x) \rmd t
=
-\int_{0}^{1} \partial_t \log p_t(x) \rmd t.
\]
\end{proposition}

The time score is $s^{\text{time}}_t(x)=\partial_t\log p_t(x)$.

\paragraph{Time score matching objective.}
Let $s^{\text{time}}_{\theta, t}(x)$ be a time score model. A natural population loss is $\calR^\dagger(s^{\text{time}}_{\theta}) \coloneqq
\int_0^1
\bbE_{X_t\sim p_t}\left[
\lambda(t)\Bigl(\partial_t\log p_t(X_t)-s^{\text{time}}_{\theta, t}(X_t)\Bigr)^2
\right]\rmd t$, 
where $\lambda:[0,1]\to\R_+$ is a weighting function. The true time score is unknown, but \citet{Choi2022densityratio} shows that (up to an additive constant that does not depend on $\theta$) this objective is equivalent to an integration-by-parts objective:
\begin{align}
\label{eq:dreinfty}
\calR(s^{\text{time}}_{\theta})\coloneqq &\bbE_{X_0\sim q}\left[\lambda(0) s^{\text{time}}_{\theta,0}(X_0)\right]
-
\bbE_{X_1\sim p}\left[\lambda(1) s^{\text{time}}_{\theta,1}(X_1)\right]\\
&\ \ \ + \int_0^1
\bbE_{X_t\sim p_t}\left[
\partial_t\bigl(\lambda(t) s^{\text{time}}_{\theta,t}(X_t)\bigr)
+\frac{1}{2}\lambda(t) s^{\text{time}}_{\theta,t}(X_t)^2
\right]\rmd t.\nonumber
\end{align}
In the integral term, $\partial_t s^{\text{time}}_{\theta, t}(x)$ denotes the \emph{partial} derivative with respect to $t$ while holding $x$ fixed. 

\paragraph{Recovering the density ratio.}
After fitting $s^{\text{time}}_{\theta,t}$, the ratio can be estimated by numerical integration as $\widehat{\log r}(x) \coloneqq \int_{1}^{0} s^{\text{time}}_{\theta, t}(x) \rmd t$ and $\widehat{r}(x)\coloneqq \exp\p{\widehat{\log r}(x)}$. 

\section{Data-ScoreMatchingRiesz}
\label{sec:datasmr}
We first define Data-ScoreMatchingRiesz, which is an application of DSM. 
Let $\sigma>0$ and define a noising kernel
\[
q_\sigma(\tilde x\mid x)=\calN(\tilde x;x,\sigma^2 I_d),
\quad
\tilde X = X + \sigma \varepsilon,\ \ \varepsilon\sim\calN(0,I_d).
\]
Let $p_\sigma$ denote the marginal density of $\tilde X$, i.e. the convolution
$p_\sigma = p_0 * \calN(0,\sigma^2 I_d)$.
Data-ScoreMatchingRiesz fits a model $s^{\text{data}}_{\theta, \sigma}(\tilde x)$ by minimizing
\[\calR^{\mathrm{DSM}\dagger}(s^{\text{data}}_{\theta, \sigma})
 \coloneqq 
\bbE_{\sigma\sim \pi}\bbE_{X\sim p_0}\bbE_{\tilde X\sim q_\sigma(\cdot\mid X)}\Bigsqb{
\lambda(\sigma)
\bigl\|
s^{\text{data}}_{\theta, \sigma}(\tilde X)
-\nabla_{\tilde x}\log q_\sigma(\tilde X\mid X)
\bigr\|_2^2},\] 
where $\pi$ is a distribution over noise levels and $\lambda(\sigma)\ge 0$ is a weighting function.
For Gaussian noise, $\nabla_{\tilde x}\log q_\sigma(\tilde x\mid x)
= -\frac{\tilde x-x}{\sigma^2}
= -\frac{\varepsilon}{\sigma}$ holds. 
Hence, an equivalent feasible form of $\calR^{\mathrm{DSM}\dagger}(s^{\text{data}}_{\theta, \sigma})$ is
\begin{align*}
&\calR^{\mathrm{DSM}}(s^{\text{data}}_{\theta, \sigma})
=
\bbE_{\sigma\sim \pi}\bbE_{X\sim p_0}\bbE_{\varepsilon\sim\calN(0,I_d)}
\Bigl[
\lambda(\sigma)
\bigl\|
s^{\text{data}}_{\theta, \sigma}(X+\sigma\varepsilon)+\varepsilon/\sigma
\bigr\|_2^2
\Bigr].\nonumber
\end{align*}
It is known that for each fixed $\sigma$, the population minimizer $\theta^*$ satisfies $s^{\text{data}}_{\theta^*, \sigma}(\cdot) \coloneqq \nabla_{\tilde x}\log p_\sigma(\cdot)$.

This is the formulation most commonly used in diffusion models \citep{Song2021denoisingdiffusion}:
one trains a single network on a continuum of noise levels $\sigma$, which can be interpreted as a ``time'' variable,
and then evaluates (or extrapolates) the score near the small-noise regime.

\section{Time-ScoreMatchingRiesz}
The (time) score matching approach above extends directly to Riesz representer estimation when the representer can be expressed through density ratios.

\begin{remark}[Densities and endpoint degeneracy]
    We use $p_t$ to denote the law of a bridge sample induced by an explicit sampling map. When this law is absolutely continuous, we write $p_t(x)$ for its density and define the time score $\partial_t\log p_t(x)$. Some bridge constructions can yield degenerate endpoint laws. This does not prevent evaluation of expectations, but it motivates either interpreting endpoint terms as limits or downweighting endpoints by choosing $\lambda$ with $\lambda(0)=\lambda(1)=0$ (or sampling $t$ from $[\varepsilon,1-\varepsilon]$).
\end{remark}

\subsection{AME estimation}
In AME estimation with continuous $D$, the Riesz representer is $\alpha^{\mathrm{AME}}_0(D,Z)= -\partial_d\log p_0(d,Z)\big|_{d=D}$. 
This Riesz representer can be interpreted as a time score.

\paragraph{Time score interpretation.}
For intuition, consider the affine bridge $D_t=\beta^{(1)}(t)+\beta^{(2)}(t)D$ with $\beta^{(1)}(0)=0$, $\beta^{(2)}(0)=1$, and derivatives $\partial_t\beta^{(1)}(0)=1$, $\partial_t\beta^{(2)}(0)=0$. Let $p_t(d,z)$ be the density of $(D_t,Z)$. A direct change-of-variables calculation yields, for $t$ where a density exists,
\begin{align*}
&\partial_t\log p_t(d,z)
=
-\frac{\partial_t\beta^{(2)}(t)}{\beta^{(2)}(t)}
-\frac{\partial_t\beta^{(1)}(t)+u \partial_t\beta^{(2)}(t)}{\beta^{(2)}(t)} 
\partial_u\log p_0(u,z)\Big|_{u=\frac{d-\beta^{(1)}(t)}{\beta^{(2)}(t)}}.
\end{align*}
Evaluating at $t=0$ gives $\partial_t\log p_t(d,z)\big|_{t=0}=-\partial_d\log p_0(d,z)$, 
so the AME representer can be viewed as a time score at $t=0$ for this local translation-like bridge. This observation connects AME estimation to the time-score framework used for ratio-based representers.

\subsection{APE estimation}
In APE estimation, the Riesz representer is the difference of two density ratios: $\alpha^{\mathrm{APE}}_0(x)=\frac{p_1(x)}{p_0(x)}-\frac{p_{-1}(x)}{p_0(x)}$. 
Thus it suffices to estimate $p_1/p_0$ and $p_{-1}/p_0$.

\paragraph{Bridge construction and time score training.}
Construct a bridge family $\{p_t\}_{t\in[-1,1]}$ for $X$ using independent draws $X_0\sim p_0$, $X_1\sim p_1$, $X_{-1}\sim p_{-1}$ and the same style of interpolation as in the ATE case:
\[
X_t \coloneqq
\begin{cases}
\beta^{(1)}(t) X_0+\beta^{(2)}(t) X_1, & t\in[0,1],\\[4pt]
\beta^{(1)}(-t) X_0+\beta^{(2)}(-t) X_{-1}, & t\in[-1,0].
\end{cases}
\]
We train $s^{\text{time}}_{\theta, t}(x)$ by minimizing
\begin{align*}
&\calR^{\mathrm{APE}}(s^{\text{time}}_{\theta})\coloneqq
\bbE_{X_{-1}\sim p_{-1}}\left[\lambda(-1) s^{\text{time}}_{\theta,-1}(X_{-1})\right] -
\bbE_{X_{1}\sim p_{1}}\left[\lambda(1) s^{\text{time}}_{\theta,1}(X_{1})\right] \\
&\ \ \ +
\int_{-1}^{1}\bbE_{X_t\sim p_t}\Biggsqb{
\partial_t\bigl(\lambda(t) s^{\text{time}}_{\theta,t}(X_t)\bigr) +\frac{1}{2}\lambda(t) s^{\text{time}}_{\theta,t}(X_t)^2}\rmd t.
\end{align*}

\paragraph{Constructing $\widehat{\alpha}^{\mathrm{APE}}$.}
After training, we obtain $\widehat{\log\frac{p_1}{p_0}}(x)\coloneqq \int_{0}^{1} s^{\text{time}}_{\theta, t}(x) \rmd t$ and $\widehat{\log\frac{p_{-1}}{p_0}}(x)\coloneqq \int_{0}^{-1} s^{\text{time}}_{\theta, t}(x) \rmd t$. Then, we estimate the Riesz representer as 
\[\widehat{\alpha}^{\mathrm{APE}}(x)
\coloneqq
\exp\p{\widehat{\log\frac{p_1}{p_0}}(x)}
-
\exp\p{\widehat{\log\frac{p_{-1}}{p_0}}(x)}.\] 

\subsection{Joint Training of the Scores}
To stabilize training, we jointly train the data score while training the time score, following the suggestion in \citet{Choi2022densityratio}, in a multitask learning manner. For details, see Appendix~\ref{app:joint-smr}.

\subsection{Estimation of the Parameter of Interest} 
The procedure for estimating the parameter of interest is as follows: (1) estimate the nuisance parameters $\eta_0 = (\gamma_0, \alpha_0)$; (2) plug in the estimates $\widehat{\eta} = (\widehat{\gamma}, \widehat{\alpha})$ into the Neyman score and solve for $\widehat{\theta}$ from $\frac{1}{n}\sum_{i=1}^n \psi\bigl(W_i;\widehat{\eta},\widehat{\theta}\bigr)=0$. 

If the estimators of the nuisance parameters do not satisfy the Donsker condition, we employ cross-fitting to control the empirical process term \citep{Klaassen1987consistentestimation,Zheng2011crossvalidatedtargeted,Chernozhukov2018doubledebiased}. 

\section{Generalization of ScoreMatchingRiesz}
\label{sec:generalization_smr}

ScoreMatchingRiesz admits a generalized formulation based on Bregman divergences, which makes the choice of the loss and the link explicit.
Under suitable choices, the implied ratio-based representer satisfies an \emph{automatic covariate balancing} property that directly reduces the error for the Neyman orthogonal score \citep{Kato2026rieszrepresenter,BrunsSmith2025augmentedbalancing}.
Due to space constraints, we focus on Time-ScoreMatchingRiesz; the same argument applies to Data-ScoreMatchingRiesz.

\subsection{Generalization via Bregman Divergence}
\label{subsec:bregman_time_score}

Let $g\colon\R\to\R$ be differentiable and strictly convex, and let
$D_g(u\|v)\coloneqq g(u)-g(v)-g'(v)(u-v)$
denote the Bregman divergence.
For the bridge $\{p_t\}_{t\in[0,1]}$ with time score $s^{\text{time}}_{0,t}(x)\coloneqq \partial_t\log p_t(x)$,
define the weighted integrated divergence 
\[\mathrm{BD}_g^\dagger\left(s^{\text{time}}_{0}\middle\| s^{\text{time}}_{\theta}\right)
\coloneqq
\int_0^1
\bbE_{X_t\sim p_t}\Bigl[\lambda(t)\,
D_g\bigl(s^{\text{time}}_{0,t}(X_t)\,\big\|\, s^{\text{time}}_{\theta,t}(X_t)\bigr)\Bigr]\rmd t,\]
where $s^{\text{time}}_{\theta,t}$ is a differentiable model and $\lambda(t)\ge 0$ is a weighting function.
If the model is well specified, minimizing $\mathrm{BD}_g^\dagger\left(s^{\text{time}}_{0}\middle\| s^{\text{time}}_{\theta}\right)$ recovers the true time score.

\begin{theorem}[Bregman time-score matching via integration by parts]
\label{thm:bregmanmatching}
Assume $g'(s^{\text{time}}_{\theta,t}(x))$ is differentiable in $t$ for each $x$.
Up to an additive constant independent of $\theta$,
$\mathrm{BD}_g^\dagger(s^{\text{time}}_{0}\| s^{\text{time}}_{\theta})$ equals
\begin{align*}
&\mathrm{BD}_g(s^{\text{time}}_{\theta})
\coloneqq \int_0^1 \bbE_{X_t\sim p_t}\Bigl[\lambda(t)\bigl\{g'(s^{\text{time}}_{\theta,t}(X_t))\, s^{\text{time}}_{\theta,t}(X_t)-g(s^{\text{time}}_{\theta,t}(X_t))\bigr\}
\nonumber\\
&\qquad\qquad
+\lambda(t)\partial_t g'(s^{\text{time}}_{\theta,t}(X_t))
+\lambda'(t)\, g'(s^{\text{time}}_{\theta,t}(X_t))
\Bigr]\rmd t
\nonumber\\
&\qquad\qquad
+\sum_{s\in \{1, 0\}}\lambda(s)\bbE_{X_s\sim q}\bigl[g'(s^{\text{time}}_{\theta,s}(X_s))\bigr].\nonumber
\end{align*}
\end{theorem}
For $g(u)=u^2/2$, $\mathrm{BD}_g(s^{\text{time}}_{\theta})$ reduces to the squared-loss objective used in Time-ScoreMatchingRiesz.

\subsection{Automatic Covariate Balancing}
\label{subsec:autobalance_neyman_main}
Let $r_0(x)\coloneqq q(x)/p(x)$ denote the endpoint ratio.
Time-ScoreMatchingRiesz constructs a log-ratio model by integrating the time score and applying an exponential link: $\ell_{\theta}(x)\coloneqq -\int_0^1 s^{\text{time}}_{\theta,t}(x)\rmd t$ and $r_\theta(x)\coloneqq \exp\p{\ell_{\theta}(x)}$.
Since $r_0$ satisfies the covariate-balancing identity $\bbE_{X\sim p}\bigl[r_0(X)\varphi(X)\bigr]=\bbE_{X\sim q}\bigl[\varphi(X)\bigr]$ for all $\varphi\in L_2(p)$, 
ratio-based representers in ScoreMatchingRiesz can be read as balancing weights.

\paragraph{Loss, link, and model choice.}
Under matched choices of a Bregman loss and a link, generalized Riesz regression yields (approximate) sample covariate balancing as a first-order condition \citep{Kato2026rieszrepresenter}.
In our setting, choosing the generator $g$ in $\mathrm{BD}_g(s^{\text{time}}_{\theta})$ and the link $\ell_{\theta}$ determines them.
This viewpoint suggests selecting the score model class and $g$ so that the induced ratio $r_\theta$ can balance the same features that drive the outcome regression.

\paragraph{Neyman error minimization.}
For the orthogonal estimator $\widehat{\theta}$,
the leading remainder is the error term $\bbE[(\widehat\alpha-\alpha_0)(\widehat\gamma-\gamma_0)]$.
When a Riesz representer estimator achieves balancing for a feature map spanning the regression function, this error term is automatically reduced, and in linear settings it can vanish exactly \citep{Kato2026rieszrepresenter,BrunsSmith2025augmentedbalancing}.
Appendix~\ref{sec:loss_link_balancing} records the corresponding balancing identities and a concise sample-level statement.

\section{Policy Path Estimation}
\label{sec:policy_path}

We introduce a \emph{policy path}---a continuum of policy effects indexed by a scalar
policy intensity $\delta$. The path provides a continuous representation of finite-shift
policy effects (APEs). Moreover, \emph{only under a known pushforward structure} does the
$\delta\to0$ limit recover the AME.

\subsection{Counterfactual Family and Pushforward.}
Let $\{P_\delta\}_{\delta\in[-\delta_{\max},\delta_{\max}]}$ be counterfactual laws of the regressor
$X\in\calX$ with densities $\{p_\delta\}$.
A policy is called a \emph{known pushforward} if there exists a \emph{known} measurable map
$T_\delta:\calX\to\calX$ such that
\[
X_\delta = T_\delta(X)\quad\text{for }X\sim P_0,
\]
equivalently $P_\delta=(T_\delta)_\# P_0$, 
where $(T_\delta)_\# P_0$ denotes the pushforward distribution (formal definition and density
formulas are given in Appendix~\ref{appdx:pushforward}).

\subsection{Policy Path as a Continuum of APEs}
Define the (symmetric) policy-path parameter
\begin{align*}
\theta_0(\delta)
\coloneqq
\bbE_{X\sim p_{+\delta}}[\gamma_0(X)]
-
\bbE_{X\sim p_{-\delta}}[\gamma_0(X)],
\ \ 
\delta\in[0,\delta_{\max}],
\end{align*}
where $\gamma_0(x)\coloneqq \bbE[Y\mid X=x]$.
For any target shift $\delta^\star$, the APE at that shift is simply the point on the path:
$\theta_0^{\mathrm{APE}}(\delta^\star)=\theta_0(\delta^\star)$.
Thus, $\delta\mapsto\theta_0(\delta)$ is a \emph{continuous} representation of finite policy effects.

When $p_{\pm\delta}\ll p_0$, we can rewrite $\theta_0(\delta)$ under $P_0$ using density ratios
$r_{\pm\delta}(x)\coloneqq p_{\pm\delta}(x)/p_0(x)$:
\begin{align*}
\theta_0(\delta)
=
\bbE\bigl[\alpha_{0,\delta}(X)\,\gamma_0(X)\bigr],
\ \ 
\alpha_{0,\delta}(x)\coloneqq r_{+\delta}(x)-r_{-\delta}(x).
\end{align*}
Hence, estimating the path reduces to estimating the \emph{path representer}
$\alpha_{0,\delta}$, i.e., (a difference of) density ratios.

\subsection{The \texorpdfstring{$\delta\to0$}{d -> infinity} Limit}
Assume $\delta\mapsto p_\delta(x)$ is differentiable at $\delta=0$ for a.e.\ $x$ and define the policy (time) score at the baseline $s^{\mathrm{pol}}_0(x)\coloneqq \partial_\delta\log p_\delta(x)\big|_{\delta=0}$. 
Then a standard differentiation argument yields $\theta_0'(0)
=
2\,\bbE\bigl[\gamma_0(X)\,s^{\mathrm{pol}}_0(X)\bigr]$. 
Crucially, $s^{\mathrm{pol}}_0$ depends on \emph{how the policy perturbs the law of $X$} and is
\emph{not} determined by the observational \emph{data score}
$s^{\mathrm{data}}_0(x)=\nabla_x\log p_0(x)$ in general.
Therefore, without additional structure, the infinitesimal limit
$\theta_0(\delta)/(2\delta)$ is not canonically tied to the AME.

\paragraph{AME limit for shift policies under known pushforward.}
Suppose the policy is a \emph{known pushforward} $P_\delta=(T_\delta)_\#P_0$ with $T_0=\mathrm{Id}$.
Then, $\theta_0(\delta)
=
\bbE\bigl[\gamma_0(T_{+\delta}(X))-\gamma_0(T_{-\delta}(X))\bigr]$ holds, 
so the local behavior of the path is governed by the geometry of $T_\delta$.
In the canonical \emph{translation (shift) policy} $X=(D,Z)$, $T_\delta(d,z)=(d+\delta,z)$, 
we have $\theta_0(\delta)=\bbE[\gamma_0(D+\delta,Z)-\gamma_0(D-\delta,Z)]$ and, under mild smoothness, $\lim_{\delta\downarrow0}\frac{\theta_0(\delta)}{2\delta}
=
\bbE[\partial_d\gamma_0(D,Z)]
=: 
\theta_0^{\mathrm{AME}}$. 
Thus, the policy path recovers AME as its $\delta\to0$ limit \emph{only in the structured (known-pushforward) regime}.
Appendix~\ref{appdx:pushforward} further shows that the same structure allows construction of
$\alpha_{0,\delta}$ (hence APEs and the full path) from the data score.

\paragraph{Coincidence of the Data Score and Time Score}
A one-parameter policy family $\{p_\delta\}$ itself provides a natural ``time'' index.
Define the time score along the policy as $s^{\mathrm{time}}_\delta(x)\coloneqq \partial_\delta\log p_\delta(x)$.
Then $s^{\mathrm{time}}_\delta$ is exactly the policy score.
For translation pushforwards,
\begin{align*}
s^{\mathrm{time}}_\delta(d,z)
=
\partial_\delta\log p_\delta(d,z)
=
-\partial_d\log p_0(d-\delta,z),
\end{align*}
so the time score along the policy \emph{coincides} with the treatment-direction data score evaluated along the known path.
This is the basic reason why, under known pushforwards, a single data-score model on $P_0$ can generate ratios for all $\delta$
via one-dimensional integration (Appendix~\ref{appdx:pushforward}).

\section{Theoretical Analysis}
\label{sec:theoretical_analysis}
This section provides theoretical analysis related to our proposed ScoreMatchingRiesz.
Because our nuisance components are constructed from either the \emph{data score}
$\nabla_x \log p(x)$ or the \emph{time score} $\partial_t \log p_t(x)$, we focus on
$L_2$-type guarantees for these scores.

\subsection{Data-ScoreMatchingRiesz}
\label{subsec:rates_oko}
This subsection records $L_2$-type score estimation rates from diffusion-model-based
(denoising) score matching, as established by \citet{Oko2023diffusionmodels}.
Throughout, let $p_0$ be a probability density on a bounded domain $\calX\subset\R^d$
(e.g., $\calX=[-1,1]^d$), and let $X\sim p_0$. Recall that for $\sigma>0$, we defined the smoothed density $p_\sigma \coloneqq p_0 * \calN(0,\sigma^2 I_d)$ (so $\tilde X_\sigma \sim p_\sigma$), and its (data) score $\nabla_x\log p_\sigma(x)$ for $x\in\calX$. Let $\widehat s^{\text{data}}_\sigma(x)$ denote an estimator of this score.

\paragraph{Additional notations and assumptions.}
For a function $g\colon\calX\to\R^d$ and a density $p$ on $\calX$, define $\|g\|_{L_2(p)}^2 \coloneqq \bbE_{X\sim p}\bigl[\|g(X)\|_2^2\bigr]$. 
For a score field $g(\cdot,\sigma)$ depending on $\sigma\in[\sigma_{\min},\sigma_{\max}]$, define the
integrated $L_2$ norm $\|g_{\cdot}\|_{\nu}^2
\coloneqq
\int_{\sigma_{\min}}^{\sigma_{\max}}
\|g_\sigma(\cdot)\|_{L_2(p_\sigma)}^2 \rmd \sigma$, where $0<\sigma_{\min}<\sigma_{\max}<\infty$. 
We write $a_n\lesssim b_n$ if $a_n\le C b_n$ for a universal constant $C>0$ and all large $n$.
We write $\mathrm{polylog}(n)$ for a factor bounded by $(\log n)^c$ for some $c>0$. Let $B^{s}_{p,q}(\calX)$ denote a Besov space on $\calX$ with smoothness $s>0$
and integrability parameters $p,q\in[1,\infty]$.
We assume $p_0\in B^{s}_{p,q}(\calX)$.

\paragraph{$L_2$ rate.} \citet{Oko2023diffusionmodels} shows the following result, which is directly applicable to Data-ScoreMatchingRiesz.

\begin{theorem}[$L_2$ rate for Data-ScoreMatchingRiesz; \citet{Oko2023diffusionmodels}]
\label{thm:oko_l2_score}
Under the assumptions of \citet{Oko2023diffusionmodels}, the Data-ScoreMatchingRiesz estimator
$\widehat s^{\text{data}}$ can be constructed so that
\begin{align*}
    &\bbE\Bigl[\ \|\widehat s^{\text{data}}_\sigma-\nabla\log p_\sigma\|_{\nu}^2\ \Bigr]
 \lesssim 
n^{-2s/(2s+d)}\cdot \mathrm{polylog}(n).
\end{align*}
Moreover, the minimax lower bound over the same Besov class matches $n^{-2s/(2s+d)}$
up to logarithmic factors.
\end{theorem}

In AME estimation with $X=(D,Z)$ and continuous $D$, the Riesz representer is
$\alpha^{\mathrm{AME}}_0(D,Z)=-\partial_d\log p_0(D,Z)$ (equivalently, $-\{s^{\text{data}}_0(D,Z)\}_d$).

\subsection{Time-ScoreMatchingRiesz}
\label{subsec:time_score_rate}
We next give an $L_2$ rate for Time-ScoreMatchingRiesz. Since the formulation varies across applications, we consider the following representative problem. 

\paragraph{Additional notations and assumptions.}
Let $q$ and $p$ be two probability densities on $\calX$ such that $q\ll p$.
Define the density ratio as $r(x)\coloneqq \frac{q(x)}{p(x)}$ for $x\in\calX$. 
Let $\{p_t\}_{t\in[0,1]}$ be a family of \emph{bridge} densities connecting $q$ and $p$: $p_0=q$ and $p_1=p$. 
We assume $t\mapsto p_t(x)$ is differentiable for a.e.\ $x$, and define the \emph{time score} $s^{\text{time}}_t(x)\coloneqq \partial_t\log p_t(x)$ for $(x,t)\in\calX\times(0,1)$. 
Write $X_t\sim p_t$.

Let $\lambda:[0,1]\to\R_+$ be a measurable weighting function and define $\lambda_{\min}\coloneqq \inf_{t\in[0,1]}\lambda(t)$. 
When $\lambda_{\min}>0$, define the weighted $L_2$ norm for a family $\{f_t\}_{t\in[0,1]}$ with $f_t\colon\calX\to\R$ by $\|f\|_{\mu_\lambda}^2
\coloneqq
\int_0^1 \bbE_{X_t\sim p_t}\bigl[\lambda(t)  f_t(X_t)^2\bigr] \rmd t$. 
We write $\mu_1$ for the special case $\lambda(t)\equiv 1$.

Let $P_n$ denote an effective complexity parameter for $\calF_n$ (e.g., number of parameters or an upper bound thereof),
used to express approximation and generalization rates.

\begin{assumption}[Overlap of $p_t$]
\label{ass:overlap}
There exist constants $0<c\le C<\infty$ such that for all $t\in[0,1]$ and for a.e.\ $x\in\calX$, $c p(x)\ \le\ p_t(x)\ \le\ C p(x)$ holds. 
\end{assumption}
Assumption~\ref{ass:overlap} is a technical device to transfer $L_2$ control along the bridge to downstream bounds for ratios and representers.
Appendix~\ref{appdx:bridge_overlap} summarizes practical bridge and weighting choices that mitigate violations of overlap.

\begin{assumption}[Smoothness and approximation rate]
\label{ass:approx}
The map $(x,t)\mapsto s^{\text{time}}_t(x)$ belongs to a Besov ball on $\calX\times[0,1]$
with smoothness $s>0$ (e.g.\ $B^{s}_{2,2}(\calX\times[0,1])$), and the model class $\calF_n$ satisfies:
there exists $s_n^\star\in\calF_n$ such that $\|s_n^\star - s^{\text{time}}\|_{\mu_\lambda}^2 \ \lesssim\ P_n^{-2s/(d+1)}$. 
\end{assumption}

\begin{assumption}[Uniform deviation and complexity]
\label{ass:gen}
There exists a (data-dependent) random variable $\Delta_n$ such that with high probability, $\sup_{s\in\calF_n}\bigl|\widehat{\calR}_n(s)-\calR(s)\bigr| \ \le\ \Delta_n$ holds, where $\Delta_n \ \lesssim\ \frac{P_n \mathrm{polylog}(n)}{n}$. 
\end{assumption}

\paragraph{$L_2$ rate.} Let $\calR(\cdot)$ denote the population time-score-matching objective in \eqref{eq:dreinfty}.
Under regularity conditions (as in Lemma~\ref{lem:quadratic_identity}),
$s^{\text{time}}_t$ is the population minimizer of $\calR$. Let $\widehat{\calR}_n(\cdot)$ be an empirical counterpart of $\calR(\cdot)$ computed from
samples from $q$ and $p$ and Monte Carlo bridge draws as in Section~\ref{sec:dreic}.
Let $\calF_n$ be a class of time-score models $s^{\text{time}}_{\theta, t}(x)$ (e.g., neural networks) that are differentiable in $t$.
Define the empirical risk minimizer as $\widehat s \in \arg\min_{s\in\calF_n} \widehat{\calR}_n(s)$. Then, the following result holds.

\begin{theorem}[$L_2$ rate for Time-ScoreMatchingRiesz]
\label{thm:dreinf_rate}
Suppose Assumptions~\ref{ass:overlap}--\ref{ass:gen} hold and $\lambda_{\min}>0$.
Then, $\|\widehat s - s^{\text{time}}\|_{\mu_\lambda}^2
\lesssim 
\inf_{s\in\calF_n}\|s-s^{\text{time}}\|_{\mu_\lambda}^2
 + \Delta_n$ holds. 
In particular, under Assumptions~\ref{ass:approx} and \ref{ass:gen}, $\|\widehat s - s^{\text{time}}\|_{\mu_\lambda}^2
  \lesssim\
P_n^{-2s/(d+1)} + \frac{P_n \mathrm{polylog}(n)}{n}$. 
Choosing $P_n \asymp n^{(d+1)/(2s+d+1)}$ 
yields the (near-)optimal rate
\[
\|\widehat s - s^{\text{time}}\|_{\mu_\lambda}
  =\
O_{\bbP}\Bigl(n^{-s/(2s+d+1)}\cdot \mathrm{polylog}(n)\Bigr).
\]
\end{theorem}

Integrating an estimated time score $\widehat s$ yields a plug-in estimator of $\log r$ and hence of $r$.
Appendix~\ref{subsec:rates_ratio} records the corresponding $L_2$ rate for the Riesz representer.

\subsection{Efficient Estimation of the Parameter of Interest}
We recall the asymptotic efficiency result for $\widehat{\theta} = \frac{1}{n}\sum_{i=1}^n \Bigp{m(W_i,\widehat{\gamma})+\widehat{\alpha}(X_i)(Y_i-\widehat{\gamma}(X_i))}$. Assume that both $\widehat{\gamma}$ and $\widehat{\alpha}$ satisfy the Donsker condition or are constructed via cross-fitting. Assume that the following conditions hold: (i) $\|\widehat{\alpha} - \alpha_0\|_2 = o_p(1)$; (ii) $\|\widehat{\gamma} - \gamma_0\|_2 = o_p(1)$; (iii) $\|\widehat{\alpha} - \alpha_0\|_2\|\widehat{\gamma} - \gamma_0\|_2 = o_p(1/\sqrt{n})$. Then, it holds that $\sqrt{n}\p{\widehat{\theta}-\theta_0} \xrightarrow{d} \calN(0,\bbE[\psi(W;\eta_0,\theta_0)^2])$ as $n\to\infty$.

\begin{corollary}[$L_2$ rate for Riesz representers]
\label{cor:score_to_dml}
In AME estimation, $\alpha^{\mathrm{AME}}_0(D,Z)=-\partial_d\log p_0(D,Z)$, so for $\widehat{\alpha}^{\mathrm{AME}}=-\{\widehat s^{\text{data}}\}_d$, $\|\widehat{\alpha}^{\mathrm{AME}}-\alpha^{\mathrm{AME}}_0\|_2
=
\|\{\widehat s^{\text{data}}\}_d-\partial_d\log p_0\|_2$ holds. 
For ratio-based targets, including ATE, APE, and each fixed-$\delta$ policy-path point, construct $\widehat{\log r}$ by integrating an estimated score field and apply clipping before exponentiation.
Then the standard propagation chain (Appendix~\ref{subsec:rates_ratio}) is $\|\widehat{\log r}-\log r\|_{L_2(p)}\lesssim\|\widehat s^{\text{time}}-s^{\text{time}}\|_{\mu_1}$, $\|\widehat r-r\|_{L_2(p)}\lesssim e^{c}\|\widehat{\log r}-\log r\|_{L_2(p)}$, and $\|\widehat\alpha-\alpha_0\|_2\lesssim\|\widehat r_+-r_+\|_2+\|\widehat r_- - r_-\|_2$, 
for a fixed clipping constant $c$ and the corresponding ratio-based representer $\alpha_0$.
Consequently, Theorems~\ref{thm:oko_l2_score} and \ref{thm:dreinf_rate} yield $\|\widehat{\alpha}-\alpha_0\|_2=o_{\bbP}(1)$, and condition (iii) holds if $\|\widehat{\alpha}-\alpha_0\|_2\|\widehat{\gamma}-\gamma_0\|_2=o_{\bbP}(n^{-1/2})$ (e.g.\ $a+b>1/2$ when the two errors are $O_{\bbP}(n^{-a})$ and $O_{\bbP}(n^{-b})$).
\end{corollary}

\begin{table}[t]
\centering
\centering
\caption{AME (true = 2.225).}
\label{tab:metrics_ame}
\scalebox{1}{
\begin{tabular}{lrrr}
\toprule
Metric & Data-SMR & Time-SMR & Riesz reg. \\
\midrule
Bias & 0.001 & -0.075 & -0.205 \\
MSE & 0.161 & 0.637 & 10.076 \\
Cov. & 91.8\% & 90.9\% & 86.9\% \\
\bottomrule
\end{tabular}}
\centering
\caption{APE (true = 4.063).}
\label{tab:metrics_ape}
\scalebox{1}{
\begin{tabular}{lrrr}
\toprule
Metric & Data-SMR & Time-SMR & Riesz reg. \\
\midrule
Bias & 0.131 & 0.268 & -0.152 \\
MSE & 0.486 & 1.117 & 74.918 \\
Cov. & 89.4\% & 75.0\% & 85.2\% \\
\bottomrule
\end{tabular}}
\caption{Performance metrics for AME and APE.}
\label{tab:metrics_side_by_side}
\end{table}

\begin{figure}[tbp]
  \centering
    \includegraphics[width=0.5\linewidth]{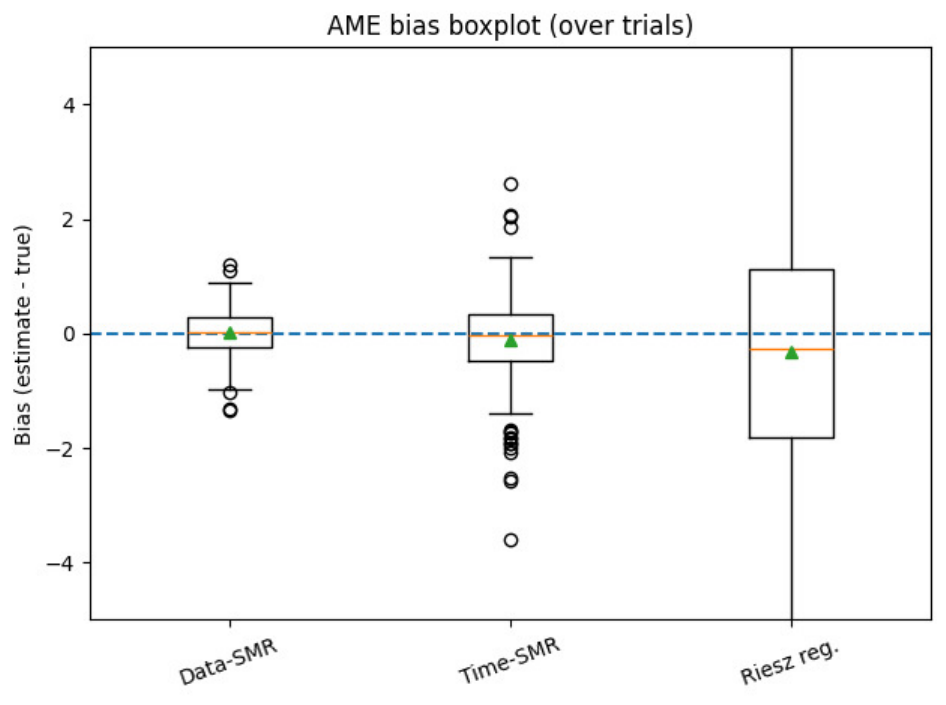}
    \caption{Results in AME estimation}
    \label{fig:left}
\end{figure}
\begin{figure}[tbp]
\centering
    \includegraphics[width=0.5\linewidth]{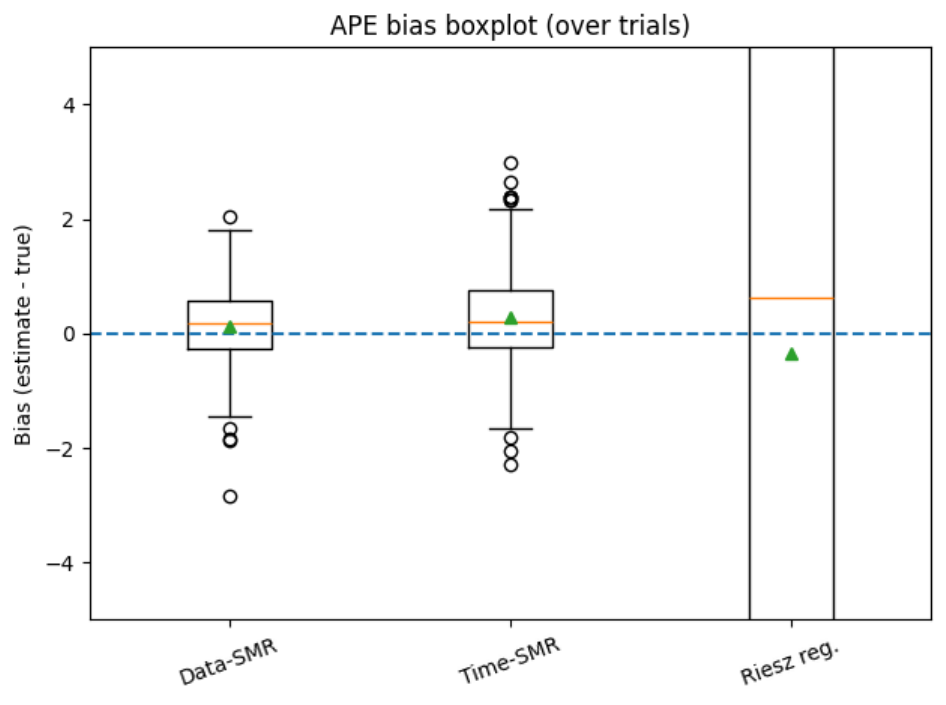}
    \caption{Results in APE estimation}
    \label{fig:right}
  \caption{Box plots of errors (biases) of the estimates}
  \label{fig:side_by_side}
\end{figure}

\section{Experiments}
\label{sec:experiments}

We compare Data-ScoreMatchingRiesz (Data-SMR) and Time-ScoreMatchingRiesz (Time-SMR),
and we use Riesz regression \citep{Chernozhukov2021automaticdebiased,Kanamori2009aleastsquares} as a baseline. Note that Riesz regression is identical to Hyv\"arinen score matching in AME estimation and LSIF in APE estimation, while its dual corresponds to stable balancing weights \citep{Zubizarreta2015stableweights}. Therefore, it is sufficient to compare our method only with Riesz regression because it is equivalent to various other methods. For details on these equivalences, see \citet{BrunsSmith2025augmentedbalancing} and \citet{Kato2026rieszrepresenter}.
All methods use the same MLP architecture for $\widehat\gamma$ and two-fold cross-fitting.
Additional results are reported in Appendices~\ref{sec:additional_sims} and \ref{app:empirical_finance}.

\paragraph{Data generating process.}
Throughout, $X=(X^{(1)},X^{(2)},X^{(3)})\in\R^3$ follows $X\sim \calN(0,\Sigma)$ with
\[
\Sigma=
\begin{pmatrix}
1 & 0.1 & 0.1\\
0.1 & 1 & 0.1\\
0.1 & 0.1 & 1
\end{pmatrix}.
\]
We set $D\coloneqq X^{(1)}$ and $Z\coloneqq (X^{(2)},X^{(3)})$.
The outcome is $Y=\mu(X)+\varepsilon$ with $\varepsilon\sim\calN(0,1)$ and $\mu(x)
=
1+x^{(1)}+0.1(x^{(1)})^2+2\sin(x^{(1)})+x^{(2)}+x^{(1)}x^{(2)}+(x^{(3)})^2+(x^{(3)})^3$. 

\paragraph{Targets.}
We estimate the AME $\theta_0^{\mathrm{AME}}=\bbE[\partial_d\gamma_0(D,Z)]$
and the APE under the shift policy at $\delta=1$,
$\theta_0^{\mathrm{APE}}=\bbE[\mu(D+1,Z)]-\bbE[\mu(D-1,Z)]$.
The ground truths are computed by population Monte Carlo.

\paragraph{Evaluation metrics.}
Across $200$ trials, we report (i) bias and MSE of $\widehat\theta$ and (ii) empirical coverage of 95\% confidence intervals.
For the shift policy, we also estimate the full policy path $\delta\mapsto\theta_0(\delta)$ on a grid
(Section~\ref{sec:policy_path}) and plot pointwise 95\% bands, the true path, and the local linear approximation
$2\delta\,\theta_0^{\mathrm{AME}}$.

\begin{figure}[tbp]
  \centering
  \includegraphics[width=0.6\linewidth]{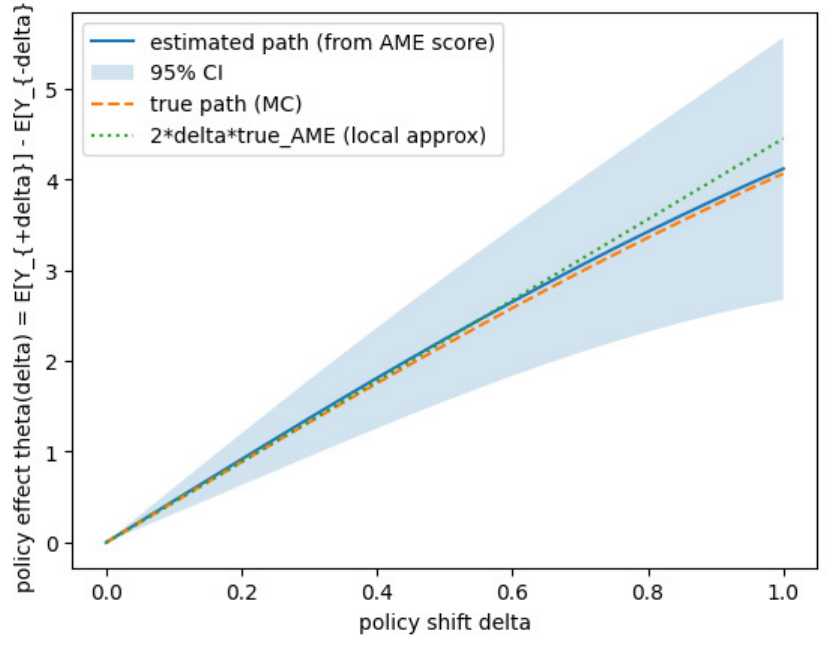}
  \caption{Policy path visualization}
  \label{fig:three}
\end{figure}

\paragraph{Results.}
Tables~\ref{tab:metrics_ame} and~\ref{tab:metrics_ape} show that Data-SMR is the most accurate and stable method, achieving the smallest MSE for both AME (0.161) and APE at $\delta=1$ (0.486) with near-nominal coverage, whereas Time-SMR is less accurate and exhibits marked undercoverage for APE (75\%). Riesz regression is substantially less stable, especially for APE where it suffers severe variance amplification (MSE 74.918). Figure~\ref{fig:three} further suggests that the policy path deviates from the local linear approximation $2\delta,\theta_0^{\mathrm{AME}}$ as $\delta$ increases, highlighting the need for stable representer estimation to recover the nonlinear curve.

\section{Conclusion}
We proposed ScoreMatchingRiesz to stably estimate the Riesz representer, which also provides a unified interpretation across causal targets. In practice, although Data-ScoreMatchingRiesz shows preferable empirical performance, there are applications where we cannot use Data-ScoreMatchingRiesz and only Time-ScoreMatchingRiesz is available; thus, we need to choose the method based on the application. We define the policy path that bridges the AME and APE, and we demonstrated that this path can be estimated using the score. These connections suggest that score matching is a general tool for causal inference.

\bibliography{arXiv2.bbl}
\bibliographystyle{tmlr}

\onecolumn

\appendix

\section{Related Work}
\label{appdx:related}
This appendix complements the brief discussion in the main text and places
ScoreMatchingRiesz in the intersection of (i) semiparametric efficiency and debiased machine learning,
(ii) Riesz-representer estimation and covariate balancing,
and (iii) score/density-ratio estimation via score matching and infinitesimal classification.

\subsection{Efficient Estimation, Orthogonal scores, and Debiased Machine Learning}
A recurring theme in semiparametric statistics is that (asymptotically) efficient estimation can be achieved by
augmenting a plug-in estimator with a correction term based on an influence function.
Classical constructions include one-step estimators and estimating-equation (Z-estimation) approaches
\citep{Levit1976onthe,Ibragimov1981statisticalestimation,Pfanzagl1982contributionsto,Bickel1998efficientadaptive,VanderVaart1998asymptoticstatistics,Newey1994theasymptotic}.
Related ideas also underpin sieve-based inference for (possibly irregular) functionals, where Riesz representers
play an explicit role in characterizing local sensitivity and efficient variance
\citep{Chen2014sieveinference,Chen2015sievesemiparametric,Chen2015sievewald}.

In missing-data and causal-inference settings, these influence-function constructions yield doubly robust / augmented inverse-probability-weighted scores
\citep{Robins1994estimationregression,Bang2005doublyrobust,Hirano2003efficientestimation}
and targeted learning estimators that explicitly target the efficient influence function
\citep{vanderLaan2006targetedmaximum,Schuler2018comparisonmethods}.
Debiased (or ``double'') machine learning extends these principles to high-dimensional and nonparametric nuisance estimation by
leveraging Neyman orthogonality and cross-fitting to control the effect of first-stage regularization
\citep{Chernozhukov2018doubledebiased,Klaassen1987consistentestimation,Zheng2011crossvalidatedtargeted}.
The \emph{automatic} debiased ML formulation emphasizes that for general linear and continuous functionals of the regression function,
efficient influence functions involve a \emph{Riesz representer} (or, in causal examples, a set of balancing weights)
that must itself be estimated \citep{Chernozhukov2022automaticdebiased}.
Our paper follows this line and focuses on the representer-estimation subproblem, proposing score-matching-based modules
that can be plugged into Neyman orthogonal scores.

\subsection{Riesz representers, Riesz regression, and Covariate Balancing}
\paragraph{Riesz regression and its variants.}
A direct approach to representer estimation is \emph{Riesz regression}, which fits $\alpha$ by minimizing a sample analog of the
Hilbert-space quadratic characterization of the Riesz representer
\citep{Chernozhukov2021automaticdebiased,Chernozhukov2022automaticdebiased}.
In AME settings, this objective coincides with Hyv\"arinen score matching and (in certain linear models) with LSIF-type density-ratio objectives
\citep{Hyvarinen2005estimationof,Kanamori2009aleastsquares}.
Alternative representer estimators include matching / nearest-neighbor methods that directly approximate the Riesz functional
\citep{Lin2023estimationbased,Kato2025nearestneighbor}.
For ATE, the representer reduces to inverse propensity weights, which are also the key ingredient in classical efficient ATE estimators
\citep{Hirano2003efficientestimation,Bang2005doublyrobust}.

\paragraph{Balancing weights as ratio-based representers.}
In many targets, the Riesz representer is a density ratio (or a simple transform of one),
so representer estimation is equivalent to constructing weights that \emph{balance} covariate moments between two distributions.
This connection links representer estimation to a large literature on covariate balancing and stable weighting,
including entropy balancing \citep{Hainmueller2012entropybalancing},
stable balancing weights \citep{Zubizarreta2015stableweights},
approximate residual balancing \citep{Athey2018approximateresidual,Benmichael2021balancingact},
kernel balancing \citep{Hazlett2020kernelbalancing},
and tailored loss minimization / covariate-balancing perspectives on efficiency \citep{Zhao2019covariatebalancing}.
Recent work shows that such weighting schemes can be interpreted through generalized Riesz regression objectives built from Bregman divergences,
clarifying the role of the loss and link function in determining which balancing weights are favored
\citep{Kato2026rieszrepresenter,BrunsSmith2025augmentedbalancing}.
Section~\ref{sec:generalization_smr} adapts this ``loss--link'' viewpoint to our score-based construction:
the time-score loss (Theorem~\ref{thm:bregmanmatching}) and the score-to-ratio link together determine the implied balancing weights,
and the resulting (approximate) balancing identities help control the Neyman remainder in orthogonal estimation.

\subsection{Score Matching and Diffusion Models}
Score matching estimates the score $\nabla_x\log p(x)$ without estimating the density normalizing constant.
Hyv\"arinen's original formulation minimizes a Fisher-divergence objective that becomes computable via integration by parts
\citep{Hyvarinen2005estimationof}.
A practical challenge is that the Hyv\"arinen objective involves a divergence (Jacobian-trace) term, which can be expensive or unstable in high dimensions.
Denoising score matching (DSM) addresses this by training on Gaussian-perturbed samples, replacing the target score by the score of a smoothed density
\citep{Vincent2011aconnection}.
DSM is the workhorse of modern score-based generative modeling and diffusion models, where one trains a single network over a continuum of noise levels
\citep{Song2020generativemodeling,Song2021denoisingdiffusion}.

While diffusion models are typically used for density estimation and sampling, their statistical theory provides off-the-shelf
nonparametric guarantees for score estimation.
In particular, \citet{Oko2023diffusionmodels} establish minimax-optimal (up to logs) convergence rates for diffusion-model score estimators,
which we leverage as representer rates for Data-ScoreMatchingRiesz (Section~\ref{subsec:rates_oko}).
Our use of DSM is therefore conceptually different from generative modeling: we employ score learning as a \emph{Riesz-representer module}
inside orthogonal score equations, and we propagate score error to the error of the representer and of the final debiased estimator.

\subsection{Density Ratio Estimation, Telescoping Bridges, and Infinitesimal Classification}
Density ratio estimation (DRE)---estimating $q/p$ from samples---is a central primitive in covariate shift, importance weighting, and off-policy evaluation
\citep{Sugiyama2012densityratio,Uehara2020offpolicy}.
Representative approaches include least-squares importance fitting (LSIF) \citep{Kanamori2009aleastsquares} and its generalizations
based on Bregman divergences \citep{Sugiyama2011densityratio}, as well as classifier-based ratio estimation
\citep{Bickel2009discriminativelearning}.
DRE also underlies debiased estimation under distribution shift, where the relevant representers are ratios between the target and source covariate laws
\citep{Kato2024doubledebiasedcovariateshift}.

A key practical issue is that highly flexible ratio models can overfit badly, producing extreme weights and unstable downstream estimators.
This phenomenon has been documented as ``train loss hacking'' \citep{Kato2021nonnegativebregman} and as the ``density chasm'' in high dimensions
\citep{Rhodes2020telescopingdensityratio}.
One mitigation strategy is to replace direct endpoint-to-endpoint ratio estimation by a product of intermediate ratios along a bridge
\citep{Rhodes2020telescopingdensityratio}.
Taking a continuum limit leads to \emph{infinitesimal classification} and the DRE-$\infty$ objective, which learns the time score
$\partial_t\log p_t(x)$ along the bridge and recovers the log ratio by integration
\citep{Choi2022densityratio}.
Our Time-ScoreMatchingRiesz adopts this mechanism as a representer estimator when the target representer is ratio-based (ATE/APE/policy paths),
and our analysis focuses on the error metrics needed for Neyman orthogonal debiased estimation rather than on ratio estimation alone.

\subsection{Policy Effects under Distributional Shifts and Policy Paths}
Counterfactual policy evaluation can often be phrased as estimating a regression functional under a shifted distribution of covariates.
In economics, these objects include policy counterfactuals and structural policy effects, and in macro-finance they arise naturally in
local projection analyses \citep{Jorda2005estimationand}.
In machine learning and econometrics, related themes appear in policy learning and off-policy evaluation, where the goal is to evaluate or optimize
the value of (possibly stochastic) policies \citep{Athey2021policylearning,Chernozhukov2025policylearning,Uehara2020offpolicy}.
Our ``policy path'' parameter is a descriptive object that traces policy effects continuously as the intervention intensity varies.

Methodologically, our paper emphasizes that the representer for a finite policy shift is typically a density ratio, and hence a balancing weight.
When the counterfactual law is a \emph{known pushforward} of the observational law, density ratios can be expressed as line integrals of the
observational data score along the known transport path, enabling reuse of a single score model across many policy shifts
\citep{Kato2025directbias,Kato2025directbias2}.
Outside the pushforward regime (e.g.\ stochastic interventions that resample $D\mid Z$), this score-to-ratio construction is unavailable and one must
learn ratios (or time scores) from endpoint samples, motivating our Time-ScoreMatchingRiesz constructions.

\section{Derivation of the Hyv\"arinen Score Matching Objective}
\label{appdx:hsm}

Concretely, expanding $\calR^\dagger$ yields
\[
\calR^\dagger(s)
=
\frac12\bbE\left[\|s(X)\|^2\right]
-\bbE\left[s(X)^\top \nabla_x\log p_0(X)\right]
+\text{const}.
\]
Assuming boundary decay so that integration by parts applies, we have
\[
\bbE\left[s(X)^\top \nabla_x\log p_0(X)\right]
=
-\bbE\left[\nabla_x\cdot s(X)\right],
\]
where $\nabla_x\cdot s(x)=\mathrm{tr}\{\nabla_x s(x)\}$ is the divergence of the vector field $s$.
Substituting this identity into the expansion yields \eqref{eq:hyvarinen_sm}.
Thus, minimizing \eqref{eq:hyvarinen_sm} over a function class estimates the score without ever estimating $p_0$ itself.

\section{Equivalence of ``Time Scores'' in DSM and DRE-\texorpdfstring{$\infty$}{infty} Under a Shared Bridge}
In DRE-$\infty$, the \emph{time score} associated with a differentiable bridge of densities
$\{p_t\}_{t\in[0,1]}$ is defined as
\[
s^{\mathrm{time}}(x,t)  \coloneqq  \partial_t \log p_t(x).
\]
This definition is purely path-based: $s^{\mathrm{time}}$ depends on the chosen bridge $\{p_t\}$.

A DSM forward noising mechanism also induces a one-parameter family of marginal densities.
For instance, Gaussian smoothing defines
\[
p_\sigma  \coloneqq  p_0 * \mathcal N(0,\sigma^2 I_d),
\qquad \sigma \ge 0,
\]
which can be viewed as a bridge from the data distribution ($\sigma=0$) toward a Gaussian reference
as $\sigma$ increases. Along this noising family, a natural ``DSM time score'' is given as
\[
s^{\mathrm{DSM}\mathchar`-\mathrm{time}}_{\sigma}(x)  \coloneqq  \partial_\sigma \log p_\sigma(x).
\]

If we choose the DRE-$\infty$ bridge to coincide with the DSM noising family up to a smooth
reparameterization, i.e.\ $p_t \equiv p_{\sigma(t)}$ for a monotone $C^1$ map $t\mapsto \sigma(t)$,
then the two time scores represent the same infinitesimal object and differ only by the deterministic
change of variables:
\[
s^{\mathrm{time}}_t(x)
=
\partial_t \log p_{\sigma(t)}(x)
=
\sigma'(t) \partial_\sigma \log p_\sigma(x)\big|_{\sigma=\sigma(t)}
=
\sigma'(t)\, s^{\mathrm{DSM}\mathchar`-\mathrm{time}}_{\sigma(t)}(x).
\]
Consequently, under a shared bridge (and possibly a deterministic reparameterization of time),
the ``time score'' in DRE-$\infty$ can be interpreted as the same time-derivative-of-log-density
object that arises from DSM.
In particular, for any $\sigma_0,\sigma_1$ with $0\le \sigma_0<\sigma_1$,
\[
\log \frac{p_{\sigma_0}(x)}{p_{\sigma_1}(x)}
=
\int_{\sigma_1}^{\sigma_0} s^{\mathrm{time}}_{\sigma}(x) \rmd\sigma
=
\int_{t_1}^{t_0} s^{\mathrm{time}}_t(x)\rmd t,
\qquad \text{where } \sigma(t_j)=\sigma_j.
\]

\section{Implementation Details: Time-ScoreMatchingRiesz}
\label{app:joint-smr}

This appendix makes the design of \emph{Time-ScoreMatchingRiesz (Time-SMR)} explicit.
Time-SMR is a multi-task score-learning module that combines
(i) \emph{Data-ScoreMatchingRiesz} (data-score learning via denoising score matching) and
(ii) \emph{Time-ScoreMatchingRiesz} (time-score learning via infinitesimal classification / time-score matching; \citealp{Choi2022densityratio}).
We specify (a) the parameter-sharing pattern, (b) the joint objective and loss weights,
and (c) the training schedule and inference procedure.

\subsection{Shared vs. Separate Networks}
\label{app:joint-smr:sharing}

Let $x \in \mathcal X \subset \mathbb R^d$ denote the regressor.
Time-SMR involves two score fields:
\[
s^{\mathrm{data}}(x,\sigma)\in \mathbb R^d,
\qquad
s^{\mathrm{time}}(x,t)\in \mathbb R,
\]
where $\sigma$ is a DSM noise level and $t$ is the bridge index used in time-score matching.

A reproducible parameterization is a \emph{shared backbone with two heads}:
\[
h_\phi(x,u) \in \mathbb R^H,
\qquad
s^{\mathrm{data}}_{\theta}(x,\sigma)=g^{\mathrm{data}}_{\theta_d}\!\big(h_\phi(x,\mathrm{emb}(\sigma))\big),
\qquad
s^{\mathrm{time}}_{\theta}(x,t)=g^{\mathrm{time}}_{\theta_t}\!\big(h_\phi(x,\mathrm{emb}(t))\big),
\]
where $\phi$ are shared parameters, $(\theta_d,\theta_t)$ are head-specific parameters, and
$\mathrm{emb}(\cdot)$ is a scalar embedding (e.g., Fourier/sinusoidal features).
This sharing choice is motivated by the joint-training perspective of \citet{Choi2022densityratio}, where
a single model can represent multiple infinitesimal score components along a continuum.

\paragraph{Alternative.}
For maximum modularity, one can use disjoint networks
$s^{\mathrm{data}}_{\theta_d}$ and $s^{\mathrm{time}}_{\theta_t}$.
All steps below remain identical, except that $\phi$ is absent.
In our experience, sharing a backbone is often more stable when one task (typically data-score learning)
is easier and regularizes the representation used by the other task.

\subsection{Losses and Joint Objective}
\label{app:joint-smr:objective}

\paragraph{Data-score loss (Data-ScoreMatchingRiesz).}
We use denoising score matching with Gaussian noise:
$\tilde x = x + \sigma \varepsilon$, $\varepsilon \sim \mathcal N(0,I_d)$, $\sigma \sim \pi$.
The per-minibatch loss is
\begin{align}
\label{eq:joint-smr:ldata}
\mathcal L_{\mathrm{data}}(\theta)
=
\mathbb E_{x\sim p_0}\mathbb E_{\sigma\sim\pi}\mathbb E_{\varepsilon}
\!\left[
\lambda_{\mathrm{data}}(\sigma)\,
\left\|
s^{\mathrm{data}}_{\theta}(x+\sigma\varepsilon,\sigma)+\varepsilon/\sigma
\right\|_2^2
\right].
\end{align}
(For AME estimation with $x=(d,z)$, one may inject noise only in the $d$-coordinate to directly learn
$\partial_d\log p_0(d,z)$; see Appendix~\ref{app:smoothing-bias} for the corresponding discussion.)

\paragraph{Time-score loss (Time-ScoreMatchingRiesz; \citealp{Choi2022densityratio}).}
Let $q$ and $p$ be endpoint densities and $\{p_t\}_{t\in[0,1]}$ a differentiable bridge.
Let $x_0\sim q$, $x_1\sim p$, and $x_t \leftarrow \textsc{Interpolate}(x_0,x_1,t)$ with $t\sim \mathrm{Unif}(0,1)$.
We minimize the integration-by-parts objective (the continuum limit underlying infinitesimal classification),
implemented by Monte Carlo as
\begin{align}
\label{eq:joint-smr:ltime-mc}
\widehat{\mathcal L}_{\mathrm{time}}(\theta)
=
\underbrace{\lambda(t)\,\partial_t s^{\mathrm{time}}_{\theta}(x_t,t)
+ \lambda'(t)\,s^{\mathrm{time}}_{\theta}(x_t,t)
+\tfrac12\lambda(t)\,s^{\mathrm{time}}_{\theta}(x_t,t)^2}_{\text{integrand term}}
+\underbrace{\lambda(0)\,s^{\mathrm{time}}_{\theta}(x_0,0)-\lambda(1)\,s^{\mathrm{time}}_{\theta}(x_1,1)}_{\text{endpoint terms}}.
\end{align}
Here $\partial_t s^{\mathrm{time}}_{\theta}(x_t,t)$ is a \emph{partial} derivative in $t$ holding $x_t$ fixed,
computed by automatic differentiation.

\paragraph{Joint objective (loss weighting).}
Time-SMR minimizes a weighted sum:
\begin{align}
\label{eq:joint-smr:ljoint}
\mathcal L_{\mathrm{joint}}(\theta)
=
\omega_{\mathrm{data}}\,\mathcal L_{\mathrm{data}}(\theta)
+
\omega_{\mathrm{time}}\,\mathcal L_{\mathrm{time}}(\theta),
\end{align}
with fixed nonnegative weights $(\omega_{\mathrm{data}},\omega_{\mathrm{time}})$.
A simple default is $\omega_{\mathrm{data}}=\omega_{\mathrm{time}}=1$.
To make results reproducible without manual tuning, one can also \emph{normalize} by the initial minibatch scales:
set $\omega_{\mathrm{data}}=1$ and
$\omega_{\mathrm{time}}=\mathbb E[\widehat{\mathcal L}_{\mathrm{data}}]/\mathbb E[\widehat{\mathcal L}_{\mathrm{time}}]$
measured on a few initial minibatches, and then keep it fixed.

\subsection{Training Schedule}
\label{app:joint-smr:schedule}

Time-SMR can be trained either (i) fully simultaneously or (ii) with a short warm-up.
We recommend the following schedule, which is simple and tends to stabilize training:

\begin{itemize}
\item \textbf{Warm-up (optional).} Train only the data-score module for $N_{\mathrm{warm}}$ iterations
(i.e., set $\omega_{\mathrm{time}}=0$).
\item \textbf{Joint phase.} After warm-up, optimize \eqref{eq:joint-smr:ljoint} with fixed
$(\omega_{\mathrm{data}},\omega_{\mathrm{time}})$.
Optionally ramp $\omega_{\mathrm{time}}$ linearly from $0$ to its final value over $N_{\mathrm{ramp}}$ iterations.
\end{itemize}

\subsection{Pseudocode}
\label{app:joint-smr:pseudocode}

\begin{algorithm}[t]
\caption{Time-ScoreMatchingRiesz: one training iteration}
\label{alg:joint-smr-train}
\begin{algorithmic}[1]
\REQUIRE Baseline samples $\mathcal D_0\sim p_0$; endpoint samples $\mathcal D_q\sim q$, $\mathcal D_p\sim p$;
noise distribution $\pi$; interpolation routine \textsc{Interpolate}; weights $\omega_{\mathrm{data}},\omega_{\mathrm{time}}$;
weight functions $\lambda_{\mathrm{data}}(\cdot)$ and $\lambda(\cdot)$; iteration index $k$; warm-up steps $N_{\mathrm{warm}}$.
\STATE Sample minibatch $\{x\}$ from $\mathcal D_0$; sample $\sigma\sim\pi$ and $\varepsilon\sim \mathcal N(0,I_d)$;
set $\tilde x \gets x+\sigma\varepsilon$.
\STATE $\widehat{\mathcal L}_{\mathrm{data}} \gets
\frac1{B}\sum \lambda_{\mathrm{data}}(\sigma)\left\|s^{\mathrm{data}}_\theta(\tilde x,\sigma)+\varepsilon/\sigma\right\|_2^2$.
\STATE Sample $t\sim \mathrm{Unif}(0,1)$; sample $x_0\sim \mathcal D_q$, $x_1\sim \mathcal D_p$;
set $x_t\gets \textsc{Interpolate}(x_0,x_1,t)$.
\STATE Compute $\widehat{\mathcal L}_{\mathrm{time}}$ via \eqref{eq:joint-smr:ltime-mc}
(using autodiff for $\partial_t s^{\mathrm{time}}_\theta(x_t,t)$).
\IF{$k\le N_{\mathrm{warm}}$}
  \STATE $\omega_{\mathrm{time}}^{(k)}\gets 0$
\ELSE
  \STATE $\omega_{\mathrm{time}}^{(k)}\gets \omega_{\mathrm{time}}$ \ \textit{(or a linear ramp)}
\ENDIF
\STATE $\widehat{\mathcal L}_{\mathrm{joint}} \gets \omega_{\mathrm{data}}\widehat{\mathcal L}_{\mathrm{data}}
+ \omega_{\mathrm{time}}^{(k)}\widehat{\mathcal L}_{\mathrm{time}}$.
\STATE Update parameters by one gradient step on $\widehat{\mathcal L}_{\mathrm{joint}}$.
\end{algorithmic}
\end{algorithm}

\begin{algorithm}[t]
\caption{Time-SMR inference: constructing representers from learned scores}
\label{alg:joint-smr-infer}
\begin{algorithmic}[1]
\REQUIRE Trained $(s^{\mathrm{data}}_{\hat\theta}, s^{\mathrm{time}}_{\hat\theta})$; target type (AME or ratio-based);
quadrature rule with $M$ grid points; clipping level $c$ (optional).
\IF{target is AME (data-score-based)}
  \STATE Choose a small evaluation noise $\sigma_{\mathrm{eval}}$ (e.g.\ the smallest trained noise).
  \STATE Set $\hat\alpha_{\mathrm{AME}}(d,z)\gets -\big(s^{\mathrm{data}}_{\hat\theta}((d,z),\sigma_{\mathrm{eval}})\big)_d$.
\ELSE
  \STATE For each $x$, compute $\widehat{\log r}(x)\gets -\int_0^1 s^{\mathrm{time}}_{\hat\theta}(x,t)\,dt$
  by numerical integration on $M$ grid points.
  \STATE Optionally clip: $\widehat{\log r}(x)\gets \mathrm{clip}(\widehat{\log r}(x),[-c,c])$.
  \STATE Set $\hat r(x)\gets \exp(\widehat{\log r}(x))$ and form $\hat\alpha$ from ratios (difference-of-ratios if needed).
\ENDIF
\end{algorithmic}
\end{algorithm}

\paragraph{Stabilizers (recommended).}
When ratios are constructed by exponentiating integrated scores, we recommend:
(i) \emph{log-ratio clipping} before exponentiation and
(ii) \emph{mean-one calibration} on the training fold (i.e., subtract $\log\mathbb E_{\mathrm{train}}[\exp(\widehat{\log r}(X))]$)
to enforce $\mathbb E_{\mathrm{train}}[\hat r(X)]=1$.
These steps reduce extreme weights and improve downstream stability.

\section{Application to ATE Estimation}
\label{appdx:ate}

\subsection{Riesz Representer in ATE Estimation}
Let $X=(D,Z)$ with $D\in\{1,-1\}$ and covariates $Z$. Let $(Y(1),Y(-1))$ be potential outcomes and $Y=\mathbbm{1}[D=1]Y(1)+\mathbbm{1}[D=-1]Y(-1)$. The ATE is
\[
\theta^{\mathrm{ATE}}_0 \coloneqq \bbE[Y(1)-Y(-1)] = \bbE\bigsqb{\gamma_0(1,Z)-\gamma_0(-1,Z)},
\]
where $\gamma_0(d,z)=\bbE[Y\mid D=d,Z=z]$ under standard identification assumptions (e.g. unconfoundedness and overlap). Let us define
\[
m^{\mathrm{ATE}}(W,\gamma)\coloneqq \gamma(1,Z)-\gamma(-1,Z).
\]
Then, the Riesz representer is given as
\[
\alpha^{\mathrm{ATE}}_0(D,Z) \coloneqq \frac{\mathbbm{1}[D=1]}{e_0(Z)}-\frac{\mathbbm{1}[D=-1]}{1-e_0(Z)},\]
where $e_0(z)\coloneqq P(D=1\mid Z=z)$ is the propensity score.

\subsection{Time-ScoreMatchingRiesz in ATE Estimation}
Let $\pi\coloneqq P(D=1)$ and denote the conditional covariate densities $p_1(z)\coloneqq p_0(z\mid D=1)$ and $p_{-1}(z)\coloneqq p_0(z\mid D=-1)$, 
and the marginal $p_0(z)=\pi p_1(z)+(1-\pi)p_{-1}(z)$. Then, we have $e_0(z)=P(D=1\mid Z=z)=\pi \frac{p_1(z)}{p_0(z)}$. 
Hence, the ATE Riesz representer is written as $\alpha^{\mathrm{ATE}}_0(D,Z) =
\mathbbm{1}[D=1]\frac{p_0(Z)}{\pi p_1(Z)}
-
\mathbbm{1}[D=-1]\frac{p_0(Z)}{(1-\pi)p_{-1}(Z)}$. 
Therefore, estimating $\alpha^{\mathrm{ATE}}_0$ reduces to estimating the density ratios $p_0/p_1$ and $p_0/p_{-1}$.

\paragraph{Data-score-based alternative.}
A Data-ScoreMatchingRiesz approach could also target $\alpha^{\mathrm{ATE}}_0$ by separately estimating the covariate scores for $p_0$, $p_1$, and $p_{-1}$ and then reconstructing the corresponding log densities. Pursuing this route requires specifying the estimation procedure, the treatment of normalizing constants in the reconstruction, and an error analysis. We therefore do not develop this alternative here.

\paragraph{Bridge construction.}
Construct a bridge family $\{p_t\}_{t\in[-1,1]}$ on the $Z$ space as follows.
Let $Z_0\sim p_0$, $Z_1\sim p_1$, and $Z_{-1}\sim p_{-1}$ be mutually independent.
Let $\beta^{(1)},\beta^{(2)}:[0,1]\to[0,1]$ be $C^2$ functions satisfying
\[
\beta^{(1)}(0)=1,\ \beta^{(2)}(0)=0,\qquad
\beta^{(1)}(1)=0,\ \beta^{(2)}(1)=1,
\]
with (for instance) $\beta^{(1)}$ nonincreasing and $\beta^{(2)}$ nondecreasing.
Define
\[
Z_t \coloneqq
\begin{cases}
\beta^{(1)}(t) Z_0+\beta^{(2)}(t) Z_1, & t\in[0,1],\\[4pt]
\beta^{(1)}(-t) Z_0+\beta^{(2)}(-t) Z_{-1}, & t\in[-1,0].
\end{cases}
\]
Let $p_t$ denote the induced law of $Z_t$.

\paragraph{Time score estimation.}
Let $s^{\text{time}}_{\theta,t}(z)$ approximate $s^{\text{time}}_t(z) = \partial_t\log p_t(z)$ on $t\in(-1,1)$. We fit $s^{\text{time}}_{\theta}$ by minimizing the integration-by-parts objective (the natural extension of Section~\ref{sec:dreic}):
\begin{align*}
&\calR^{\mathrm{ATE}}(s^{\text{time}}_{\theta}) \coloneqq
\bbE_{Z_{-1}\sim p_{-1}}\left[\lambda(-1) s^{\text{time}}_{\theta, -1}(Z_{-1})\right]\\
&\ \ \ \ \ \ \ \ \ \ \ \ \ \ \ \ \ \ \ \ \ \  -
\bbE_{Z_{1}\sim p_{1}}\left[\lambda(1) s^{\text{time}}_{\theta, +1}(Z_{1})\right] +\\
&\int_{-1}^{1}
\bbE_{Z_t\sim p_t}\left[
\partial_t\bigl(\lambda(t) s^{\text{time}}_{\theta, t}(Z_t)\bigr)
+\frac{1}{2}\lambda(t) s^{\text{time}}_{\theta, t}(Z_t)^2
\right]\rmd t.
\end{align*}
As discussed earlier, $\partial_t s^{\text{time}}_{\theta,t}(z)$ is a partial derivative in $t$ at fixed $z$.

\paragraph{Constructing $\widehat{\alpha}^{\mathrm{ATE}}$.}
After training, we obtain $\widehat{\log\frac{p_0}{p_1}}(z)\coloneqq \int_{1}^{0} s^{\text{time}}_{\theta,t}(z) \rmd t$ and $\widehat{\log\frac{p_0}{p_{-1}}}(z)\coloneqq \int_{0}^{-1} s^{\text{time}}_{\theta,t}(z) \rmd t$. Then, we estimate the Riesz representer as
\begin{align*}
&\widehat{\alpha}^{\mathrm{ATE}}(D,Z)
\coloneqq\\
&\mathbbm{1}[D=1]\frac{\exp\p{\widehat{\log\frac{p_0}{p_1}}(Z)}}{\pi}
-
\mathbbm{1}[D=-1]\frac{\exp\p{\widehat{\log\frac{p_0}{p_{-1}}}(Z)}}{1-\pi}.
\end{align*}

\begin{remark}[Alternative (logit) parameterization]
One can also estimate $\log\bigl(p_1(z)/p_{-1}(z)\bigr)$ via $\log\frac{p_1(z)}{p_{-1}(z)} \approx \int_{-1}^{1} s^{\text{time}}_{\theta,t}(z) \rmd t$
which yields a logistic representation of the propensity score:
\[
e_0(z)
\approx
\frac{1}{1+\exp\p{-\p{\log\frac{\pi}{1-\pi}+\int_{-1}^{1} s^{\text{time}}_{\theta,t}(z) \rmd t}}}.
\]
\end{remark}

\section{Policy Path Estimation Algorithm}
\label{appdx:policy_path_algo}
Algorithm~\ref{alg:policy_path} summarizes the estimator in Section~\ref{sec:policy_path}.

\begin{algorithm}[h!]
\caption{Cross-fitted policy path estimation under shift policies}
\label{alg:policy_path}
\begin{enumerate}
\item Input: data $\{(X_i,Y_i)\}_{i=1}^n$, grid $\{\delta_j\}_{j=1}^J$, $K$ folds.
\item For each fold $k$:
\begin{enumerate}
\item Fit $\widehat\gamma^{(-k)}$ on training folds; compute residuals $Y_i-\widehat\gamma^{(-k)}(X_i)$ for $i$ in fold $k$.
\item Fit a score model on training folds to obtain $\widehat{\alpha}^{\mathrm{AME},(-k)}$.
\item For each $\delta_j$:
\begin{enumerate}
\item Compute $\widehat{\log r}^{(-k)}_{\pm\delta_j}(X_i)$ for test points by integrating $\widehat{\alpha}^{\mathrm{AME},(-k)}$ along the $D$-path.
\item Calibrate $\widehat r^{(-k)}_{\pm\delta_j}$ to have mean one on the training fold; set $\widehat\alpha^{(-k)}_{\delta_j}=\widehat r^{(-k)}_{+\delta_j}-\widehat r^{(-k)}_{-\delta_j}$.
\item Form orthogonal scores
\[
\psi_{i,\delta_j}=
\widehat\gamma^{(-k)}(D_i+\delta_j,Z_i)-\widehat\gamma^{(-k)}(D_i-\delta_j,Z_i)
+\widehat\alpha^{(-k)}_{\delta_j}(X_i)\bigl(Y_i-\widehat\gamma^{(-k)}(X_i)\bigr).
\]
\end{enumerate}
\end{enumerate}
\item Output $\widehat\theta(\delta_j)=n^{-1}\sum_{i=1}^n \psi_{i,\delta_j}$ and pointwise standard errors from $\{\psi_{i,\delta_j}\}_{i=1}^n$.
\end{enumerate}
\end{algorithm}

\section{Pushforward Policies: Definitions and Data-Score-Based Construction of APEs and Policy Paths}
\label{appdx:pushforward}
\label{sec:ape_pushforward}

This appendix formalizes pushforward policies and explains why, when the pushforward map is known,
both (i) APE representers and (ii) the full policy path can be constructed from the observational
\emph{data score}. The key point is that the counterfactual density $p_\delta$ is then determined
by $p_0$ and a known map.

\subsection{Pushforward of a distribution}
Let $(\calX,\mathcal{B})$ be a measurable space, $P$ a probability measure on it, and
$T:\calX\to\calX$ a measurable map. The \emph{pushforward} (a.k.a.\ image measure) of $P$ by $T$ is
the measure $T_\#P$ defined by
\[
(T_\#P)(A)\coloneqq P(T^{-1}(A)),
\qquad A\in\mathcal{B}.
\]
Equivalently, if $X\sim P$, then $T(X)\sim T_\#P$.

\paragraph{Density under a diffeomorphism.}
When $\calX\subseteq\R^d$, $P$ admits a density $p$ w.r.t.\ Lebesgue measure, and
$T$ is a $C^1$ diffeomorphism, the pushforward $T_\#P$ admits a density
\begin{align}
\label{eq:cov_pushforward_density}
p_T(x)=p(T^{-1}(x))\left|\det\nabla T^{-1}(x)\right|.
\end{align}
Equation \eqref{eq:cov_pushforward_density} is the standard change-of-variables formula.

\subsection{Known pushforward policy families}
A counterfactual policy family $\{P_\delta\}_{\delta\in\R}$ is a \emph{known pushforward} of $P_0$ if
there exists a \emph{known} family of measurable maps $\{T_\delta:\calX\to\calX\}_{\delta\in\R}$ such that
\[
P_\delta=(T_\delta)_\#P_0,
\qquad\text{i.e.}\qquad
X_\delta=T_\delta(X)\ \ \text{for }X\sim P_0.
\]
Assume $P_0$ has a density $p_0$ and that each $T_\delta$ is a $C^1$ diffeomorphism.
Then $P_\delta$ has density
\begin{align}
\label{eq:pdelta_pushforward}
p_\delta(x)
=
p_0(T_\delta^{-1}(x))\left|\det\nabla T_\delta^{-1}(x)\right|.
\end{align}
Define the density ratio
\[
r_\delta(x)\coloneqq \frac{p_\delta(x)}{p_0(x)}.
\]
By \eqref{eq:pdelta_pushforward},
\begin{align}
\label{eq:ratio_pushforward_general}
\log r_\delta(x)
=
\underbrace{\log p_0(T_\delta^{-1}(x))-\log p_0(x)}_{\text{data-score term}}
+
\underbrace{\log\left|\det\nabla T_\delta^{-1}(x)\right|}_{\text{Jacobian term}}.
\end{align}

\subsection{From the data score to density ratios under known pushforwards}
Let the \emph{data score} be $s^{\mathrm{data}}_0(x)\coloneqq\nabla_x\log p_0(x)$.
The first term in \eqref{eq:ratio_pushforward_general} can be written as a line integral of the data score
along the known inverse path $u\mapsto T_u^{-1}(x)$:
\begin{align}
\label{eq:logratio_linescore_general}
\log p_0(T_\delta^{-1}(x))-\log p_0(x)
=
\int_0^\delta
\Bigl\langle s^{\mathrm{data}}_0(T_u^{-1}(x)),\ \partial_u T_u^{-1}(x)\Bigr\rangle
\,\rmd u.
\end{align}
Combining \eqref{eq:ratio_pushforward_general} and \eqref{eq:logratio_linescore_general} yields an explicit
data-score-based ratio formula:
\begin{align}
\label{eq:ratio_from_datascore_general}
\log r_\delta(x)
=
\int_0^\delta
\Bigl\langle s^{\mathrm{data}}_0(T_u^{-1}(x)),\ \partial_u T_u^{-1}(x)\Bigr\rangle
\,\rmd u
+
\log\left|\det\nabla T_\delta^{-1}(x)\right|.
\end{align}
Thus, once we estimate the data score $s^{\mathrm{data}}_0$ on $P_0$, the entire family
$\{r_\delta\}_{\delta}$ is available by (i) evaluating the learned score along the known path and
(ii) adding the known Jacobian term.

\paragraph{Translation (shift) policy.}
Let $X=(D,Z)$ and $T_\delta(d,z)=(d+\delta,z)$.
Then $T_\delta^{-1}(d,z)=(d-\delta,z)$ and $\left|\det\nabla T_\delta^{-1}\right|\equiv1$, so
\begin{align}
\label{eq:shift_ratio_from_score}
\log r_\delta(d,z)
=
\log p_0(d-\delta,z)-\log p_0(d,z)
=
-\int_0^\delta \partial_d\log p_0(d-u,z)\,\rmd u.
\end{align}
Recalling the AME representer $\alpha_0^{\mathrm{AME}}(d,z)=-\partial_d\log p_0(d,z)$, we obtain
\begin{align}
\label{eq:shift_ratio_from_ame_rr}
\log r_\delta(d,z)
=
\int_0^\delta \alpha_0^{\mathrm{AME}}(d-u,z)\,\rmd u,
\qquad
r_\delta(d,z)=\exp\left(\int_0^\delta \alpha_0^{\mathrm{AME}}(d-u,z)\,\rmd u\right).
\end{align}
This is the basic ``score-to-ratio'' identity used by Data-ScoreMatchingRiesz in the known-pushforward regime.

\subsection{APE estimation under known pushforwards from the data score}
Fix two counterfactual indices $\delta_1,\delta_2$ and define the APE
\[
\theta_0(\delta_1,\delta_2)
\coloneqq
\bbE_{X\sim p_{\delta_1}}[\gamma_0(X)]-\bbE_{X\sim p_{\delta_2}}[\gamma_0(X)].
\]
When $p_{\delta_j}\ll p_0$, the APE has the density-ratio form
\begin{align*}
\theta_0(\delta_1,\delta_2)
=
\bbE\Bigl[\alpha_{0,\delta_1,\delta_2}(X)\,\gamma_0(X)\Bigr],
\qquad
\alpha_{0,\delta_1,\delta_2}(x)\coloneqq r_{\delta_1}(x)-r_{\delta_2}(x).
\end{align*}
Under a known pushforward, each ratio $r_{\delta_j}$ can be constructed from the data score via
\eqref{eq:ratio_from_datascore_general} (or \eqref{eq:shift_ratio_from_ame_rr} for translations),
so the APE representer $\alpha_{0,\delta_1,\delta_2}$ is identified from the observational score and known policy maps.

\paragraph{Orthogonal-score estimation (generic form).}
Under a known pushforward $X_\delta=T_\delta(X)$, the APE admits the moment representation
\[
\theta_0(\delta_1,\delta_2)
=
\bbE\!\left[\gamma_0(T_{\delta_1}(X))-\gamma_0(T_{\delta_2}(X))\right]
\equiv
\bbE\!\left[m_{\delta_1,\delta_2}(W,\gamma_0)\right],
\qquad
m_{\delta_1,\delta_2}(W,\gamma)\coloneqq \gamma(T_{\delta_1}(X))-\gamma(T_{\delta_2}(X)).
\]
Let $\alpha_{0,\delta_1,\delta_2}(x)=r_{\delta_1}(x)-r_{\delta_2}(x)$ denote the corresponding Riesz representer,
so that $\bbE[m_{\delta_1,\delta_2}(W,\gamma)]=\bbE[\alpha_{0,\delta_1,\delta_2}(X)\gamma(X)]$ for all $\gamma$.
An orthogonal-score estimator is then
\[
\widehat\theta(\delta_1,\delta_2)
=
\frac1n\sum_{i=1}^n
\Bigl[
m_{\delta_1,\delta_2}(W_i,\widehat\gamma)
+
\widehat\alpha_{\delta_1,\delta_2}(X_i)\bigl(Y_i-\widehat\gamma(X_i)\bigr)
\Bigr],
\]
with $\widehat\alpha_{\delta_1,\delta_2}$ constructed from the learned score via
$\widehat r_\delta$ and $\widehat\alpha_{\delta_1,\delta_2}=\widehat r_{\delta_1}-\widehat r_{\delta_2}$.
Cross-fitted mean-one calibration and log clipping are recommended for stability (Appendix~\ref{appdx:calibration}).

\subsection{Policy-path estimation under known pushforwards}
The policy path in the main text is the symmetric APE curve
\[
\theta_0(\delta)\equiv \theta_0(+\delta,-\delta),
\qquad
\alpha_{0,\delta}(x)=r_{+\delta}(x)-r_{-\delta}(x).
\]
Under known pushforwards, we can estimate the \emph{entire curve} $\delta\mapsto\theta_0(\delta)$ by
reusing a \emph{single} data-score estimate of $p_0$:
for each $\delta$ on a grid, compute $\widehat r_{\pm\delta}$ from the learned score (plus known Jacobian),
form $\widehat\alpha_{0,\delta}=\widehat r_{+\delta}-\widehat r_{-\delta}$, and plug into the orthogonal score.

\paragraph{Shift policy and the AME limit (revisited).}
For translation policies, \eqref{eq:shift_ratio_from_ame_rr} implies that the full path representer
$\alpha_{0,\delta}$ is constructed from the AME representer by one-dimensional score integration.
Moreover, the path itself satisfies
\[
\theta_0(\delta)=\bbE[\gamma_0(D+\delta,Z)-\gamma_0(D-\delta,Z)],
\]
hence $\theta_0'(0)=2\bbE[\partial_d\gamma_0(D,Z)]=2\theta_0^{\mathrm{AME}}$ under mild smoothness
(Section~\ref{sec:policy_path}). This identity generally fails without a known pushforward structure.

\paragraph{Implementation.}
A cross-fitted algorithm that estimates $\delta\mapsto\theta_0(\delta)$ on a grid under shift policies
is given in Appendix~\ref{appdx:policy_path_algo}.
The only pushforward-specific step is the construction of $\widehat r_{\pm\delta}$ from the learned data score,
using \eqref{eq:shift_ratio_from_ame_rr} (or \eqref{eq:ratio_from_datascore_general} for general diffeomorphisms).

\section{Proof of Theorem~\ref{thm:bregmanmatching}}
\label{appdx:proof:thm:bregmanmatching}
\begin{proof}
Expand the definition of $\mathrm{BD}_g^\dagger$ and drop the term
\[
\int_0^1\int \lambda(t)\, g(\partial_t\log p_t(x))\, p_t(x)\, \rmd x\, \rmd t,
\]
which does not depend on $\theta$. The remaining $\theta$-dependent part is
\begin{align*}
\mathrm{BD}_g^\dagger\left(\partial_t\log p_t  \middle\|  s^{\text{time}}_\theta\right)
&=
\int_0^1
\bbE_{X_t\sim p_t}\left[
\lambda(t)\Bigl(g'(s^{\text{time}}_\theta(X_t,t))\, s^{\text{time}}_\theta(X_t,t)-g(s^{\text{time}}_\theta(X_t,t))\Bigr)
\right]\rmd t\\
&\quad
-\int_0^1\int
\lambda(t)\, g'(s^{\text{time}}_{\theta, t}(x))\, \partial_t\log p_t(x)\, p_t(x)\, \rmd x\, \rmd t.
\end{align*}
Using $\partial_t\log p_t(x)\, p_t(x)=\partial_t p_t(x)$, the second term becomes
\[
-\int_0^1\int \lambda(t)\, g'(s^{\text{time}}_{\theta, t}(x))\, \partial_t p_t(x)\, \rmd x\, \rmd t.
\]
Apply integration by parts in $t$ (for each fixed $x$):
\begin{align*}
-\int_0^1 \lambda(t)\, g'(s^{\text{time}}_{\theta, t}(x))\, \partial_t p_t(x)\, \rmd t
&=
-\Bigl[\lambda(t)\, g'(s^{\text{time}}_{\theta, t}(x))\, p_t(x)\Bigr]_{t=0}^{t=1}\\
&\quad
+\int_0^1 \partial_t\Bigl(\lambda(t)\, g'(s^{\text{time}}_{\theta, t}(x))\Bigr)\, p_t(x)\, \rmd t\\
&=
\lambda(0)\, g'(s^{\text{time}}_\theta(x,0))\, p_0(x)
-\lambda(1)\, g'(s^{\text{time}}_\theta(x,1))\, p_1(x)\\
&\quad
+\int_0^1 \Bigl(\lambda(t)\, \partial_t g'(s^{\text{time}}_{\theta, t}(x))
+\lambda'(t)\, g'(s^{\text{time}}_{\theta, t}(x))\Bigr)\, p_t(x)\, \rmd t.
\end{align*}
Integrating over $x$ and combining terms yields exactly $\mathrm{BD}_g(s^{\text{time}}_\theta)$
as stated in Theorem~\ref{thm:bregmanmatching}, up to constants independent of $\theta$.
\end{proof}

\section{Additional Distribution-Estimation Metrics for Diffusion Models}
\label{appdx:oko_other_metrics}

This appendix section records classical distribution-estimation guarantees for diffusion models
(in metrics other than the $L_2$ score norms used in the main text), as established by
\citet{Oko2023diffusionmodels}.

\paragraph{Total variation distance.}
For probability measures $\mu,\nu$ on $\calX$ with densities $p,q$, define
\[
\mathrm{TV}(\mu,\nu)\coloneqq \sup_{A\subseteq\calX}|\mu(A)-\nu(A)|
=\frac12\int_{\calX}|p(x)-q(x)| \rmd x.
\]

\paragraph{Wasserstein-1 distance.}
Define
\[
W_1(\mu,\nu)\coloneqq \inf_{\pi\in\Pi(\mu,\nu)}\int_{\calX\times\calX}\|x-y\|_2 \pi(\rmd x,\rmd y),
\]
where $\Pi(\mu,\nu)$ is the set of couplings of $(\mu,\nu)$.

\begin{theorem}[Total variation rate; \citet{Oko2023diffusionmodels}]
\label{thm:oko_tv_appdx}
Under the assumptions of \citet{Oko2023diffusionmodels}, the diffusion-model density estimator $\widehat \mu$
constructed by score matching satisfies
\[
\bbE\bigl[\mathrm{TV}(\widehat \mu, p_0)\bigr]
  \lesssim\
n^{-s/(2s+d)} \cdot \mathrm{polylog}(n).
\]
Moreover, the minimax lower bound over the Besov class matches
$n^{-s/(2s+d)}$ up to logarithmic factors.
\end{theorem}

\begin{theorem}[Wasserstein-1 rate; \citet{Oko2023diffusionmodels}]
\label{thm:oko_w1_appdx}
Under the assumptions of \citet{Oko2023diffusionmodels}, for any fixed $\delta>0$ the diffusion-model
estimator can be constructed so that
\[
\bbE\bigl[W_1(\widehat \mu, p_0)\bigr]
  \lesssim\
n^{-(s+1-\delta)/(2s+d)}.
\]
A minimax lower bound over Besov classes is $n^{-(s+1)/(2s+d)}$ (hence the rate above is
nearly minimax optimal).
\end{theorem}

\section{Proof of Theorem~\ref{thm:dreinf_rate}}
\label{sec:rates}

We now derive a generic nonparametric convergence rate for the \emph{time-score matching}
estimator used in ScoreMatchingRiesz (DRE-$\infty$).

\subsection{Setup: bridge distributions and time scores}

Let $q$ and $p$ be two densities on $\calX\subseteq\R^d$ with $q \ll p$,
and define the density ratio $r(x)\coloneqq q(x)/p(x)$.
Following the DRE-$\infty$ construction, introduce a continuum of bridge laws $\{p_t\}_{t\in[0,1]}$
connecting $q$ and $p$ with $p_0=q$ and $p_1=p$, and define the \emph{time score}
\[
s_{0,t}(x)\coloneqq \partial_t \log p_t(x).
\]
The key identity is the continuum telescoping representation
\[
\log r(x) = \log p_0(x)-\log p_1(x)
= \int_{1}^{0} s_{0,t}(x)\,\rmd t
= -\int_{0}^{1} s_{0,t}(x)\,\rmd t.
\]

We consider the weighted $L_2$-norm on functions $f\colon\calX\times[0,1]\to\R$:
\[
\|f\|_{\mu_\lambda}^2
\coloneqq
\int_0^1 \bbE_{X_t\sim p_t}\bigl[\lambda(t) f(X_t,t)^2\bigr]\rmd t,
\qquad
\lambda(t)>0.
\]

\subsection{Population objective and quadratic identity}

Let $s\colon\calX\times[0,1]\to\R$ be a differentiable candidate time-score function.
Consider the integration-by-parts population objective
\begin{align}
\calR(s)
&\coloneqq
\bbE_{X_0\sim q}\bigl[\lambda(0)s(X_0,0)\bigr]
-
\bbE_{X_1\sim p}\bigl[\lambda(1)s(X_1,1)\bigr]
\nonumber\\
&\quad+
\int_0^1
\bbE_{X_t\sim p_t}\Bigl[
\partial_t\bigl(\lambda(t)s(X_t,t)\bigr)
+\frac12 \lambda(t)s(X_t,t)^2
\Bigr]\rmd t.
\label{eq:dreinf_pop_obj}
\end{align}

\begin{lemma}[Quadratic identity]
\label{lem:quadratic_identity}
Assume $t\mapsto p_t(x)$ is differentiable for a.e.\ $x$, $s$ is differentiable in $t$,
and boundary terms are well-defined.
Then, up to an additive constant independent of $s$,
\[
\calR(s)
=
\calR(s_0)
+\frac12 \|s-s_0\|_{\mu_\lambda}^2.
\]
In particular, $s_0$ is the unique minimizer of $\calR(\cdot)$ over any convex class.
\end{lemma}

\begin{proof}
Expand the weighted squared loss:
\[
\frac12\|s-s_0\|_{\mu_\lambda}^2
=
\frac12\int_0^1 \bbE_{X_t\sim p_t}\bigl[\lambda(t)s^2\bigr]\rmd t
-\int_0^1 \bbE_{X_t\sim p_t}\bigl[\lambda(t) s  s_0\bigr]\rmd t
+\frac12\int_0^1 \bbE_{X_t\sim p_t}\bigl[\lambda(t)s_0^2\bigr]\rmd t.
\]
The last term does not depend on $s$ and can be absorbed into a constant.

For the cross term, use $s_{0,t}(x)=\partial_t\log p_t(x)$ so that
$s_{0,t}(x)p_t(x)=\partial_t p_t(x)$, hence
\[
\bbE_{X_t\sim p_t}\bigl[\lambda(t)s(X_t,t)\, s_{0,t}(X_t)\bigr]
=
\int \lambda(t)\, s(x,t)\, \partial_t p_t(x)\, \rmd x.
\]
Integrate by parts in $t$ (for each fixed $x$) to obtain
\begin{align*}
\int_0^1 \lambda(t)\, s(x,t)\, \partial_t p_t(x)\, \rmd t
&=
\Bigl[\lambda(t)\, s(x,t)\, p_t(x)\Bigr]_{t=0}^{t=1}
-\int_0^1 \partial_t\bigl(\lambda(t)s(x,t)\bigr)\, p_t(x)\, \rmd t.
\end{align*}
Integrating over $x$ yields
\begin{align*}
\int_0^1 \bbE_{X_t\sim p_t}\bigl[\lambda(t)s  s_0\bigr]\rmd t
&=
\bbE_{X_1\sim p}\bigl[\lambda(1)s(X_1,1)\bigr]
-\bbE_{X_0\sim q}\bigl[\lambda(0)s(X_0,0)\bigr]\\
&\quad-
\int_0^1 \bbE_{X_t\sim p_t}\Bigl[\partial_t\bigl(\lambda(t)s(X_t,t)\bigr)\Bigr]\rmd t.
\end{align*}
Substituting this expression into the expansion of $\frac12\|s-s_0\|_{\mu_\lambda}^2$ shows that
$\calR(s)$ in \eqref{eq:dreinf_pop_obj} equals $\calR(s_0)+\frac12\|s-s_0\|_{\mu_\lambda}^2$,
up to constants independent of $s$.
\end{proof}

\subsection{Proof of Theorem~\ref{thm:dreinf_rate}}

\begin{proof}
By definition of $\widehat s$ and the basic inequality,
\[
\widehat{\calR}_n(\widehat s)  \le  \widehat{\calR}_n(s)
\qquad\forall s\in\calF_n.
\]
Add and subtract the population objective:
\begin{align*}
\calR(\widehat s)
&\le
\widehat{\calR}_n(\widehat s) + \Delta_n
\le
\widehat{\calR}_n(s) + \Delta_n
\le
\calR(s) + 2\Delta_n
\qquad\forall s\in\calF_n,
\end{align*}
hence
\[
\calR(\widehat s) - \calR(s_0)
  \le\
\inf_{s\in\calF_n}\bigl\{\calR(s)-\calR(s_0)\bigr\} + 2\Delta_n.
\]
Using Lemma~\ref{lem:quadratic_identity},
\[
\calR(\widehat s)-\calR(s_0)=\frac12\|\widehat s-s_0\|_{\mu_\lambda}^2,
\qquad
\calR(s)-\calR(s_0)=\frac12\|s-s_0\|_{\mu_\lambda}^2,
\]
which gives
\[
\|\widehat s-s_0\|_{\mu_\lambda}^2
  \le\
\inf_{s\in\calF_n}\|s-s_0\|_{\mu_\lambda}^2 + 4\Delta_n.
\]
This proves the oracle inequality.

Under Assumption~\ref{ass:approx}, the approximation term is $\lesssim P_n^{-2s/(d+1)}$,
and under Assumption~\ref{ass:gen}, $\Delta_n\lesssim P_n \mathrm{polylog}(n)/n$.
Balancing the two terms yields the stated rate after taking square roots.
\end{proof}

\section{Downstream \texorpdfstring{$L_2$}{L2} Bounds: Log Ratios, Ratios, and Ratio-Based Representers}
\label{subsec:rates_ratio}

This section records how $L_2$ error in an estimated score field propagates to
$L_2$ error of integrated log ratios, ratios, and ratio-based representers.
It applies to Time-ScoreMatchingRiesz, and it also applies whenever $\widehat{\log r}$ is constructed as an integral of an estimated score along a known path.

\paragraph{Plug-in log ratio estimator and clipping.}
Define the plug-in log ratio estimator
\[
\widehat{\log r}(x)  \coloneqq  \int_1^0 \widehat s_t(x) \rmd t
=
-\int_0^1 \widehat s_t(x) \rmd t.
\]
When exponentiating, we may apply clipping:
for a constant $c>0$, define $\mathrm{clip}(u,[-c,c])\coloneqq \max\{-c,\min\{u,c\}\}$ and set
\[
\widehat r(x)\coloneqq \exp(\mathrm{clip}(\widehat{\log r}(x),[-c,c])).
\]

\begin{corollary}[Log-ratio and ratio bounds in $L_2(p)$ (with clipping)]
\label{cor:logratio_rate}
Under Assumption~\ref{ass:overlap} and $\lambda(\delta)\equiv 1$, we have
\[
\|\widehat{\log r}-\log r\|_{L_2(p)}
  \lesssim\
\|\widehat s-s_0\|_{\mu_1},
\]
where $\|g\|_{L_2(p)}^2\coloneqq \bbE_{X\sim p}[g(X)^2]$.
If we additionally clip $\widehat{\log r}$ to $[-c,c]$ and set
$\widehat r(x)\coloneqq \exp(\mathrm{clip}(\widehat{\log r}(x),[-c,c]))$,
then
\[
\|\widehat r-r\|_{L_2(p)}
  \lesssim\
e^{c} \|\widehat{\log r}-\log r\|_{L_2(p)}.
\]
Consequently, under the rate in Theorem~\ref{thm:dreinf_rate},
\[
\|\widehat r-r\|_{L_2(p)}
=
O_{\bbP}\Bigl(n^{-s/(2s+d+1)}\cdot \mathrm{polylog}(n)\Bigr).
\]
\end{corollary}

\begin{corollary}[APE and Policy-Path Representer Bound in $L_2(p_0)$]
\label{cor:ape_representer_rate}
Let $p_0$ be a baseline density on $\calX$ and let $p_+$ and $p_-$ be two counterfactual densities on $\calX$
that are absolutely continuous with respect to $p_0$.
Define the corresponding density ratios
\[
r_+(x)\coloneqq \frac{p_+(x)}{p_0(x)},
\qquad
r_-(x)\coloneqq \frac{p_-(x)}{p_0(x)},
\]
and the ratio-based representer
\[
\alpha_0(x)\coloneqq r_+(x)-r_-(x).
\]
Let $\widehat r_+$ and $\widehat r_-$ be estimators obtained by score integration (possibly using different bridges),
and define
\[
\widehat\alpha(x)\coloneqq \widehat r_+(x)-\widehat r_-(x).
\]
If each ratio estimator satisfies $\|\widehat r_{\pm}-r_{\pm}\|_{L_2(p_0)}=O_{\bbP}(\varepsilon_n)$, then
\[
\|\widehat\alpha-\alpha_0\|_{L_2(p_0)}
=O_{\bbP}(\varepsilon_n).
\]
In particular, if Theorem~\ref{thm:dreinf_rate} applies to both ratio estimators (with $p=p_0$ in Corollary~\ref{cor:logratio_rate}),
then one may take $\varepsilon_n=n^{-s/(2s+d+1)}\mathrm{polylog}(n)$.
\end{corollary}

\subsection{Proof of Corollary~\ref{cor:logratio_rate}}
\begin{proof}[Proof of Corollary~\ref{cor:logratio_rate}]
For any $x$,
\[
\widehat{\log r}(x)-\log r(x)
=
-\int_0^1 \bigl(\widehat s_t(x)-s_{0,t}(x)\bigr) \rmd t.
\]
By Cauchy--Schwarz,
\[
\bigl(\widehat{\log r}(x)-\log r(x)\bigr)^2
\le
\int_0^1 \bigl(\widehat s_t(x)-s_{0,t}(x)\bigr)^2 \rmd t.
\]
Take expectation under $X\sim p$ and use Assumption~\ref{ass:overlap} with $p\le c^{-1}p_t$:
\begin{align*}
\|\widehat{\log r}-\log r\|_{L_2(p)}^2
&=
\bbE_{X\sim p}\Bigl[\bigl(\widehat{\log r}(X)-\log r(X)\bigr)^2\Bigr]\\
&\le
\int_0^1 \bbE_{X\sim p}\Bigl[\bigl(\widehat s_t(X)-s_{0,t}(X)\bigr)^2\Bigr]\rmd t\\
&\le
\frac{1}{c}\int_0^1 \bbE_{X_t\sim p_t}\Bigl[\bigl(\widehat s_t(X_t)-s_{0,t}(X_t)\bigr)^2\Bigr]\rmd t
=
\frac{1}{c}\|\widehat s-s_0\|_{\mu_1}^2.
\end{align*}
This proves the first inequality.

For the ratio, on the clipped range $[-c,c]$, the map $u\mapsto e^u$ is $e^c$-Lipschitz, hence
\[
|\widehat r(x)-r(x)|
\le
e^c |\widehat{\log r}(x)-\log r(x)|
\]
and the $L_2(p)$ bound follows by squaring and integrating.
\end{proof}

\subsection{Proof of Corollary~\ref{cor:ape_representer_rate}}
\begin{proof}[Proof of Corollary~\ref{cor:ape_representer_rate}]
Immediate from the triangle inequality:
\[
\|\widehat\alpha-\alpha_0\|_{L_2(p_0)}
\le
\|\widehat r_+-r_+\|_{L_2(p_0)}+\|\widehat r_--r_-\|_{L_2(p_0)}.
\]
\end{proof}

\paragraph{Optional remark (finite many $\delta$'s).}
If the score model is trained conditionally in a finite set of policy parameters
$\delta\in\{\delta_1,\dots,\delta_M\}$, one can apply a union bound to obtain simultaneous
rates over all $\delta_m$ at the price of an additional $\log M$ factor in the complexity term.

\section{From Data-ScoreMatchingRiesz to the AME Representer: Smoothing Bias and DML Conditions}
\label{app:smoothing-bias}

This appendix addresses the concern that Data-ScoreMatchingRiesz learns a \emph{smoothed} score
while the AME Riesz representer involves the \emph{unsmoothed} score of $p_0$.
We write the discussion in terms of Data-ScoreMatchingRiesz / Time-ScoreMatchingRiesz (not DSM/DRE-$\infty$).

\subsection{What Data-ScoreMatchingRiesz estimates}
Let $p_0$ be the density of $X\in\mathbb R^d$.
For $\sigma>0$, define the Gaussian-smoothed density $p_\sigma=p_0 * \varphi_\sigma$ and its data score
$s_\sigma(x):=\nabla_x \log p_\sigma(x)$.
Data-ScoreMatchingRiesz produces $\widehat s_\sigma(\cdot)$ that targets $s_\sigma(\cdot)$.
For AME with $X=(D,Z)$, the Riesz representer is
\[
\alpha_0^{\mathrm{AME}}(d,z)=-\partial_d \log p_0(d,z)=-(s_0(d,z))_d.
\]
Our implementation uses a small evaluation noise $\sigma_{\mathrm{eval}}$ and sets
$\widehat \alpha_{\mathrm{AME}}(d,z):=-(\widehat s_{\sigma_{\mathrm{eval}}}(d,z))_d$.

\subsection{Error decomposition: estimation error + smoothing bias}
\label{app:smoothing-bias:decomp}

A direct decomposition is
\begin{align}
\label{eq:smoothing-bias:decomp}
\|\widehat \alpha_{\mathrm{AME}}-\alpha^{\mathrm{AME}}_0\|_{L_2(P_0)}
\;\le\;
\underbrace{\|(\widehat s_{\sigma_{\mathrm{eval}}}- s_{\sigma_{\mathrm{eval}}})_d\|_{L_2(P_0)}}_{\text{(A) score estimation error}}
+
\underbrace{\|(s_{\sigma_{\mathrm{eval}}}- s_{0})_d\|_{L_2(P_0)}}_{\text{(B) smoothing bias}}.
\end{align}

\paragraph{(A) Estimation error controlled by Oko et al.\ (2023).}
Existing diffusion/score-matching theory provides $L_2$-type guarantees for estimating $s_\sigma$.
In particular, under Besov-type smoothness, \citet{Oko2023diffusionmodels} gives (near-)minimax rates for score estimation of
Gaussian-smoothed densities, which we directly use as a bound on term (A).

\paragraph{(B) Smoothing bias requires a separate (and standard) control.}
Term (B) is a deterministic approximation error induced by Gaussian smoothing.
Under mild regularity (smoothness of $p_0$ and a lower bound away from zero),
$s_{\sigma}(x)\to s_0(x)$ as $\sigma\downarrow 0$ and the bias decays polynomially in $\sigma$.

\begin{lemma}[Smoothing bias of the data score (illustrative statement)]
\label{lem:smoothing-bias}
Assume $p_0$ is supported on a bounded domain, satisfies $\inf_x p_0(x)\ge c>0$,
and has smoothness $s>1$ (e.g., $p_0\in B^{s}_{2,2}$ with bounded first derivative).
Then there exists $\beta>0$ (typically $\beta\approx s-1$) and a constant $C$ such that, for all sufficiently small $\sigma$,
\[
\|s_{\sigma}-s_0\|_{L_2(P_0)} \le C\,\sigma^{\beta}.
\]
Consequently, the AME representer bias satisfies
$\|(\,s_{\sigma}-s_{0}\,)_d\|_{L_2(P_0)} \le C\,\sigma^{\beta}$.
\end{lemma}

Lemma~\ref{lem:smoothing-bias} is a standard consequence of kernel-approximation bounds for $p_\sigma$ and $\nabla p_\sigma$,
combined with the identity $s_\sigma=\nabla p_\sigma/p_\sigma$ and the lower bound on $p_0$.

\subsection{Implications for DML rate conditions and choice of \texorpdfstring{$\sigma_{\mathrm{eval}}$}{sigma}}
\label{app:smoothing-bias:dml}

Combining \eqref{eq:smoothing-bias:decomp} with Lemma~\ref{lem:smoothing-bias} yields
\[
\|\widehat \alpha_{\mathrm{AME}}-\alpha^{\mathrm{AME}}_0\|_{L_2(P_0)}
\;\lesssim\;
\underbrace{\|(\widehat s_{\sigma_{\mathrm{eval}}}- s_{\sigma_{\mathrm{eval}}})_d\|_{L_2(P_0)}}_{\text{controlled by \citealp{Oko2023diffusionmodels}}}
+
\underbrace{\sigma_{\mathrm{eval}}^{\beta}}_{\text{smoothing bias}}.
\]
For debiased ML with an orthogonal score, a sufficient condition is the product rate
\[
\|\widehat \alpha-\alpha_0\|_{L_2(P_0)}\cdot \|\widehat \gamma-\gamma_0\|_{L_2(P_0)} = o_p(n^{-1/2}).
\]
Thus, $\sigma_{\mathrm{eval}}$ should be chosen so that the smoothing bias term is negligible at the DML scale.
For example, if $\|\widehat \gamma-\gamma_0\|_{L_2}=O_p(n^{-b})$ for some $b>0$, it suffices that
$\sigma_{\mathrm{eval}}^{\beta}=o(n^{-1/2+b})$.
This is the usual \emph{undersmoothing} principle: decrease $\sigma_{\mathrm{eval}}$ with $n$
to kill the smoothing bias faster than the DML remainder.

\subsection{Why this issue is specific to Data-ScoreMatchingRiesz (and not Time-ScoreMatchingRiesz)}
Time-ScoreMatchingRiesz does not introduce Gaussian smoothing of $p_0$; it estimates a \emph{time score}
$s^{\mathrm{time}}_t(x)=\partial_t\log p_t(x)$ along an explicit bridge.
The representer is then obtained by integrating the learned time score and exponentiating to form ratios.
Accordingly, the analogue of \eqref{eq:smoothing-bias:decomp} is an \emph{error propagation} bound
from time-score estimation error to log-ratio and ratio error.
These bounds are controlled by $L_2$ guarantees for $s^{\mathrm{time}}$ together with stabilization
(e.g., clipping and mean-one calibration) when exponentiating.

\section{Remarks on the Implementation of ScoreMatchingRiesz}

\subsection{Infinitesimal classification viewpoint}
Time-ScoreMatchingRiesz is closely related to \emph{infinitesimal classification} for DRE.
\citet{Choi2022densityratio} interprets time score matching as the continuum limit of telescoping density-ratio estimation along a bridge of distributions $\{p_t\}_{t\in[0,1]}$, where one learns the instantaneous change rate $\partial_t\log p_t(x)$.

For small $\Delta t$, the Bayes-optimal classifier that discriminates samples from $p_t$ and $p_{t+\Delta t}$ admits a first-order expansion around $1/2$, whose leading term is proportional to the time score.
Moreover, letting $\Delta t\to0$ in the binary cross-entropy objective yields a squared-error objective equivalent to time score matching.
Thus, rather than training many separate classifiers, we learn a single time-score field and recover endpoint log-ratios by integrating it along the bridge.

We exploit the same mechanism for Riesz representer estimation.
In many causal targets, the representer can be expressed via endpoint density ratios; our bridge construction converts global ratio estimation into local time-score learning, whose aggregation yields the density ratio and hence the representer used in orthogonal-score estimation.

\subsection{Ratio stabilization: cross-fitted calibration and clipping}
\label{appdx:calibration}
When ratios are constructed by exponentiating estimated log ratios, both multiplicative constants and numerical instability can degrade downstream estimation.
We apply two stabilizers, computed \emph{only on the training fold} and then used for test-fold evaluation:
\begin{enumerate}[topsep=0pt, itemsep=0pt, leftmargin=*]
\item \textbf{Cross-fitted mean-one calibration.}
Given $\widehat{\ell}(x)=\widehat{\log r}(x)$, define
\[
\widehat{\ell}^{\,\mathrm{cal}}(x)
:= \widehat{\ell}(x) - \log \bbE_{\text{train}}\!\left[\exp\{\widehat{\ell}(X)\}\right],
\qquad
\widehat r^{\,\mathrm{cal}}(x):=\exp\{\widehat{\ell}^{\,\mathrm{cal}}(x)\},
\]
so that $\bbE_{\text{train}}[\widehat r^{\,\mathrm{cal}}(X)]=1$.
\item \textbf{Log clipping before exponentiation.}
For a moderate constant $c>0$, clip
\[
\widehat{\ell}^{\,\mathrm{clip}}(x) := \mathrm{clip}\!\left(\widehat{\ell}^{\,\mathrm{cal}}(x),[-c,c]\right),
\qquad
\widehat r(x):=\exp\{\widehat{\ell}^{\,\mathrm{clip}}(x)\}.
\]
\end{enumerate}

\subsection{Bridge design and overlap in practice}
\label{appdx:bridge_overlap}
Assumption~\ref{ass:overlap} is a convenient sufficient condition for analyzing Time-ScoreMatchingRiesz, but it can be demanding when endpoint distributions are far apart.
The following choices make overlap along the bridge more plausible and stabilize the implied weights:
\begin{enumerate}[topsep=0pt, itemsep=0pt, leftmargin=*]
\item \textbf{Smooth interpolation schedule $\beta$.}
For interpolation bridges $X_t=\beta^{(1)}(t)X_0+\beta^{(2)}(t)X_1$, avoid rapid changes near $t=0$ and $t=1$ to prevent unstable endpoint behavior.
A simple choice is the cubic schedule
\[
\beta^{(2)}(t)=3t^2-2t^3,\qquad \beta^{(1)}(t)=1-\beta^{(2)}(t),
\]
which preserves $\beta^{(2)}(0)=0$, $\beta^{(2)}(1)=1$ while flattening endpoint derivatives.
\item \textbf{Endpoint downweighting via $\lambda$.}
Downweight endpoints by choosing $\lambda(t)$ small near $t=0$ and $t=1$ (e.g., $\lambda(t)=t(1-t)$), or equivalently sample $t$ away from endpoints and treat endpoint expectations separately.
\item \textbf{If overlap is weak.}
If exponentiated ratios become heavy-tailed, add mild Gaussian noise along the bridge for interior $t$ (keeping noise close to zero near endpoints) and apply ratio-stage stabilization via log clipping and mean-one calibration (Appendix~\ref{appdx:calibration}).
\end{enumerate}

\paragraph{A simple diagnostic.}
Report (i) the fraction of test-fold points affected by log-ratio clipping and (ii) a high quantile of the implied weights.
Large values typically indicate weak overlap under the chosen bridge and/or insufficient stabilization.

\subsection{Stochastic policy intervention (non-pushforward APE)}
\label{appdx:stochastic_policy}
This appendix specifies the stochastic policy intervention used to illustrate a regime where data-score-based APE construction is not available.
Write $X=(D,Z)$ with $Z\in\R^2$ and let $P_0$ denote the observational law.
The policy keeps the marginal law of $Z$ fixed at $p_0(z)$ but replaces the conditional law of $D\mid Z=z$ by a new Gaussian distribution:
\[
D^{(+)}\mid Z=z \sim \calN\bigl(m_{+}(z),s_{+}(z)^2\bigr),\qquad
D^{(-)}\mid Z=z \sim \calN\bigl(m_{-}(z),s_{-}(z)^2\bigr),
\]
where $m_{\pm}$ and $s_{\pm}$ are nonlinear functions of $z$ (chosen to induce substantial but smooth policy changes while maintaining overlap).
The resulting counterfactual densities satisfy $p_{\pm}(d,z)=p_0(z)\,p_{\pm}(d\mid z)$ and are generally not representable as $p_0(T^{-1}(d,z))$ for any known deterministic map $T$.
Consequently, $\nabla\log p_0(d,z)$ does not determine $p_{\pm}/p_0$, so diffusion-style score matching on $P_0$ alone cannot identify
$\alpha^{\mathrm{APE}}(x)=p_+(x)/p_0(x)-p_-(x)/p_0(x)$.
By contrast, DRE-$\infty$ applies directly by treating $(q,p)=(p_{\pm},p_0)$ and using simulated samples from $p_{\pm}$ together with observed samples from $p_0$.

\subsection{Practical cautions for policy-path estimation}
Policy-path estimation involves repeated ratio construction across $\delta$, which can magnify error:
\begin{itemize}[topsep=0pt, itemsep=0pt, leftmargin=*]
\item \textbf{Exponentiation and overlap.}
Exponentiating integrated scores can amplify small systematic errors, and limited overlap between $p_0$ and $p_{\pm\delta}$ typically worsens as $|\delta|$ grows.
\item \textbf{Stabilization is often essential.}
Mean-one calibration and conservative log clipping can materially improve stability of $\theta(\delta)$ estimation (Appendix~\ref{appdx:calibration}).
\item \textbf{Bridge and weighting choices matter (time scores).}
Downweighting endpoints and avoiding degenerate endpoint behavior are practical safeguards when estimating ratios along a bridge (Appendix~\ref{appdx:bridge_overlap}).
\end{itemize}
These issues are especially visible when estimating an entire curve $\delta\mapsto\theta(\delta)$, rather than a single APE at a fixed $\delta$.

\section{Bregman--Riesz regression: feasible Bregman risks for representer estimation}
\label{sec:kato-bregman-riesz}

This appendix records a Bregman-divergence formulation of \emph{generalized Riesz regression} due to
\citet{Kato2026rieszrepresenter}. The formulation is useful for viewing a range of representer estimators
(including classical Riesz regression and score-matching-type objectives) through a single optimization lens.

\paragraph{Riesz representer identity.}
Recall that the Riesz representer $\alpha_0\in L_2(P_X)$ associated with the linear functional
$\gamma\mapsto \bbE[m(W,\gamma)]$ satisfies
\begin{align}
  \bbE\!\left[\alpha_0(X)\,\gamma(X)\right]
  \;=\;
  \bbE\!\left[m(W,\gamma)\right],
  \qquad\forall \gamma\in\Gamma,
  \label{eq:riesz-def}
\end{align}
for a suitable class $\Gamma\subseteq L_2(P_X)$.

\paragraph{From an oracle Bregman risk to a feasible objective.}
Let $g:\mathbb{R}\to\mathbb{R}$ be convex and differentiable, and define the scalar Bregman divergence
\begin{align*}
  D_g(u,v)\coloneqq g(u)-g(v)-g'(v)(u-v).
\end{align*}
A natural ``oracle'' criterion is the Bregman risk
\begin{align}
  \mathcal{B}_g^\dagger(\alpha)
  \coloneqq
  \bbE\!\left[D_g\!\left(\alpha_0(X),\alpha(X)\right)\right],
  \label{eq:oracle-bregman-risk}
\end{align}
which is minimized at $\alpha=\alpha_0$ (uniquely if $g$ is strictly convex).
Although \eqref{eq:oracle-bregman-risk} involves the unknown $\alpha_0$, \citet{Kato2026rieszrepresenter} shows that it is equivalent---up to an additive constant independent of $\alpha$---to the \emph{feasible} objective
\begin{align}
  \mathcal{B}_g(\alpha)
  \coloneqq
  \bbE\!\Big[
    -g(\alpha(X))
    + g'(\alpha(X))\,\alpha(X)
    - m\!\bigl(W,\, g'(\alpha)\bigr)
  \Big],
  \label{eq:kato-feasible-bregman-risk}
\end{align}
where $g'(\alpha)$ denotes the composition $x\mapsto g'(\alpha(x))$.
Indeed, expanding \eqref{eq:oracle-bregman-risk} and using \eqref{eq:riesz-def} with $\gamma=g'(\alpha)$ yields
\[
\mathcal{B}_g^\dagger(\alpha)=\bbE[g(\alpha_0(X))]+\mathcal{B}_g(\alpha),
\]
so minimizing \eqref{eq:kato-feasible-bregman-risk} targets $\alpha_0$ without evaluating $\alpha_0$.

Empirical risk minimization for a sample analog of \eqref{eq:kato-feasible-bregman-risk}, with regularization and a model class for $\alpha$, yields \emph{generalized Riesz regression} \citep{Kato2026rieszrepresenter}.
The generator $g$ (loss) and the parameterization of $\alpha$ (link/model) jointly determine which representers are favored.
This ``loss--link'' viewpoint is used in Section~\ref{sec:generalization_smr} when interpreting score-based constructions.

\paragraph{Quadratic generator and classical Riesz regression.}
For $g(u)=\tfrac12u^2$ we have $g'(u)=u$ and \eqref{eq:kato-feasible-bregman-risk} reduces to
\begin{align}
  \mathcal{B}_{\mathrm{SQ}}(\alpha)
  =
  \bbE\!\Big[\tfrac12 \alpha(X)^2 - m(W,\alpha)\Big],
  \label{eq:riesz-regression}
\end{align}
the population objective underlying (regularized) Riesz regression in automatic debiasing
\citep{Chernozhukov2021automaticdebiased,Chernozhukov2022automaticdebiased}.
In derivative-functionals such as AME, where the representer is a (negative) score, \eqref{eq:riesz-regression}
is also the population Hyv\"arinen score-matching objective; see Section~\ref{sec:dreic} and Appendix~\ref{appdx:hsm}.

\paragraph{Nonquadratic generators and classical weighting objectives.}
When the target representer is ratio-based (ATE, APE, policy-path points), $\alpha_0$ is (a signed combination of)
density ratios and therefore a balancing weight.
With exponential/logistic-type links, KL-type generators recover calibrated weighting objectives that are classical in
covariate balancing and density-ratio estimation (e.g., tailored loss minimization and entropy-style balancing)
\citep{Zhao2019covariatebalancing,Sugiyama2012densityratio,Kato2026rieszrepresenter}.
Appendix~\ref{sec:loss_link_balancing} records the relevant balancing identities and their role in controlling the Neyman remainder.

\paragraph{Relation to Bregman and generalized score matching.}
Beyond the quadratic case, there is a broader literature on Bregman-divergence formulations of score/ratio estimation.
For unnormalized models, \citet{Gutman2011bregmandivergence} embed score matching in a larger Bregman framework;
\citet{Lyu2009innterpretationand} generalizes score matching via linear operators, yielding generalized Fisher divergences.
These perspectives align with the message above: representer estimation and score/ratio learning can often be phrased as minimizing a feasible Bregman risk.
\section{Losses, Link Functions, and Covariate Balancing}
\label{sec:loss_link_balancing}

This appendix complements Section~\ref{sec:generalization_smr} by making explicit the balancing interpretation of
ratio-based Riesz representers and by clarifying how the choice of (i) a loss (through the generator $g$ in
Theorem~\ref{thm:bregmanmatching}) and (ii) a link map (score-to-weight transformation) determines the resulting
balancing weights. The discussion is aligned with generalized Riesz regression \citep{Kato2026rieszrepresenter}.

\subsection{Balancing Identities Implied by Ratio-Based Representers}
\label{sec:balancing_identity}

Let $q$ and $p$ be two densities on $\calX$ with $q\ll p$, and define the density ratio
$r(x)\coloneqq q(x)/p(x)$.
Then, for any measurable test function $\varphi\colon\calX\to\R$ such that the expectations exist,
\begin{align}
\label{eq:population_balance_general}
\bbE_{X\sim p}\bigl[r(X) \varphi(X)\bigr]
=
\bbE_{X\sim q}\bigl[\varphi(X)\bigr].
\end{align}
Thus, an exact ratio $r$ is a moment-transport weight that maps expectations under $p$ to those under $q$.

\paragraph{ATE balancing.}
Let $X=(D,Z)$ with $D\in\{1,-1\}$ and propensity $e_0(z)=\bbP(D=1\mid Z=z)$.
The ATE representer is
\[
\alpha^{\mathrm{ATE}}_0(D,Z)
=
\frac{\mathbbm{1}[D=1]}{e_0(Z)}-\frac{\mathbbm{1}[D=-1]}{1-e_0(Z)}.
\]
Then \eqref{eq:population_balance_general} implies that, for any $\varphi\colon\calZ\to\R$,
\begin{align*}
\bbE\Bigl[\mathbbm{1}[D=1]\frac{1}{e_0(Z)} \varphi(Z)\Bigr]
&=
\bbE[\varphi(Z)],
\\
\bbE\Bigl[\mathbbm{1}[D=-1]\frac{1}{1-e_0(Z)} \varphi(Z)\Bigr]
&=
\bbE[\varphi(Z)].
\end{align*}

\paragraph{APE balancing.}
For APE under distribution shift, the representer is a difference of ratios.
Let $p_0$ be the baseline density and $p_{1}$ and $p_{-1}$ two counterfactual densities on $\calX$ with
$r_{\pm 1}(x)\coloneqq p_{\pm 1}(x)/p_0(x)$.
Then $\alpha^{\mathrm{APE}}_0(x)=r_1(x)-r_{-1}(x)$ and, for any $\varphi\colon\calX\to\R$,
\[
\bbE_{X\sim p_0}\bigl[r_{1}(X) \varphi(X)\bigr]
=
\bbE_{X\sim p_1}\bigl[\varphi(X)\bigr],
\qquad
\bbE_{X\sim p_0}\bigl[r_{-1}(X) \varphi(X)\bigr]
=
\bbE_{X\sim p_{-1}}\bigl[\varphi(X)\bigr].
\]
The same identities apply to each fixed-$\delta$ point on the policy path, where $\alpha_{0,\delta}=r_{+\delta}-r_{-\delta}$.

\subsection{Loss and Link Functions in ScoreMatchingRiesz}
\label{sec:loss_link_scorematchingriesz}

\paragraph{Link functions (mapping scores to weights).}
ScoreMatchingRiesz estimates log ratios (or log odds) by integrating score fields and then applies a link map.

\begin{itemize}[topsep=0pt, itemsep=0pt, partopsep=0pt, leftmargin=*]
\item \textbf{Log link (ratios).}
For endpoints $(q,p)$ connected by a bridge $\{p_t\}_{t\in[0,1]}$, the endpoint log ratio satisfies
$\log\{q(x)/p(x)\}=\int_{1}^{0} \partial_t\log p_t(x)\,dt$ (Section~\ref{sec:dreic}).
Time-ScoreMatchingRiesz therefore sets
\begin{align*}
\widehat r(x)\coloneqq \exp\bigl(\widehat{\log r}(x)\bigr),
\qquad
\widehat{\log r}(x)\coloneqq \int_{1}^{0} s^{\text{time}}_{\theta, t}(x) \rmd t,
\end{align*}
(optionally with clipping and mean-one calibration; Appendix~\ref{appdx:calibration}).
The exponential map enforces $\widehat r(x)>0$ by construction.

\item \textbf{Logit link (propensities).}
In ATE with $D\in\{1,-1\}$, write $\pi\coloneqq\bbP(D=1)$ and $p_{\pm1}(z)=p_0(z\mid D=\pm1)$.
Then
\[
\mathrm{logit}\, e_0(z)
=
\log\frac{e_0(z)}{1-e_0(z)}
=
\log\frac{\pi}{1-\pi}+\log\frac{p_1(z)}{p_{-1}(z)}.
\]
Using Time-ScoreMatchingRiesz to estimate $\log\{p_1(z)/p_{-1}(z)\}$ by score integration yields a logistic-link estimator
\begin{align*}
\widehat e(z)
=
\frac{1}{1+\exp\Bigl\{-\Bigl(\log\frac{\pi}{1-\pi}+\widehat{\log\frac{p_1}{p_{-1}}}(z)\Bigr)\Bigr\}}.
\end{align*}
\end{itemize}

\paragraph{Loss functions (how the score model is trained).}
The role of the loss in Time-ScoreMatchingRiesz is played by the time-score fitting objective.
Theorem~\ref{thm:bregmanmatching} shows that the squared-loss objective is a special case of a Bregman family indexed by $g$.
Changing $g$ changes which local discrepancies in the score field are penalized more heavily and thus affects the implied weights after exponentiation.

\paragraph{Normalization as a balancing constraint.}
A minimal balancing condition for ratios is $\bbE_{X\sim p}[r(X)]=1$, which is \eqref{eq:population_balance_general} with $\varphi\equiv 1$.
Because ratios recovered from exponentiated integrals can be shifted by an additive constant in $\widehat{\log r}$,
we impose cross-fitted mean-one calibration on the training fold and apply conservative log clipping before exponentiation when needed
(Appendix~\ref{appdx:calibration}).

\subsection{Automatic Covariate Balancing and Neyman Error}
\label{sec:autobalance_neyman_appendix}

\paragraph{Automatic covariate balancing (generalized Riesz regression).}
Let $\phi:\calX\to\R^p$ be a feature map and consider a representer model $\alpha_\beta(x)$ with a link matched to a Bregman generator $g$
(e.g., exponential tilting for KL-type losses).
For generalized Riesz regression (Appendix~\ref{sec:kato-bregman-riesz}), the first-order optimality/KKT conditions imply approximate sample balancing:
\begin{align}
\label{eq:acb_kkt}
\left|
\frac{1}{n}\sum_{i=1}^n \Bigl\{\alpha_{\widehat\beta}(X_i)\phi_j(X_i)-m(W_i,\phi_j)\Bigr\}
\right|
\le \lambda,
\qquad j=1,\dots,p,
\end{align}
with equality under exact balance when $\lambda=0$ \citep{Kato2026rieszrepresenter}.
When $\alpha_0$ is a density ratio, \eqref{eq:acb_kkt} is a direct sample covariate-balancing statement.

\paragraph{Implication for the Neyman error.}
For the orthogonal estimator
$\widehat{\theta}=\bbE_n[m(W,\widehat\gamma)+\widehat\alpha(X)\{Y-\widehat\gamma(X)\}]$,
the leading remainder is driven by products of nuisance errors.
A convenient sample diagnostic is the imbalance term
$\bbE_n[\widehat\alpha(X)\widehat\gamma(X)-m(W,\widehat\gamma)]$.
If $\widehat\gamma$ is (approximately) in the span of $\{\phi_j\}$, then \eqref{eq:acb_kkt} controls this imbalance and reduces the Neyman remainder;
in linear settings it can vanish under exact balance ($\lambda=0$), yielding ``automatic'' orthogonalization
\citep{Kato2026rieszrepresenter,BrunsSmith2025augmentedbalancing}.

\paragraph{Connection to augmented balancing weights.}
In linear settings, augmented balancing weights can be rewritten as a single regression estimator with coefficients that interpolate between
a base regression and unregularized least squares, clarifying how weight regularization interacts with regression regularization
\citep{BrunsSmith2025augmentedbalancing}.

\section{Additional Synthetic Simulation Designs}
\label{sec:additional_sims}

In addition to the baseline Gaussian design in Section~\ref{sec:experiments}, we consider three additional synthetic designs (Cases~1--3).
Cases~1--2 use shift policies (known pushforwards), so the APE/policy-path representer can be constructed from the data score of $P_0$
(Appendix~\ref{sec:ape_pushforward}).
Case~3 uses a stochastic intervention that changes $D\mid Z$ and is not a pushforward; in this case the data-score construction is unavailable
for APE/policy-path estimation, and we rely on Time-SMR. In Case~3, we also compare Time-SMR with and without joint training.

Across all cases, we follow the protocol in Section~\ref{sec:experiments}:
two-fold cross-fitting, the same MLP architecture for $\widehat{\gamma}$,
and trial-wise bias, MSE, and empirical 95\% coverage (truth approximated by population Monte Carlo).
For shift policies, we also plot $\delta\mapsto\theta_0(\delta)$ on a grid $\delta\in[0,\delta_{\max}]$.

\begin{table}[t]
\centering
\caption{Availability of APE/policy-path representer construction in the additional designs.
Data-SMR refers to the data-score-based pushforward construction in Appendix~\ref{sec:ape_pushforward};
Time-SMR refers to endpoint-sample-based ratio learning via time scores (DRE-$\infty$).}
\scalebox{0.7}{
\begin{tabular}{lcc}
\hline
Design & Data-SMR & Time-SMR \\
\hline
Case~1: High-dimensional Gaussian $P_0$, sparse outcome, shift policy & yes (pushforward) & yes \\
Case~2: High-dimensional Gaussian mixture $P_0$, random-feature outcome, shift policy & yes (pushforward) & yes \\
Case~3: Stochastic intervention (resampling $D\mid Z$) & no (non-pushforward) & yes \\
\hline
\end{tabular}
}
\label{tab:additional_sims_applicability}
\end{table}

Across all cases, we follow the evaluation protocol in Section~\ref{sec:experiments}:
two-fold cross-fitting, the same MLP architecture for $\widehat{\gamma}$,
and trial-wise bias, mean squared error (MSE), and empirical 95\% coverage relative to population Monte Carlo approximations.
For shift policies, we also report policy-path plots on a grid $\delta\in[0,\delta_{\max}]$.

\subsection{Case~1: High-Dimensional Gaussian Regressors with Sparse Outcome under a Shift Policy}
\label{sec:sim_hd_sparse_shift}

This design increases the regressor dimension and makes the outcome regression sparse.

\paragraph{Regressors.}
Let $X=(D,Z)\in\R^p$ with $p=50$.
Draw
\[
X\sim P_0\equiv\calN(0,\Sigma),
\]
where $\Sigma$ is Toeplitz (correlations decay with coordinate distance).
We set $D=X^{(1)}$ and $Z=(X^{(2)},\dots,X^{(p)})$.

\paragraph{Outcome.}
Generate
\[
Y=\mu(X)+\varepsilon,\qquad \varepsilon\sim\calN(0,1),
\]
where $\mu$ depends on $D$ and only $k=10$ coordinates of $Z$ through nonlinear additive terms and interactions.
The remaining coordinates enter only through a small nuisance component.

\paragraph{Policy and targets.}
We consider the deterministic shift
\[
T_\delta(d,z)=(d+\delta,z),\qquad X_\delta=T_\delta(X),\quad X\sim P_0,
\]
and the symmetric policy path
\[
\theta_0(\delta)=\bbE[\mu(D+\delta,Z)]-\bbE[\mu(D-\delta,Z)],\qquad \delta\in[0,\delta_{\max}].
\]
Because $P_\delta=(T_\delta)_\#P_0$ is a known pushforward, the APE/path ratios can be constructed from the observational data score
(Appendix~\ref{sec:ape_pushforward}). Time-SMR is also applicable using endpoint samples obtained by shifting $D$.

\begin{table}[t]
\centering
\caption{Case~1 results.}
\label{tab:case1_results}

\begin{tabular}{lrrr}
\toprule
\multicolumn{4}{l}{\textbf{AME} (true = 2.214)}\\
\midrule
Metric & Data-SMR & Time-SMR & Riesz reg. \\
\midrule
Bias & 0.299 & -0.198 & 22.828 \\
MSE & 3.272 & 10.736 & 2015.537 \\
Cov. (95\%) & 88.0\% & 84.7\% & 79.2\% \\
\bottomrule
\end{tabular}
\begin{tabular}{lrrr}
\toprule
\multicolumn{4}{l}{\textbf{APE} (true = 4.042)}\\
\midrule
Metric & Data-SMR & Time-SMR & Riesz reg. \\
\midrule
Bias & 3.465 & -1.162 & 27.828 \\
MSE & 133.853 & 3.872 & 3066.051 \\
Cov. (95\%) & 85.6\% & 16.2\% & 77.7\% \\
\bottomrule
\end{tabular}
\end{table}

Table~\ref{tab:case1_results} shows that Data-SMR yields the smallest AME MSE, while Riesz regression is highly unstable.
For APE, Time-SMR attains a much smaller MSE than Data-SMR but exhibits severe undercoverage, suggesting overly narrow confidence intervals.
Data-SMR has larger bias/MSE for APE but substantially closer-to-nominal coverage.

\subsection{Case~2: High-Dimensional Gaussian Mixture Regressors with Random-Feature Outcomes under a Shift Policy}
\label{sec:sim_hd_mog_rf_shift}

This design makes $P_0$ multimodal and uses a dense outcome regression.

\paragraph{Regressors.}
Let $X=(D,Z)\in\R^p$ with $p=50$.
Draw
\[
X\sim P_0 \equiv \frac{1}{K}\sum_{k=1}^K \calN(\mu^{(k)},\Sigma),
\]
with shared Toeplitz covariance $\Sigma$ and separated means $\mu^{(k)}$.
We use $K=4$.

\paragraph{Outcome.}
Generate
\[
Y=\mu(X)+\varepsilon,\qquad \varepsilon\sim\calN(0,1),
\]
where $\mu$ includes low-order terms in $D$ and a dense random-feature component
$x \mapsto v^\top \tanh(Wx+b)$ (200 random features in our experiments), plus interactions.

\paragraph{Policy and targets.}
We use the same shift $T_\delta(d,z)=(d+\delta,z)$ and the same AME/APE targets as in Case~1.
Data-SMR is applicable for APE via the pushforward construction, and Time-SMR is applicable via endpoint samples.

\begin{table}[t]
\centering
\caption{Case~2 results.}
\label{tab:case2_results}

\begin{tabular}{lrrr}
\toprule
\multicolumn{4}{l}{\textbf{AME} (true = 0.396)}\\
\midrule
Metric & Data-SMR & Time-SMR & Riesz reg. \\
\midrule
Bias & 0.128  & 0.187 & 1.424 \\
MSE & 0.092 & 0.715 & 106.067 \\
Cov. (95\%) & 83.9\% & 86.7\% & 83.5\% \\
\bottomrule
\end{tabular}

\begin{tabular}{lrrr}
\toprule
\multicolumn{4}{l}{\textbf{APE} (true = 0.843)}\\
\midrule
Metric & Data-SMR & Time-SMR & Riesz reg. \\
\midrule
Bias & 0.206  & 0.212 & 0.669 \\
MSE & 0.687  & 0.097 & 128.748 \\
Cov. (95\%) & 86.2\% & 20.1\% & 82.9\% \\
\bottomrule
\end{tabular}
\end{table}

Table~\ref{tab:case2_results} shows a similar pattern: Data-SMR performs best for AME, while Riesz regression is unstable.
For APE, Time-SMR has the smallest MSE but exhibits severe undercoverage.
Data-SMR has larger MSE but markedly better coverage.

\subsection{Case~3: Non-Pushforward Stochastic Policies via Resampling \texorpdfstring{$D\mid Z$}{dz}}
\label{sec:sim_stochastic_policy}

This design uses stochastic interventions that are not known pushforwards of $P_0$.
As a result, the APE/path representer cannot be constructed from the observational data score alone.

\paragraph{Observational law and outcome.}
Use the baseline Gaussian regressor design from Section~\ref{sec:experiments}:
\[
X=(D,Z)\sim \calN(0,\Sigma),\qquad D=X^{(1)},\ Z=(X^{(2)},X^{(3)}),
\]
and generate $Y=\mu(X)+\varepsilon$ with the same $\mu$ and $\varepsilon\sim\calN(0,1)$.

\paragraph{Stochastic policies.}
Fix $\delta_{\max}>0$.
For each $\delta\in[0,\delta_{\max}]$, define $P_{+\delta}$ and $P_{-\delta}$ by drawing $Z\sim p_0(z)$ and resampling
\[
D^{(\pm,\delta)}\mid Z=z \sim \calN\bigl(m_{\pm,\delta}(z), s_{\pm}(z)^2\bigr),
\qquad X^{(\pm,\delta)}=(D^{(\pm,\delta)},Z).
\]
We set
\[
m_{\pm,\delta}(z)=\bbE_{P_0}[D\mid Z=z]\pm \delta\, b(z),
\qquad
b(z)=\tanh(z^{(1)})+\tfrac{1}{2}\sin(z^{(2)}),
\]
and
\[
s_{\pm}(z)=\sqrt{\Var_{P_0}(D\mid Z=z)}\Bigl(0.8+0.4\,\sigma(\pm z^{(1)})\Bigr),
\qquad
\sigma(u)=\frac{1}{1+e^{-u}}.
\]
Then $p_{\pm,\delta}(d,z)=p_0(z)\,p_{\pm,\delta}(d\mid z)$.

\paragraph{Non-pushforward implication.}
Because the policy changes the conditional law of $D\mid Z$, there is generally no deterministic map $T_\delta$ such that
$X^{(\pm,\delta)}=T_\delta(X)$ with $X\sim P_0$.
Thus the ratio $p_{\pm,\delta}/p_0$ is not determined by $p_0$ (or its data score) alone, and the data-score-based APE construction is unavailable.
Time-SMR applies by treating $(q,p)=(p_{\pm,\delta},p_0)$ and using simulated samples from $p_{\pm,\delta}$ together with observed samples from $p_0$.

\paragraph{APE and results.}
Define
\[
\theta_0(\delta)=\bbE_{X\sim p_{+,\delta}}[\gamma_0(X)]-\bbE_{X\sim p_{-,\delta}}[\gamma_0(X)],
\qquad \delta\in[0,\delta_{\max}],
\]
and approximate the truth by Monte Carlo draws from $p_{+,\delta}$ and $p_{-,\delta}$.

\begin{table}[t]
\centering
\caption{Case~3 results.}
\label{tab:case3_results}

\begin{tabular}{lrr}
\toprule
\multicolumn{3}{l}{\textbf{AME} (true = 2.225)}\\
\midrule
Metric & Time-SMR (without joint training) & Time-SMR (with joint training)\\
\midrule
Bias & 0.012 & 0.012 \\
MSE & 0.147 & 0.147 \\
Cov. (95\%) & 86.0\% & 86.0\% \\
\bottomrule
\end{tabular}

\begin{tabular}{lrr}
\toprule
\multicolumn{3}{l}{\textbf{APE} (true = 1.324)}\\
\midrule
Metric & Time-SMR (without joint training) & Time-SMR (with joint training) \\
\midrule
Bias & -0.949 & -0.511 \\
MSE & 7.392 & 1.254 \\
Cov. (95\%) & 82.0\% & 74.0\% \\
\bottomrule
\end{tabular}
\end{table}

Table~\ref{tab:case3_results} illustrates the non-pushforward regime, where the data-score-based APE/policy-path construction is unavailable.
AME performance is essentially unchanged by joint training.
For APE, joint training reduces bias and MSE, but coverage remains below nominal for both variants (74--82\%).

\section{Empirical Application: Policy Paths of Monetary Policy Shifts on Equity Returns}
\label{app:empirical_finance}

This appendix describes an empirical application of ScoreMatchingRiesz using publicly available U.S.\ financial and macroeconomic data.
The empirical objective is \emph{methodological}: we demonstrate how a single score-based Riesz pipeline produces (i) an AME,
(ii) a finite-shift APE, and (iii) an entire \emph{policy path} $\delta \mapsto \theta_h(\delta)$, and we compare these outputs
to a standard \emph{Riesz regression} baseline in terms of point estimates, confidence intervals, and policy-path visualizations.
No ``true'' effects are assumed or required.
Because the data form a dependent time series, the i.i.d.\ theory in the main text does not directly cover this setting, and we interpret the results as a methodological demonstration.
We do not rely on train--test splits; all estimation uses the full sample.

\subsection{What We Estimate and Why: Local Projections Meet Policy Paths}
\label{app:empirical_finance:goal}

Empirical work in macro-finance often seeks to quantify how an observable policy variable (e.g.\ an interest rate move)
is associated with future asset returns. A standard approach is the \emph{local projection} (LP) method \citep{Jorda2005estimationand},
which estimates horizon-specific responses by regressing future outcomes on current policy moves and controls.
In the simplest linear LP, one estimates a coefficient $\beta_h$ from
\[
Y_{t,h} = \beta_h D_t + \text{controls} + \text{error},
\]
where $Y_{t,h}$ is a forward outcome over horizon $h$, and $D_t$ is a policy move at time $t$.

Our goal is to move beyond a linear coefficient and estimate a \emph{nonlinear, policy-shift response curve}
that maps the size of a policy shift $\delta$ to the implied change in outcomes.
This curve is the \emph{policy path}:
\[
\delta \longmapsto \theta_h(\delta),
\]
where $\theta_h(\delta)$ is a symmetric effect of shifting the policy move by $+\delta$ versus $-\delta$.
The path nests both local and global objects:
\begin{itemize}
\item The \textbf{AME} corresponds to the \emph{local slope} of the path at the origin.
\item The \textbf{APE} corresponds to the path evaluated at a particular policy shift $\delta_{\mathrm{policy}}$ (e.g.\ $25$ bps).
\end{itemize}
This design makes it natural to compare different Riesz-representer estimators:
ScoreMatchingRiesz constructs the path by estimating an AME score and integrating it to obtain density ratios,
while Riesz regression estimates the representer directly by minimizing a Riesz loss.

\subsection{Data Sources and Preprocessing}
\label{app:empirical_finance:data}

\paragraph{Data sources (replicable, public).}
We combine:
(i) macroeconomic, monetary, and financial indicators from the Federal Reserve Economic Data (FRED) database,
and (ii) asset prices from Yahoo Finance (downloaded via \texttt{yfinance}).
We choose these sources to ensure replicability with a short script.

\paragraph{Sampling frequency and alignment.}
All series are converted to monthly frequency and aligned on end-of-month dates.
Daily asset prices are converted to month-end adjusted close prices and then to monthly log returns.
Daily macro-financial series (e.g.\ volatility indices) are aggregated to monthly averages.
We merge all series on the monthly index and drop months with missing values after constructing lags and forward outcomes.

\paragraph{Variables.}
Let $t=1,\dots,T$ index months. We define:
\begin{itemize}
\item \textbf{Policy variable.} $D_t$ is the \emph{monthly change} in an interest-rate proxy, measured in percentage points.
A baseline choice is
\[
D_t \coloneqq \Delta \mathrm{FF}_t,
\]
where $\mathrm{FF}_t$ is an effective federal funds rate series. We use changes rather than levels to reduce persistence
and to align with a ``policy-move'' interpretation.
\item \textbf{Market outcome.} Let $P_t$ denote the end-of-month adjusted close price of an equity index proxy (e.g.\ an S\&P 500 ETF).
Define the one-month log return $r_t\coloneqq \log P_t-\log P_{t-1}$.
We construct horizon-$h$ forward outcomes as cumulative returns:
\begin{align}
Y_{t,h}  \coloneqq  \sum_{j=1}^h r_{t+j},
\qquad h\in\{1,\dots,H\}.
\label{eq:lp_outcome_cum}
\end{align}
Optionally, one can use excess returns $r^e_t=r_t-r^f_t$ by subtracting a short-rate proxy.
\item \textbf{Controls and lag structure.} $Z_t$ collects macro-financial controls and lags, in the spirit of LP specifications:
\[
Z_t = \bigl(\text{macro controls at } t,\ \text{lags of } (D,Z)\bigr).
\]
The macro control set can include inflation, unemployment, industrial production growth, term spreads, and volatility proxies.
We include $p$ lags of key controls and $D_t$ (e.g.\ $p=12$ months) to capture state dependence and dynamics.
\item \textbf{Factor controls (PCA).}
To summarize common movements in a panel of asset returns, we extract a small number of latent factors by PCA.
Let $R_t\in\mathbb{R}^M$ be a panel of monthly log returns for a set of liquid ETFs (e.g.\ sector ETFs).
After standardizing each return series, we compute the first $K$ principal components:
\begin{align*}
F_t  =  \mathrm{PCA}_K(R_t)\in\mathbb{R}^K,
\end{align*}
and append $F_t$ (and optionally its lags) to $Z_t$.
This introduces a low-dimensional factor structure consistent with asset-pricing intuition while keeping the nuisance learning flexible.
\end{itemize}

\paragraph{Regressor vector.}
We define the regressor
\[
X_t  \coloneqq  (D_t, Z_t)\in\mathbb{R}^d,
\]
and the horizon-$h$ ``observation'' as $W_{t,h}\coloneqq (X_t, Y_{t,h})$.

\subsection{Targets and Their Riesz Representers: AME, APE, and Policy Paths}
\label{app:empirical_finance:targets}

For each horizon $h$, define the conditional mean (LP regression function)
\begin{align*}
\gamma_{0,h}(x) \coloneqq \mathbb{E}[Y_{t,h}\mid X_t=x],
\qquad x\in\mathbb{R}^d.
\end{align*}
We study three related parameters.

\subsubsection{AME as a Linear Functional and Its Representer}
\label{app:empirical_finance:ame}

The horizon-$h$ average marginal effect (AME) is
\begin{align*}
\theta^{\mathrm{AME}}_{0,h}
 \coloneqq 
\mathbb{E}\bigl[\partial_d \gamma_{0,h}(D_t,Z_t)\bigr],
\end{align*}
where $\partial_d$ denotes the partial derivative with respect to the policy coordinate $d$.
Under standard integration-by-parts conditions (e.g.\ sufficient smoothness and boundary decay),
this functional admits a Riesz representer $\alpha^{\mathrm{AME}}_0\in L_2(P_X)$ satisfying
\[
\mathbb{E}\bigl[\partial_d \gamma(D,Z)\bigr]  =  \mathbb{E}\bigl[\alpha^{\mathrm{AME}}_0(D,Z) \gamma(D,Z)\bigr]
\quad\text{for all }\gamma\in L_2(P_X),
\]
with the explicit form
\begin{align}
\alpha^{\mathrm{AME}}_0(D,Z)  =  - \partial_d \log p_0(D,Z),
\label{eq:ame_rr_score_finance}
\end{align}
where $p_0$ is the joint density of $X=(D,Z)$ under the stationary distribution of the observed process.
This identity motivates score-based estimation of the AME representer.

\subsubsection{Policy Shifts, APE, and the Policy Path}
\label{app:empirical_finance:path}

We consider \emph{smooth policy shifts} that translate the policy coordinate:
\[
\tau_\delta(d,z) \coloneqq (d+\delta,z),
\qquad \delta\in\mathbb{R}.
\]
Let $p_\delta$ be the distribution of $X_\delta=\tau_\delta(X)$ when $X\sim p_0$.
Then $p_\delta(d,z)=p_0(d-\delta,z)$ (ignoring boundary effects).

\paragraph{Policy path.}
For horizon $h$, define the symmetric policy-path parameter
\begin{align}
\theta_{0,h}(\delta)
 \coloneqq 
\mathbb{E}_{X\sim p_{\delta}}\bigl[\gamma_{0,h}(X)\bigr]
-
\mathbb{E}_{X\sim p_{-\delta}}\bigl[\gamma_{0,h}(X)\bigr].
\label{eq:path_def_finance}
\end{align}
By a change of variables, \eqref{eq:path_def_finance} is equivalently written as an expectation under $p_0$:
\begin{align}
\theta_{0,h}(\delta)
 = 
\mathbb{E}\bigl[\gamma_{0,h}(D_t+\delta,Z_t)-\gamma_{0,h}(D_t-\delta,Z_t)\bigr].
\label{eq:path_as_mdelta}
\end{align}
This shows that the policy path is a \emph{linear functional} of $\gamma_{0,h}$.

\paragraph{APE as a point on the path.}
Fix a policy magnitude of interest $\delta_{\mathrm{policy}}>0$ (e.g.\ $0.25$ percentage points, i.e.\ $25$ bps).
We define the finite-shift APE at horizon $h$ as
\begin{align}
\theta^{\mathrm{APE}}_{0,h}
 \coloneqq 
\theta_{0,h}(\delta_{\mathrm{policy}}).
\label{eq:ape_def_finance}
\end{align}

\paragraph{Path--AME link (local approximation).}
Differentiating \eqref{eq:path_as_mdelta} at $\delta=0$ yields
\[
\theta_{0,h}'(0)
=
\mathbb{E}\bigl[\partial_d\gamma_{0,h}(D,Z)\bigr]
+
\mathbb{E}\bigl[\partial_d\gamma_{0,h}(D,Z)\bigr]
=
2\theta^{\mathrm{AME}}_{0,h}.
\]
Hence
\begin{align}
\theta^{\mathrm{AME}}_{0,h} = \tfrac12 \theta_{0,h}'(0),
\qquad
\theta_{0,h}(\delta)\approx 2\delta\cdot\theta^{\mathrm{AME}}_{0,h}\ \text{ for small }\delta.
\label{eq:ame_path_relation}
\end{align}
This motivates visual checks that compare the estimated path to the local linear approximation $2\delta \widehat\theta^{\mathrm{AME}}_h$ near the origin.

\subsubsection{Riesz Representer for the Policy Path and Its Density-Ratio Form}
\label{app:empirical_finance:path_rr}

Define the moment map for the policy path:
\begin{align}
m_\delta(W_{t,h},\gamma)
 \coloneqq 
\gamma(D_t+\delta,Z_t)-\gamma(D_t-\delta,Z_t).
\label{eq:m_delta}
\end{align}
This depends only on $X_t$ and is linear in $\gamma$.
By the Riesz representation theorem, there exists $\alpha_{0,\delta}\in L_2(P_X)$ such that
\begin{align}
\mathbb{E}\bigl[m_\delta(W_{t,h},\gamma)\bigr]
=
\mathbb{E}\bigl[\alpha_{0,\delta}(X_t) \gamma(X_t)\bigr]
\quad\text{for all }\gamma\in L_2(P_X).
\label{eq:riesz_path_def}
\end{align}
Importantly, because $m_\delta$ depends only on $X_t$, the representer $\alpha_{0,\delta}$ is \emph{common across horizons} $h$.
Thus, in both of our approaches below, representer estimation is done once (per $\delta$) and reused across all horizons.

\paragraph{Density-ratio representation under translation shifts.}
Define the importance weight
\[
r_\delta(x) \coloneqq \frac{p_\delta(x)}{p_0(x)}.
\]
Then
\[
\mathbb{E}_{X\sim p_\delta}[\gamma(X)]
=
\mathbb{E}_{X\sim p_0}\bigl[r_\delta(X) \gamma(X)\bigr],
\]
and therefore \eqref{eq:path_def_finance} implies
\begin{align*}
\alpha_{0,\delta}(x)  =  r_\delta(x)-r_{-\delta}(x).
\end{align*}
Under $p_\delta(d,z)=p_0(d-\delta,z)$, this ratio has the explicit form
\[
r_\delta(d,z) = \frac{p_0(d-\delta,z)}{p_0(d,z)}.
\]

\paragraph{From AME scores to ratios (the key ScoreMatchingRiesz identity).}
By the fundamental theorem of calculus,
\begin{align}
\log r_\delta(d,z)
&=
\log p_0(d-\delta,z)-\log p_0(d,z)
\notag\\
&=
-\int_0^\delta \partial_u\log p_0(d-u,z)\,du.
\label{eq:log_ratio_from_score}
\end{align}
Combining \eqref{eq:ame_rr_score_finance} and \eqref{eq:log_ratio_from_score} shows that estimating the AME score
$\partial_d\log p_0(d,z)$ enables construction of \emph{density ratios for all} $\delta$ by one-dimensional numerical integration,
and hence yields the entire policy path.

\subsection{Estimation I: ScoreMatchingRiesz for AME, APE, and the Policy Path}
\label{app:empirical_finance:estimation_smr}

We now describe the proposed estimator, emphasizing the ``one score model $\Rightarrow$ all $\delta$'' reuse.

\subsubsection{Step 1: Estimate the Treatment Score by Data-ScoreMatchingRiesz}
\label{app:empirical_finance:smr_dsm}

We estimate the treatment-direction score $\partial_d\log p_0(d,z)$ via Data-ScoreMatchingRiesz, which follows denoising score matching (DSM),
injecting Gaussian noise only in the policy coordinate.
Let $\varepsilon\sim\mathcal{N}(0,1)$ be independent and draw $\tau\sim \mathrm{Unif}[0,1]$.
Define
\[
\tilde D = D + \sigma(\tau)\varepsilon,
\]
where $\sigma(\tau)$ is a decreasing noise schedule (large noise at small $\tau$, small noise at $\tau$ near $1$).
We fit a neural model $s_\theta(\tilde d,z,\tau)$ minimizing
\begin{align*}
\mathcal{R}^{\mathrm{DSM}}(\theta)
=
\mathbb{E}\left[
\lambda(\tau)\left(
s_\theta(D+\sigma(\tau)\varepsilon, Z, \tau)
+
\frac{\varepsilon}{\sigma(\tau)}
\right)^2
\right].
\end{align*}
For each fixed $\tau$, the population minimizer equals the score of the $\sigma(\tau)$-smoothed density in the $\tilde d$ direction.
We then evaluate at a small-noise time $\tau_{\mathrm{eval}}\approx 1$ to approximate the unsmoothed score and define
\begin{align*}
\widehat{\alpha}^{\mathrm{AME}}(D,Z)
 \coloneqq 
- s_{\widehat\theta}(D,Z,\tau_{\mathrm{eval}}).
\end{align*}
Because this step uses only the regressor process $\{X_t\}$, it is performed once and reused for all horizons $h$.

\subsubsection{Step 2: Construct Ratios and Path Representers for a Grid of Policy Shifts}
\label{app:empirical_finance:smr_ratios}

Fix a grid $\mathcal{D}=\{\delta_1,\dots,\delta_M\}$ of nonnegative policy shifts (e.g.\ from $0$ to $1$ percentage point).
For each $\delta\in\mathcal{D}$, we estimate the log ratio using \eqref{eq:log_ratio_from_score} with the learned score:
\begin{align}
\widehat{\log r_\delta}(d,z)
 \coloneqq 
-\int_0^\delta s_{\widehat\theta}(d-u,z,\tau_{\mathrm{eval}})\,du,
\qquad
\widehat r_\delta(d,z)=\exp(\widehat{\log r_\delta}(d,z)).
\label{eq:ratio_hat}
\end{align}
The policy-path representer is then
\begin{align}
\widehat{\alpha}^{\mathrm{path}}_\delta(x)
 \coloneqq 
\widehat r_\delta(x)-\widehat r_{-\delta}(x).
\label{eq:alpha_path_hat_smr}
\end{align}
In implementation, the integral in \eqref{eq:ratio_hat} is computed by a trapezoidal rule on a fine grid in $u$.
To stabilize heavy tails, we optionally clip $\widehat{\log r_\delta}$ before exponentiation and/or clip $\widehat{\alpha}^{\mathrm{path}}_\delta$.

\subsubsection{Step 3: Horizon-Specific Regression and Orthogonal Scores}
\label{app:empirical_finance:smr_orthogonal}

For each horizon $h$, we estimate $\gamma_{0,h}$ by a flexible regressor $\widehat\gamma_h$ trained on the full sample
$\{(X_t,Y_{t,h})\}_{t=1}^T$.
Denote residuals $\widehat U_{t,h}\coloneqq Y_{t,h}-\widehat\gamma_h(X_t)$.

\paragraph{AME (direct).}
We estimate $\theta^{\mathrm{AME}}_{0,h}$ by the orthogonal-score estimator
\begin{align}
\widehat{\theta}^{\mathrm{AME}}_{h,\mathrm{SMR}}
=
\frac1T\sum_{t=1}^T
\left\{
\partial_d \widehat\gamma_h(X_t)
+
\widehat{\alpha}^{\mathrm{AME}}(X_t) \widehat U_{t,h}
\right\}.
\label{eq:ame_est_smr}
\end{align}
The derivative $\partial_d \widehat\gamma_h(X_t)$ is computed by automatic differentiation.

\paragraph{Policy path and APE.}
For each $\delta\in\mathcal{D}$, we estimate the policy path by
\begin{align}
\widehat{\theta}_{h,\mathrm{SMR}}(\delta)
=
\frac1T\sum_{t=1}^T
\Bigl[
\widehat\gamma_h(D_t+\delta,Z_t)-\widehat\gamma_h(D_t-\delta,Z_t)
+
\widehat{\alpha}^{\mathrm{path}}_\delta(X_t) \widehat U_{t,h}
\Bigr].
\label{eq:path_est_smr}
\end{align}
The APE is $\widehat{\theta}^{\mathrm{APE}}_{h,\mathrm{SMR}}=\widehat{\theta}_{h,\mathrm{SMR}}(\delta_{\mathrm{policy}})$.
Equation \eqref{eq:path_est_smr} yields the entire curve $\delta\mapsto\widehat\theta_{h,\mathrm{SMR}}(\delta)$ in one pass once ratios are available.

\subsection{Estimation II: Riesz Regression Baseline for AME, APE, and the Policy Path}
\label{app:empirical_finance:estimation_rr}

We compare ScoreMatchingRiesz to a Riesz-regression baseline that estimates the policy-path representer
\emph{directly} from the defining Riesz property \eqref{eq:riesz_path_def}, without imposing a density-ratio structure.

\subsubsection{Riesz Regression Objective for the Policy-Path Representer}
\label{app:empirical_finance:rr_objective}

For a fixed $\delta$, define the functional $m_\delta$ as in \eqref{eq:m_delta}.
The Riesz representer $\alpha_{0,\delta}$ is the unique minimizer (in $L_2(P_X)$) of the population quadratic objective
\[
\mathcal{L}_\delta(\alpha)  =  \mathbb{E}\bigl[\alpha(X)^2\bigr] - 2\mathbb{E}\bigl[m_\delta(W,\alpha)\bigr] + \text{const.},
\]
where the additive constant does not depend on $\alpha$.
The corresponding sample objective (with ridge regularization) is
\begin{align}
\widehat{\alpha}_{\delta,\mathrm{RR}}
\in
\arg\min_{\alpha\in\mathcal{A}}
\left\{
\mathbb{E}_T\bigl[\alpha(X_t)^2\bigr]
-
2 \mathbb{E}_T\bigl[\alpha(D_t+\delta,Z_t)-\alpha(D_t-\delta,Z_t)\bigr]
+\lambda \|\alpha\|_{\mathcal{A}}^2
\right\},
\label{eq:rr_loss_path}
\end{align}
where $\mathbb{E}_T[\cdot]=T^{-1}\sum_{t=1}^T(\cdot)$, $\mathcal{A}$ is a function class (e.g.\ a sieve or a neural net),
and $\lambda>0$ is a regularization parameter.

\paragraph{Sieve implementation (closed form).}
In our implementation we use a linear sieve $\alpha(x)=\beta^\top\phi(x)$ with feature map $\phi(x)\in\mathbb{R}^p$
(e.g.\ linear or low-degree polynomial features, or random features).
Plugging into \eqref{eq:rr_loss_path} yields the ridge-regularized quadratic problem with closed-form solution
\begin{align*}
\widehat\beta_\delta
=
\left(\widehat\Sigma + \lambda I\right)^{-1}\widehat b_\delta,
\qquad
\widehat\Sigma=\mathbb{E}_T[\phi(X_t)\phi(X_t)^\top],
\quad
\widehat b_\delta=\mathbb{E}_T[\phi(D_t+\delta,Z_t)-\phi(D_t-\delta,Z_t)].
\end{align*}
We then set $\widehat{\alpha}^{\mathrm{path}}_{\delta,\mathrm{RR}}(x)=\widehat\beta_\delta^\top\phi(x)$.
As with the score-based method, this representer depends only on $\{X_t\}$ and is common across horizons $h$.

\subsubsection{Orthogonal Policy-Path Estimator Using the Riesz Regression Representer}
\label{app:empirical_finance:rr_orthogonal}

Given the same horizon-specific regression $\widehat\gamma_h$ and residuals $\widehat U_{t,h}$,
we estimate the policy path by
\begin{align}
\widehat{\theta}_{h,\mathrm{RR}}(\delta)
=
\frac1T\sum_{t=1}^T
\Bigl[
\widehat\gamma_h(D_t+\delta,Z_t)-\widehat\gamma_h(D_t-\delta,Z_t)
+
\widehat{\alpha}^{\mathrm{path}}_{\delta,\mathrm{RR}}(X_t) \widehat U_{t,h}
\Bigr],
\qquad \delta\in\mathcal{D}.
\label{eq:path_est_rr}
\end{align}
The APE baseline is $\widehat{\theta}^{\mathrm{APE}}_{h,\mathrm{RR}}=\widehat{\theta}_{h,\mathrm{RR}}(\delta_{\mathrm{policy}})$.

\paragraph{AME from the path (baseline and cross-check).}
Because $\theta^{\mathrm{AME}}_{0,h}=\tfrac12\theta_{0,h}'(0)$ by \eqref{eq:ame_path_relation},
we compute a path-based AME estimate for the Riesz regression baseline using a small $\delta_{\mathrm{AME}}$:
\begin{align}
\widehat{\theta}^{\mathrm{AME}}_{h,\mathrm{RR}}
 \coloneqq 
\frac{1}{2\delta_{\mathrm{AME}}} \widehat{\theta}_{h,\mathrm{RR}}(\delta_{\mathrm{AME}}).
\label{eq:ame_from_path_rr}
\end{align}
We report \eqref{eq:ame_from_path_rr} alongside the direct AME estimate \eqref{eq:ame_est_smr} from ScoreMatchingRiesz,
and we also check the local linear approximation $2\delta \widehat\theta^{\mathrm{AME}}_h$ against the estimated paths near $\delta=0$.

\subsection{Inference for Time-Series Data: Pointwise HAC Confidence Bands}
\label{app:empirical_finance:inference}

Because $\{(X_t,Y_{t,h})\}$ is a time series, we compute heteroskedasticity-and-autocorrelation consistent (HAC)
standard errors for both methods, treating the orthogonal-score summands as estimated influence functions.

\paragraph{Estimated influence functions.}
Define the centered summands for the policy path:
\begin{align*}
\widehat\psi_{t,h,\mathrm{SMR}}(\delta)
&\coloneqq
\widehat\gamma_h(D_t+\delta,Z_t)-\widehat\gamma_h(D_t-\delta,Z_t)
+
\widehat{\alpha}^{\mathrm{path}}_{\delta}(X_t) \widehat U_{t,h}
-
\widehat{\theta}_{h,\mathrm{SMR}}(\delta),
\widehat\psi_{t,h,\mathrm{RR}}(\delta)
&\coloneqq
\widehat\gamma_h(D_t+\delta,Z_t)-\widehat\gamma_h(D_t-\delta,Z_t)
+
\widehat{\alpha}^{\mathrm{path}}_{\delta,\mathrm{RR}}(X_t) \widehat U_{t,h}
-
\widehat{\theta}_{h,\mathrm{RR}}(\delta).
\end{align*}
These sequences have mean approximately zero by construction. We estimate the variance of $T^{-1}\sum_t \widehat\psi_{t,h}(\delta)$
by a Newey--West estimator:
\begin{align}
\widehat{\Omega}_h(\delta)
=
\widehat\Gamma_0(\delta)
+
2\sum_{\ell=1}^{L_h} w_\ell \widehat\Gamma_\ell(\delta),
\qquad
\widehat\Gamma_\ell(\delta)=\frac{1}{T}\sum_{t=\ell+1}^T \widehat\psi_{t,h}(\delta)\widehat\psi_{t-\ell,h}(\delta),
\qquad
w_\ell=1-\frac{\ell}{L_h+1},
\label{eq:newey_west}
\end{align}
and set the standard error to $\widehat{\mathrm{se}}_h(\delta)=\sqrt{\widehat{\Omega}_h(\delta)/T}$.
We use a horizon-dependent lag truncation $L_h$ (e.g.\ $L_h=\max\{6,h\}$ for monthly data) and report pointwise 95\% intervals
\begin{align*}
\widehat\theta_h(\delta)\pm 1.96 \widehat{\mathrm{se}}_h(\delta),
\end{align*}
computed separately for ScoreMatchingRiesz and for Riesz regression.
These are \emph{pointwise} bands in $\delta$ (no multiple-testing adjustment).

\paragraph{Remarks on nuisance estimation.}
We do not split the sample into train and test subsets. All nuisance functions are trained on the full sample.
Because the i.i.d.\ assumptions underlying the main theoretical results do not directly apply to dependent time-series data, we view the HAC bands as a pragmatic approximation commonly used in empirical macro-finance when first-stage learners are regularized, and we use them primarily for comparing representer estimators.
As a robustness check, one can implement block cross-fitting by splitting the time index into contiguous folds and evaluating the orthogonal scores on held-out blocks.

\subsection{Reporting and Visualization: Comparing Estimators and Confidence Bands}
\label{app:empirical_finance:reporting}

Our empirical outputs consist of:

\paragraph{Numerical comparison (AME and APE).}
For each horizon $h$, we report a table containing:
\begin{itemize}
\item $\widehat\theta^{\mathrm{AME}}_{h,\mathrm{SMR}}$ from \eqref{eq:ame_est_smr} with HAC 95\% CI.
\item $\widehat\theta^{\mathrm{AME}}_{h,\mathrm{RR}}$ from \eqref{eq:ame_from_path_rr} with HAC 95\% CI (using the HAC CI for $\widehat\theta_{h,\mathrm{RR}}(\delta_{\mathrm{AME}})$ rescaled by $1/(2\delta_{\mathrm{AME}})$).
\item $\widehat\theta^{\mathrm{APE}}_{h,\mathrm{SMR}}=\widehat\theta_{h,\mathrm{SMR}}(\delta_{\mathrm{policy}})$ with HAC 95\% CI.
\item $\widehat\theta^{\mathrm{APE}}_{h,\mathrm{RR}}=\widehat\theta_{h,\mathrm{RR}}(\delta_{\mathrm{policy}})$ with HAC 95\% CI.
\end{itemize}
This provides a direct comparison of point estimates and uncertainty at economically meaningful policy magnitudes.

\paragraph{Policy-path plots.}
For selected horizons (e.g.\ $h\in\{1,3,6,12\}$ months), we plot the entire policy path for both methods:
\begin{itemize}
\item The ScoreMatchingRiesz path $\delta\mapsto\widehat\theta_{h,\mathrm{SMR}}(\delta)$ with its pointwise 95\% band.
\item The Riesz regression path $\delta\mapsto\widehat\theta_{h,\mathrm{RR}}(\delta)$ with its pointwise 95\% band.
\end{itemize}
Optionally, we overlay the local linear approximation $2\delta \widehat\theta^{\mathrm{AME}}_{h,\mathrm{SMR}}$
to highlight the degree of nonlinearity and the coherence between AME and the global path.

\begin{figure}[t]
\centering
\caption{
Illustration of the policy path $\delta\mapsto\widehat\theta_h(\delta)$ at a fixed horizon $h$.
The solid lines show point estimates from ScoreMatchingRiesz and from the Riesz regression baseline.
Shaded areas are pointwise HAC 95\% confidence bands.
A dotted line may show the local approximation $2\delta \widehat\theta^{\mathrm{AME}}_h$ near $\delta=0$.
}
\label{fig:policy_path_comparison}
\end{figure}

\paragraph{Interpretation of differences.}
The two approaches differ only in \emph{how} the representer is estimated:
\begin{itemize}
\item ScoreMatchingRiesz estimates the AME score once and constructs $\widehat{\alpha}^{\mathrm{path}}_\delta$ for all $\delta$ via the integral identity \eqref{eq:log_ratio_from_score}.
This enforces a density-ratio structure across $\delta$ and couples the entire path through a single learned score.
\item Riesz regression estimates $\widehat{\alpha}^{\mathrm{path}}_{\delta,\mathrm{RR}}$ directly by the Riesz loss \eqref{eq:rr_loss_path}, separately for each $\delta$ (or via a closed-form sieve),
without imposing the density-ratio structure.
\end{itemize}
Comparing the resulting paths and confidence bands helps diagnose whether imposing the score-to-ratio structure stabilizes estimation,
particularly for larger policy shifts where representers can become heavy-tailed.

\begin{table}
\caption{Policy path table (wide): rows are grouped by $\delta$ with three subrows (theta, se, CI).}
\label{tab:policy_path_wide_3rows}
\centering
\scalebox{0.5}{
\begin{tabular}{llllllllllllll}
\toprule
 &  & $h=1$ & $h=2$ & $h=3$ & $h=4$ & $h=5$ & $h=6$ & $h=7$ & $h=8$ & $h=9$ & $h=10$ & $h=11$ & $h=12$ \\
$\delta$ &  &  &  &  &  &  &  &  &  &  &  &  &  \\
\midrule
\multirow[t]{3}{*}{-1.00} & theta & -0.004 & 0.041 & 0.099 & -0.009 & -0.045 & -0.064 & -0.091 & -0.112 & -0.074 & -0.016 & -0.041 & -0.064 \\
 & se & 0.000 & 0.002 & 0.003 & 0.003 & 0.006 & 0.003 & 0.004 & 0.003 & 0.003 & 0.005 & 0.006 & 0.006 \\
 & $[\text{CI}_\text{low}, \text{CI}_\text{high}]$ & [-0.005, -0.004] & [0.037, 0.044] & [0.092, 0.106] & [-0.015, -0.002] & [-0.058, -0.033] & [-0.071, -0.057] & [-0.099, -0.083] & [-0.118, -0.107] & [-0.079, -0.069] & [-0.027, -0.006] & [-0.053, -0.029] & [-0.076, -0.053] \\
\cline{1-14}
\multirow[t]{3}{*}{-0.75} & theta & -0.003 & 0.029 & 0.074 & -0.007 & -0.033 & -0.056 & -0.074 & -0.088 & -0.059 & -0.016 & -0.037 & -0.054 \\
 & se & 0.000 & 0.001 & 0.003 & 0.003 & 0.005 & 0.003 & 0.003 & 0.002 & 0.002 & 0.004 & 0.005 & 0.004 \\
 & $[\text{CI}_\text{low}, \text{CI}_\text{high}]$ & [-0.003, -0.002] & [0.027, 0.032] & [0.069, 0.080] & [-0.012, -0.002] & [-0.042, -0.023] & [-0.061, -0.050] & [-0.080, -0.068] & [-0.093, -0.084] & [-0.063, -0.055] & [-0.024, -0.009] & [-0.046, -0.027] & [-0.063, -0.046] \\
\cline{1-14}
\multirow[t]{3}{*}{-0.50} & theta & -0.001 & 0.018 & 0.049 & -0.005 & -0.021 & -0.043 & -0.053 & -0.062 & -0.041 & -0.014 & -0.029 & -0.040 \\
 & se & 0.000 & 0.001 & 0.002 & 0.002 & 0.003 & 0.002 & 0.002 & 0.002 & 0.001 & 0.002 & 0.003 & 0.003 \\
 & $[\text{CI}_\text{low}, \text{CI}_\text{high}]$ & [-0.002, -0.001] & [0.017, 0.020] & [0.046, 0.053] & [-0.009, -0.002] & [-0.028, -0.014] & [-0.047, -0.038] & [-0.057, -0.049] & [-0.065, -0.059] & [-0.044, -0.038] & [-0.019, -0.009] & [-0.035, -0.022] & [-0.046, -0.034] \\
\cline{1-14}
\multirow[t]{3}{*}{-0.25} & theta & -0.000 & 0.009 & 0.024 & -0.003 & -0.010 & -0.025 & -0.028 & -0.032 & -0.021 & -0.008 & -0.017 & -0.022 \\
 & se & 0.000 & 0.000 & 0.001 & 0.001 & 0.002 & 0.002 & 0.001 & 0.001 & 0.001 & 0.001 & 0.002 & 0.002 \\
 & $[\text{CI}_\text{low}, \text{CI}_\text{high}]$ & [-0.001, -0.000] & [0.008, 0.010] & [0.023, 0.026] & [-0.004, -0.001] & [-0.014, -0.006] & [-0.028, -0.022] & [-0.031, -0.026] & [-0.034, -0.031] & [-0.023, -0.020] & [-0.011, -0.006] & [-0.021, -0.013] & [-0.025, -0.019] \\
\cline{1-14}
\multirow[t]{3}{*}{0.00} & theta & 0.000 & 0.000 & 0.000 & 0.000 & 0.000 & 0.000 & 0.000 & 0.000 & 0.000 & 0.000 & 0.000 & 0.000 \\
 & se & 0.000 & 0.000 & 0.000 & 0.000 & 0.000 & 0.000 & 0.000 & 0.000 & 0.000 & 0.000 & 0.000 & 0.000 \\
 & $[\text{CI}_\text{low}, \text{CI}_\text{high}]$ & [0.000, 0.000] & [0.000, 0.000] & [0.000, 0.000] & [0.000, 0.000] & [0.000, 0.000] & [0.000, 0.000] & [0.000, 0.000] & [0.000, 0.000] & [0.000, 0.000] & [0.000, 0.000] & [0.000, 0.000] & [0.000, 0.000] \\
\cline{1-14}
\multirow[t]{3}{*}{0.25} & theta & -0.000 & -0.008 & -0.024 & 0.004 & 0.007 & 0.024 & 0.031 & 0.034 & 0.024 & 0.012 & 0.017 & 0.026 \\
 & se & 0.000 & 0.001 & 0.001 & 0.001 & 0.001 & 0.001 & 0.001 & 0.001 & 0.001 & 0.001 & 0.002 & 0.002 \\
 & $[\text{CI}_\text{low}, \text{CI}_\text{high}]$ & [-0.000, 0.000] & [-0.009, -0.007] & [-0.026, -0.022] & [0.003, 0.006] & [0.005, 0.009] & [0.021, 0.026] & [0.029, 0.033] & [0.032, 0.036] & [0.022, 0.025] & [0.009, 0.015] & [0.014, 0.020] & [0.022, 0.029] \\
\cline{1-14}
\multirow[t]{3}{*}{0.50} & theta & -0.000 & -0.015 & -0.046 & 0.009 & 0.014 & 0.053 & 0.066 & 0.070 & 0.048 & 0.026 & 0.039 & 0.054 \\
 & se & 0.000 & 0.001 & 0.002 & 0.002 & 0.003 & 0.002 & 0.002 & 0.002 & 0.002 & 0.002 & 0.003 & 0.003 \\
 & $[\text{CI}_\text{low}, \text{CI}_\text{high}]$ & [-0.001, -0.000] & [-0.017, -0.013] & [-0.050, -0.042] & [0.005, 0.012] & [0.008, 0.019] & [0.049, 0.057] & [0.062, 0.069] & [0.066, 0.073] & [0.045, 0.051] & [0.021, 0.031] & [0.032, 0.045] & [0.047, 0.060] \\
\cline{1-14}
\multirow[t]{3}{*}{0.75} & theta & -0.001 & -0.021 & -0.067 & 0.013 & 0.019 & 0.084 & 0.101 & 0.106 & 0.073 & 0.041 & 0.063 & 0.084 \\
 & se & 0.000 & 0.001 & 0.003 & 0.003 & 0.004 & 0.003 & 0.003 & 0.002 & 0.002 & 0.003 & 0.005 & 0.005 \\
 & $[\text{CI}_\text{low}, \text{CI}_\text{high}]$ & [-0.001, -0.000] & [-0.023, -0.018] & [-0.072, -0.061] & [0.008, 0.018] & [0.011, 0.027] & [0.078, 0.089] & [0.095, 0.107] & [0.101, 0.110] & [0.068, 0.077] & [0.034, 0.047] & [0.053, 0.073] & [0.074, 0.093] \\
\cline{1-14}
\multirow[t]{3}{*}{1.00} & theta & -0.001 & -0.025 & -0.086 & 0.018 & 0.023 & 0.116 & 0.137 & 0.142 & 0.098 & 0.057 & 0.090 & 0.115 \\
 & se & 0.000 & 0.002 & 0.004 & 0.004 & 0.005 & 0.004 & 0.004 & 0.003 & 0.003 & 0.004 & 0.007 & 0.007 \\
 & $[\text{CI}_\text{low}, \text{CI}_\text{high}]$ & [-0.002, -0.001] & [-0.028, -0.022] & [-0.093, -0.079] & [0.011, 0.025] & [0.013, 0.034] & [0.109, 0.123] & [0.130, 0.145] & [0.136, 0.148] & [0.092, 0.104] & [0.048, 0.065] & [0.077, 0.102] & [0.102, 0.128] \\
\cline{1-14}
\bottomrule
\end{tabular}
}
\end{table}

\subsection{Implementation Details for Table \ref{tab:policy_path_wide_3rows}}
\label{app:empirical_finance:impl_table}

Table \ref{tab:policy_path_wide_3rows} reports the estimated \emph{one-sided} policy-shift path
\[
\theta_{0,h}(\delta,0) \coloneqq 
\mathbb{E}\left[\gamma_{0,h}(D_t+\delta,Z_t)-\gamma_{0,h}(D_t,Z_t)\right],
\]
evaluated on a grid of shifts $\delta$ (in percentage points) and horizons $h\in\{1,\dots,12\}$ months.
This is the object produced by the estimation routine in our implementation.
The table is organized in wide format with \emph{rows grouped by} $\delta$ and \emph{three subrows per} $\delta$:
the point estimate $\widehat\theta_h(\delta,0)$, its HAC standard error, and the pointwise 95\% interval.

\paragraph{Shift grid and horizons.}
We use a symmetric shift grid
\[
\delta\in\{-1.00,-0.75,-0.50,-0.25,0,0.25,0.50,0.75,1.00\},
\]
so $\delta=0.25$ corresponds to a 25 basis point move, and $H=12$.
For each horizon $h$, the LP-style outcome is the $h$-month cumulative (excess) return defined in \eqref{eq:lp_outcome_cum}.
As is standard with forward cumulative outcomes, we drop the last $H$ observations to avoid missing forward sums.

\paragraph{Nuisance estimation and reuse across horizons.}
The estimates in Table \ref{tab:policy_path_wide_3rows} are generated by the ScoreMatchingRiesz pipeline:
\begin{itemize}
\item A \textbf{single} data score network is fit once on the regressor process $\{X_t=(D_t,Z_t)\}$.
In code, the Data-ScoreMatchingRiesz model is an MLP with two hidden layers (64,64) and ELU activations,
trained for 4{,}000 steps with minibatch size 256.
Gaussian noise is injected only in the (standardized) policy coordinate with $\sigma\sim\mathrm{Unif}[0.05,0.5]$,
and the weighting uses $\lambda(\sigma)=\sigma^2$.
\item For each horizon $h$, a separate outcome regression $\widehat\gamma_h(x)$ is trained on the full sample
with an MLP (two hidden layers (128,128), ELU activations, 2{,}000 epochs).
The derivative $\partial_d\widehat\gamma_h(X_t)$ used for AME is computed by automatic differentiation.
\item For each $\delta$ on the grid, the learned score is integrated numerically to produce $\widehat{\log r_\delta}(x)$
and hence $\widehat r_\delta(x)$, using a trapezoidal rule with 64 integration steps.
These ratios enter the orthogonal score for the one-sided shift effect $\theta_h(\delta,0)$ via the correction term
$\mathbb{E}[(r_\delta(X_t)-1)\cdot (Y_{t,h}-\widehat\gamma_h(X_t))]$.
\end{itemize}
Because the data score is trained only once and depends only on $X_t$, it is reused across all horizons $h$ and all shifts $\delta$,
ensuring that the entire policy path surface $(\delta,h)\mapsto\widehat\theta_h(\delta,0)$ is generated coherently.

\paragraph{HAC inference.}
For each $(\delta,h)$, the standard error in Table \ref{tab:policy_path_wide_3rows} is a Newey--West HAC standard error
computed from the centered orthogonal-score summands (estimated influence-function proxies),
using a horizon-dependent truncation rule (in the implementation, $L_h=\max\{6,h\}$ for monthly data).
The reported intervals are pointwise 95\% CIs and do not adjust for multiple testing across $\delta$.

\paragraph{From one-sided paths to APEs and symmetric paths.}
While Table \ref{tab:policy_path_wide_3rows} reports $\widehat\theta_h(\delta,0)$, it directly yields economically interpretable contrasts.
Any two-point APE is recovered by differencing rows:
\[
\widehat\theta^{\mathrm{APE}}_h(\delta_1,\delta_2)=\widehat\theta_h(\delta_1,0)-\widehat\theta_h(\delta_2,0).
\]
In particular, the canonical ``25bp hike vs.\ 25bp cut'' contrast is
$\widehat\theta_h(0.25,0)-\widehat\theta_h(-0.25,0)$.
A symmetric path can likewise be formed by $\widehat\theta_h(\delta)=\widehat\theta_h(\delta,0)-\widehat\theta_h(-\delta,0)$.

\subsection{Empirical Patterns in Table \ref{tab:policy_path_wide_3rows}}
\label{app:empirical_finance:discussion_table}

We stress again that this appendix is \emph{methodological}: $D_t=\Delta\mathrm{FF}_t$ is an observed policy-rate change
and is not claimed to be an exogenous monetary policy shock.
Accordingly, we interpret Table \ref{tab:policy_path_wide_3rows} as documenting conditional predictive associations under the chosen controls,
and we use it primarily to illustrate how the proposed pipeline produces a rich policy-path object with uncertainty quantification.

\paragraph{(i) Basic implementation checks.}
The $\delta=0$ row is exactly zero at all horizons, as required by the definition of the one-sided shift effect
$\theta_h(0,0)=0$ and as a quick diagnostic that the table entries align with the intended target.

\paragraph{(ii) Horizon-dependent sign patterns and a sign reversal around $h=4$.}
A striking feature of Table \ref{tab:policy_path_wide_3rows} is that the sign of $\widehat\theta_h(\delta,0)$ depends strongly on the horizon.
For short horizons $h=2$ and $h=3$, the estimates are positive for negative shifts and negative for positive shifts.
For example, at $h=3$,
\[
\widehat\theta_3(-0.25,0)=0.0244
\quad\text{whereas}\quad
\widehat\theta_3(0.25,0)=-0.0236,
\]
and the corresponding CIs exclude zero.
From $h=4$ onward, the signs flip: negative shifts are associated with negative effects, and positive shifts with positive effects.
For instance, at $h=12$,
\[
\widehat\theta_{12}(-0.25,0)=-0.0219
\quad\text{whereas}\quad
\widehat\theta_{12}(0.25,0)=0.0256,
\]
again with tight CIs.

This horizon-dependent sign reversal is consistent with the economic fact that policy-rate changes are endogenous to the macro-financial state:
rate cuts tend to occur in deteriorating conditions (which also predict lower medium-run equity returns),
while rate hikes tend to occur in strengthening conditions (which predict higher medium-run returns).
Even with rich controls and lags, residual ``information effects'' and policy endogeneity can dominate the medium-run predictive patterns.
From the methodological perspective, the key point is that the policy path makes these sign changes visible rather than compressing them
into a single linear coefficient.

\paragraph{(iii) APE contrasts: ``hike vs.\ cut'' flips sign with horizon.}
Because Table \ref{tab:policy_path_wide_3rows} is one-sided, the finite-shift APE for a 25bp hike versus a 25bp cut is
\[
\widehat\theta_h(0.25,0)-\widehat\theta_h(-0.25,0).
\]
Using the table entries, this contrast is negative at short horizons and positive at longer horizons.
For example:
\[
h=2\colon  -0.0084 - 0.0086 \approx -0.0170,
\qquad
h=12\colon  0.0256 - (-0.0219) \approx 0.0475.
\]
Thus, the same policy magnitude can be associated with opposite-signed cumulative return changes depending on the horizon,
underscoring why reporting an entire $\delta$--$h$ surface is informative in macro-finance settings.

\begin{figure}[th]
    \centering
    \includegraphics[width=0.8\linewidth]{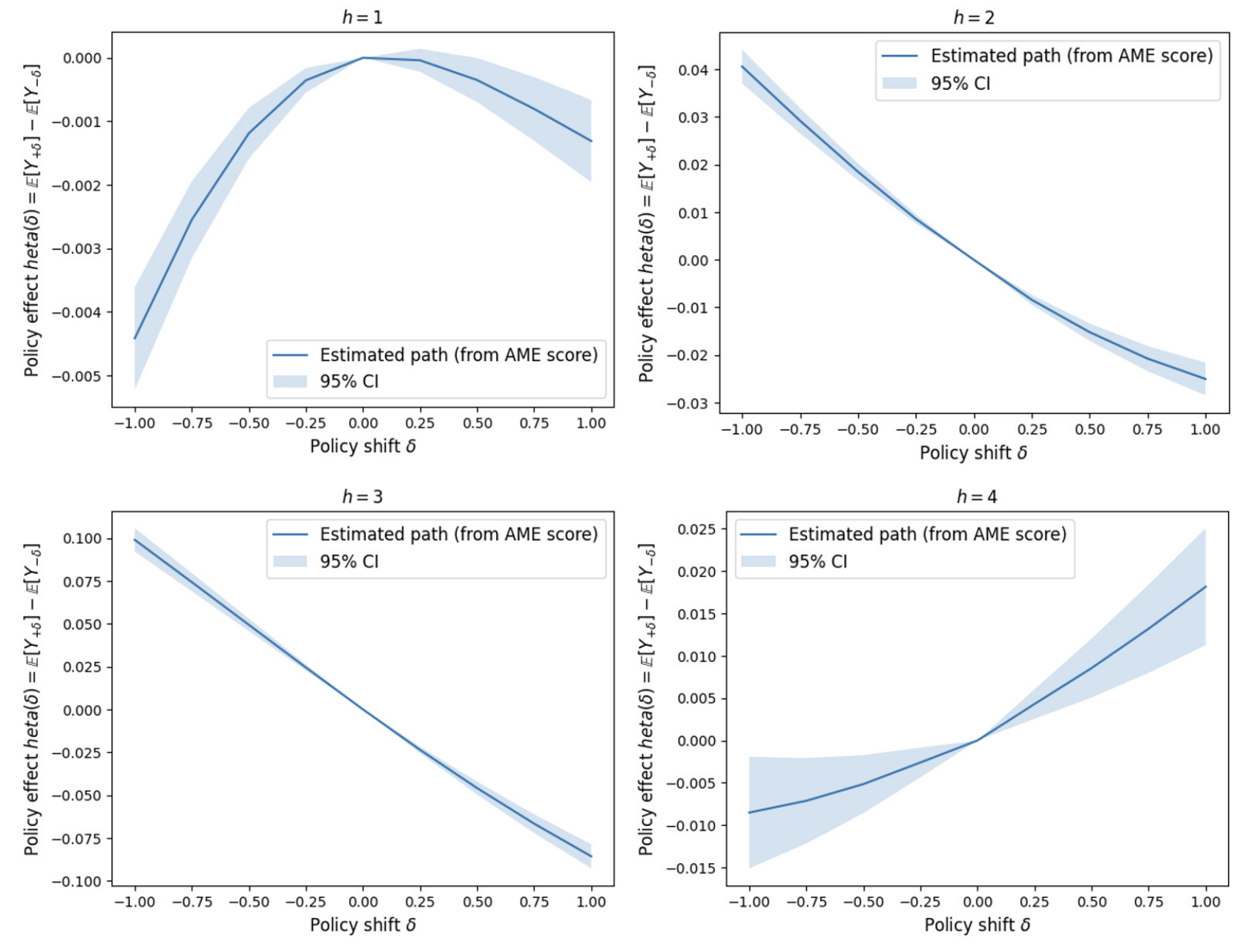}
    \caption{Policy path with $h=1,2,3,4$.}
    \label{fig:lp_figure1}
\end{figure}

\begin{figure}[th]
    \centering
    \includegraphics[width=0.8\linewidth]{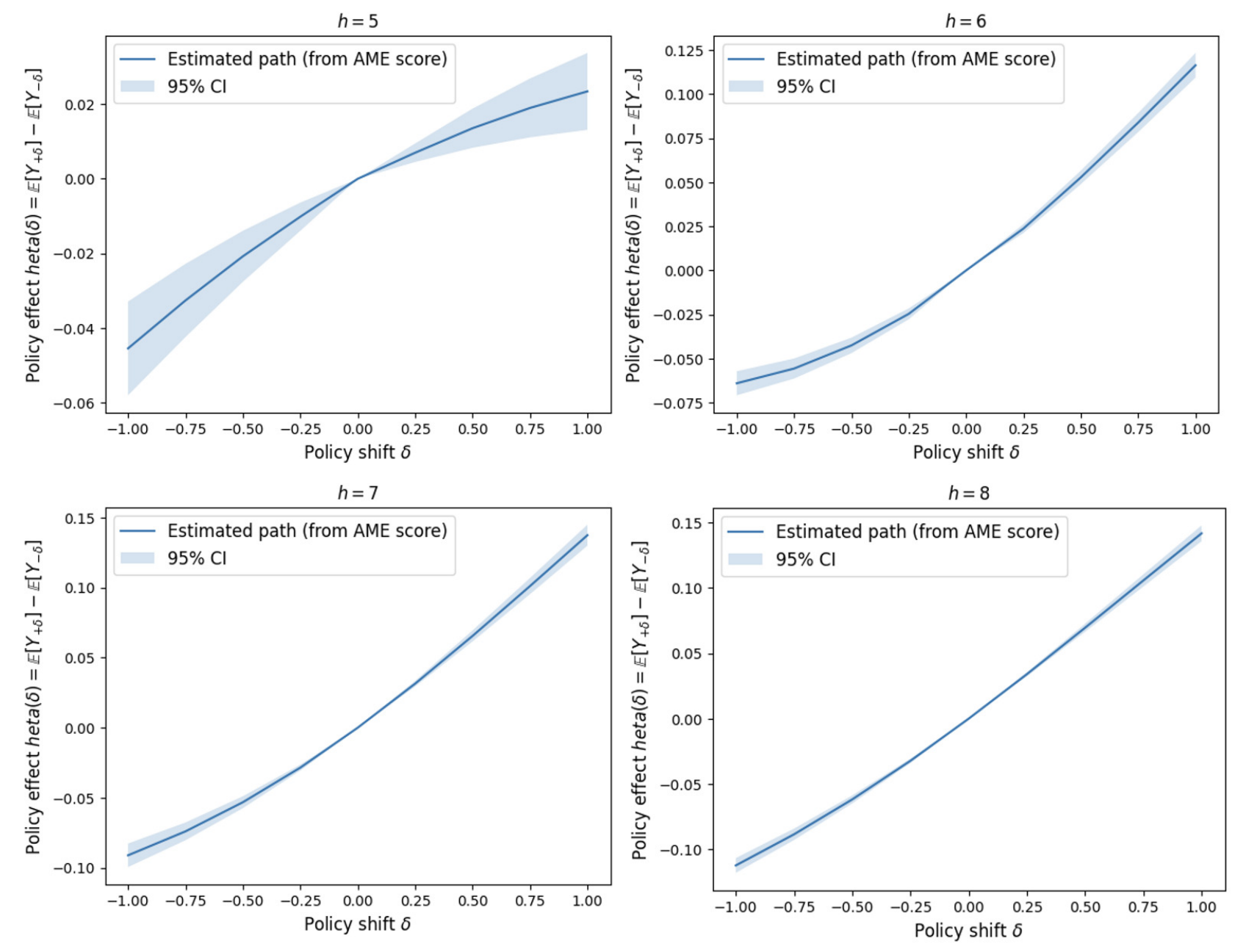}
    \caption{Policy path with $h=5,6,7,8$.}
    \label{fig:lp_figure2}
\end{figure}

\begin{figure}[th]
    \centering
    \includegraphics[width=0.8\linewidth]{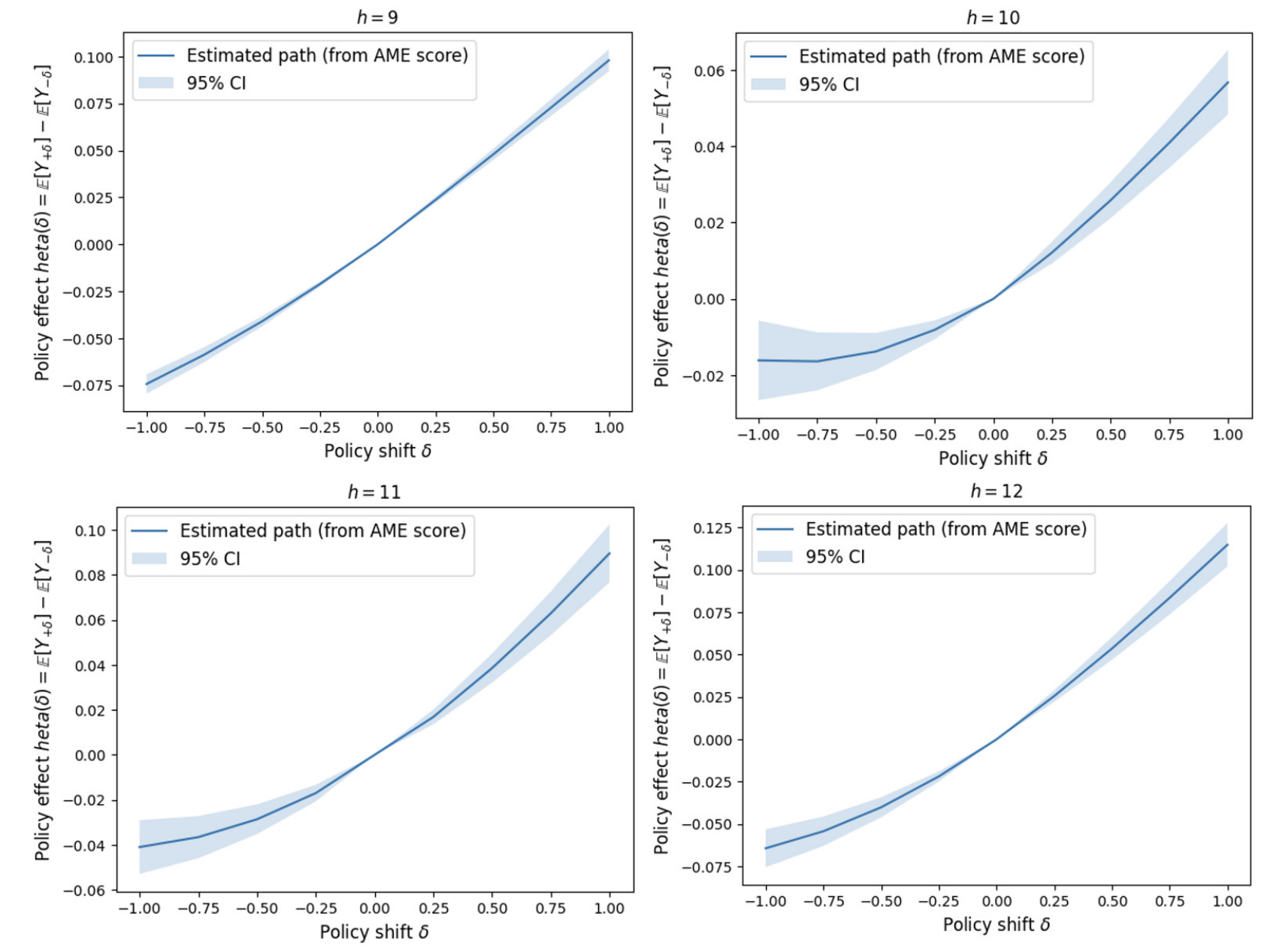}
    \caption{Policy path with $h=9,10,11,12$.}
    \label{fig:lp_figure3}
\end{figure}

\paragraph{(iv) Approximate linearity in $\delta$ and asymmetry between easing and tightening.}
Within many horizons, $\widehat\theta_h(\delta,0)$ scales roughly proportionally with $\delta$,
suggesting that in the range $|\delta|\le 1$ the estimated policy path is close to linear.
For example, at $h=12$ the positive-shift entries increase nearly linearly:
$0.0256$ (25bp), $0.0536$ (50bp), $0.0836$ (75bp), $0.1149$ (100bp).
However, the path is not perfectly odd or symmetric:
the magnitude for $\delta=1.00$ is $0.1149$, while for $\delta=-1.00$ it is $-0.0643$.
This asymmetry is economically plausible (policy easing episodes can occur in more adverse states than tightening episodes)
and statistically expected when the shifted distributions $p_\delta$ and $p_{-\delta}$ have different degrees of overlap with the baseline. We plot the policy pathes in Figures~\ref{fig:lp_figure1}--\ref{fig:lp_figure3}. 

\paragraph{(v) Precision and where uncertainty grows.}
Standard errors are small relative to effect magnitudes for most horizons and nonzero shifts,
and most pointwise CIs exclude zero, especially for $|\delta|\ge 0.25$ and $h\ge 2$.
At the same time, uncertainty generally increases with both horizon and $|\delta|$, which is consistent with:
(i) stronger serial dependence and accumulated noise in $h$-month cumulative outcomes, and
(ii) the increasing difficulty of extrapolation for larger distributional shifts.
From a representer perspective, this is precisely the region where density-ratio-based corrections can become heavy-tailed;
the table therefore doubles as a practical diagnostic of which parts of the policy path are empirically well-supported.

\paragraph{Takeaway for the methodological objective.}
Overall, Table \ref{tab:policy_path_wide_3rows} illustrates the central claim of ScoreMatchingRiesz in a real-data setting:
a single learned treatment score, reused across horizons, delivers a full policy-path object with coherent pointwise HAC uncertainty.
The resulting $\delta$--$h$ surface reveals horizon-dependent sign changes, near-linear scaling with $\delta$,
and asymmetries between tightening and easing---all of which would be obscured by reporting only an AME or a single APE.

\end{document}